\documentclass[aps,prd,10pt, twocolumn,superscriptaddress,noshowpacs,nofootinbib, preprintnumbers, longbibliography,bibnotes,hyperref]{revtex4-1}

\usepackage{amsmath}
\usepackage[T1]{fontenc}
\usepackage{amsfonts}
\usepackage{amssymb}
\usepackage{bbold}
\usepackage{epsfig}
\usepackage{graphicx}
\usepackage{bm}
\usepackage{array}
\usepackage{hyperref}
\usepackage{listings}
\usepackage{color}
\usepackage{float}
\usepackage[normalem]{ulem}

\usepackage{tikz,xcolor,hyperref}

\definecolor{lime}{HTML}{A6CE39}
\DeclareRobustCommand{\orcidicon}{\hspace{-1mm}
	\begin{tikzpicture}
	\draw[lime, fill=lime] (0,0) 
	circle [radius=0.16] 
	node[white] {{\fontfamily{qag}\selectfont \tiny \,ID}};
	\draw[white, fill=white] (-0.0525,0.095) 
	circle [radius=0.007];
	\end{tikzpicture}
	\hspace{-3mm}
}

\foreach \x in {A, ..., Z}{\expandafter\xdef\csname orcid\x\endcsname{\noexpand\href{https://orcid.org/\csname orcidauthor\x\endcsname}
			{\noexpand\orcidicon}}
}

\begin{document}

\title{Resonant Production of Light Sterile Neutrinos in Compact Binary Merger Remnants}

\author{Gar\dh ar Sigur\dh arson\orcidA{}}
%\email{jcs272@alumni.ku.dk}
\affiliation{Niels Bohr International Academy \& DARK, Niels Bohr Institute,\\University of Copenhagen, Blegdamsvej 17, 2100 Copenhagen, Denmark}

\author{Irene Tamborra\orcidB{}}
%\email{tamborra@nbi.ku.dk}
\affiliation{Niels Bohr International Academy \& DARK, Niels Bohr Institute,\\University of Copenhagen, Blegdamsvej 17, 2100 Copenhagen, Denmark}

\author{Meng-Ru Wu\orcidC{}}
%\email{mwu@gate.sinica.edu.tw}
\affiliation{Institute of Physics, Academia Sinica, Taipei, 11529, Taiwan}
\affiliation{Institute of Astronomy and Astrophysics, Academia Sinica, Taipei, 10617, Taiwan}

\date{\today}

\begin{abstract}
The existence of eV-mass sterile neutrinos is not ruled out because of persistent experimental anomalies. Upcoming multi-messenger detections of neutron-star merger remnants could provide indirect constraints on  the existence of these particles. We explore the active-sterile flavor conversion phenomenology in a two-flavor scenario ($1$ active $+1$ sterile species) as a function of the  sterile neutrino mixing parameters, neutrino emission angle from the accretion torus, and temporal evolution of the merger remnant. The torus geometry and the  neutron richness of the remnant are responsible for the occurrence of multiple resonant active-sterile conversions. The number of resonances strongly depends on the neutrino emission direction above or inside the remnant torus and leads to large production of sterile neutrinos (and no antineutrinos) in the proximity of the polar axis as well as more sterile antineutrinos  than neutrinos in the equatorial region. 
As the black hole torus evolves in time, the shallower baryon density is responsible for more adiabatic flavor conversion, leading to larger regions of the mass-mixing parameter space being affected by flavor mixing. 
Our findings imply that the production of sterile states could have indirect implications on the disk cooling rate, its outflows, and related electromagnetic observables which remain to be assessed.
\end{abstract}

\maketitle

%%%%%%%%%%%%%%%%%%%%%%%%%%%%%%%%%%
%%%%%%%%%%%%%%%%%%%%%%%%%%%%%%%%%%
\section{Introduction}
\label{sec:intro}
%%%%%%%%%%%%%%%%%%%%%%%%%%%%%%%%%%
%%%%%%%%%%%%%%%%%%%%%%%%%%%%%%%%%%

The coalescence of a neutron star (NS) with another NS or a black hole (BH) leads to the formation of a compact binary merger. Compact binary mergers lose angular momentum through the emission of gravitational waves. This conjecture was recently confirmed  through the detection of the  gravitational-wave event GW170817~\cite{Margutti:2020xbo,LIGOScientific:2017vwq,LIGOScientific:2017zic,LIGOScientific:2017ync,Goldstein:2017mmi,Savchenko:2017ffs,Margutti:2017cjl,Troja:2017nqp}. Electromagnetic follow-up observations across multiple wavebands of GW170817  confirmed that NS merger remnants are factories of the elements heavier than iron and harbor short gamma-ray bursts~\cite{Kasen:2017sxr,Drout:2017ijr,Cowperthwaite:2017dyu,Villar:2017wcc,Shibata:2017xdx,Metzger:2019zeh}.

While no neutrino has been observed from gravitational wave sources yet~\cite{Veske:2020yjt,IceCube:2020xks,Super-Kamiokande:2018dbf}, thermal neutrinos are copiously produced in binary NS mergers, with the neutrino luminosities reaching up to $10^{54}$~erg/s within $\mathcal{O}(100)$~ms~\cite{Foucart:2015vpa, Ruffert:1996by}.  Neutrinos dominate the cooling of the NS merger remnant and  
affect the ejecta composition, while neutrino pair annihilation above the BH accretion disk contributes to power   the short gamma-ray burst jet~\cite{Ruffert:1996by,Metzger:2014ila,Foucart:2015vpa,Perego:2014fma,Wanajo:2014wha,Just:2014fka,Fujibayashi:2020qda,Kullmann:2021gvo,Narayan:1992iy,Berger:2013jza,Just:2015dba}.

Despite intense work, the treatment of neutrino transport in hydrodynamical simulations of binary NS mergers is still approximated because of the technical challenges linked to the  required three-dimensional general-relativistic magnetohydrodynamical modeling of the source. In addition,  neutrinos are treated as radiation, neglecting the occurrence of flavor conversion. However, the  protonization of the merger remnant (i.e., the excess of electron antineutrinos with respect to electron neutrinos) presumably leads to the occurrence of the matter-neutrino resonance due to the cancellation of the  matter potential involving interactions of neutrinos with electrons and the neutrino-neutrino potential~\cite{Malkus:2014iqa,Malkus:2012ts,Wu:2015fga,Zhu:2016mwa,Frensel:2016fge,Tian:2017xbr,Shalgar:2017pzd}. Recent work has focused on exploring the implications of $\nu$--$\nu$ interactions on the synthesis of the elements heavier than iron in the neutrino-driven outflow and the physics of neutrino-cooled accretion disks~\cite{Tamborra:2020cul,Wu:2017qpc,Wu:2017drk,George:2020veu,Just:2022flt,Li:2021vqj,Fernandez:2022yyv}.

The expected large number of  binary NS merger remnant observations  will offer unprecedented opportunities to characterize the population of binary NS mergers as well as the physics of NSs and their  nuclear equation of state~\cite{AlvesBatista:2021gzc,Burns:2019byj,Burns:2019tqz,Aggarwal:2020olq}. At the same time, upcoming multi-messenger observations of binary NS merger remnants and short gamma-ray bursts could provide constraints on physics beyond the Standard Model, see e.g.~Refs.~\cite{Diamond:2021ekg,Berezhiani:1999qh,Berezhiani:2002ks} for some examples.  An interesting and unexplored scenario in this regard concerns  extra sterile neutrino families with eV mass~\cite{Dasgupta:2021ies,Abazajian:2012ys}.

The existence of sterile families of neutrinos has not been confirmed yet. However, to date, it is challenging to interpret a number of experimental results  within the standard three neutrino flavor framework~\cite{Boser:2019rta,Acero:2022wqg}.
Earlier hints on the existence of a fourth sterile neutrino family were provided by the LSND experiment and partly confirmed by MiniBooNE~\cite{LSND:2001aii,MiniBooNE:2020pnu}.  Along the same direction, reactor neutrino data could have been explained by invoking the existence of an eV-mass sterile neutrino; these puzzling effects concerning  reactor neutrino fluxes seem to be now fully understood~\cite{Mueller:2011nm,Huber:2011wv,Giunti:2021kab,Kopeikin:2021ugh,Berryman:2020agd}, despite remaining uncertainties  on  the reactor energy spectra~\cite{PROSPECT:2020raz,Berryman:2021yan}. Additional anomalies were also found by the Gallium experiments GALLEX and SAGE~\cite{Giunti:2010zu}, and recently confirmed by BEST~\cite{Barinov:2021asz,Barinov:2021mjj}. As a consequence, global fits invoking the existence of extra sterile neutrino families easily accommodate some datasets, but are somewhat in tension with others~\cite{Dentler:2018sju,Dentler:2018sju,Arguelles:2021meu,Giunti:2022btk}. Cosmological data do not rule out  the existence of light sterile neutrinos~\cite{Hagstotz:2020ukm,Hannestad:2012ky,Chu:2018gxk,Archidiacono:2020yey}. We refer the interested reader to Refs.~\cite{Boser:2019rta,Acero:2022wqg,Dasgupta:2021ies} for recent reviews on the topic and details on the best-fit mixing parameters preferred by the  datasets quoted above.

The phenomenology of light sterile neutrinos in core-collapse supernovae (SNe) has been widely investigated;  these particles could have an impact on the synthesis of the elements heavier than iron as well as on  shock revival~\cite{Nunokawa:1997ct,Tamborra:2011is,Wu:2013gxa,Pllumbi:2014saa,Xiong:2019nvw}. 
Their existence could also strongly affect  the expected neutrino signal from the next galactic SN~\cite{Esmaili:2014gya,Tang:2020pkp}.
However, despite  similarities in terms of neutrino number densities and energetics,  the active-sterile flavor conversion physics and its indirect consequences on the multi-messenger emission have not been explored in the context of binary NS mergers.

In this paper, we rely on the output from one of the binary remnant simulations of Ref.~\cite{Just:2014fka} and, for the first time, investigate the production of sterile particles from active states through resonant neutrino-matter interactions, and eventual reconversion into active states. Our work is organized as follows. In Sec.~\ref{sec:simulation}, we  introduce our benchmark binary NS merger remnant model and its main features. Section~\ref{sec:flavor_conversion} focuses on the physics of active-sterile flavor conversions in binary NS merger remnants. The flavor conversion phenomenology is explored in Sec.~\ref{sec:mixing_scan} as a function of the active-sterile mixing parameters, while the production of sterile particles as the BH torus evolves as a function of time is outlined in Sec.~\ref{sec:time}. 
Finally, a summary of our findings is reported in Sec.~\ref{sec:outlook}.

%%%%%%%%%%%%%%%%%%%%%%%%%%%%%%%%%%
%%%%%%%%%%%%%%%%%%%%%%%%%%%%%%%%%%
\section{Binary neutron star merger remnant model}
\label{sec:simulation}
%%%%%%%%%%%%%%%%%%%%%%%%%%%%%%%%%%
%%%%%%%%%%%%%%%%%%%%%%%%%%%%%%%%%%
\begin{figure*}[t]
    \centering
    \includegraphics[width=\textwidth]{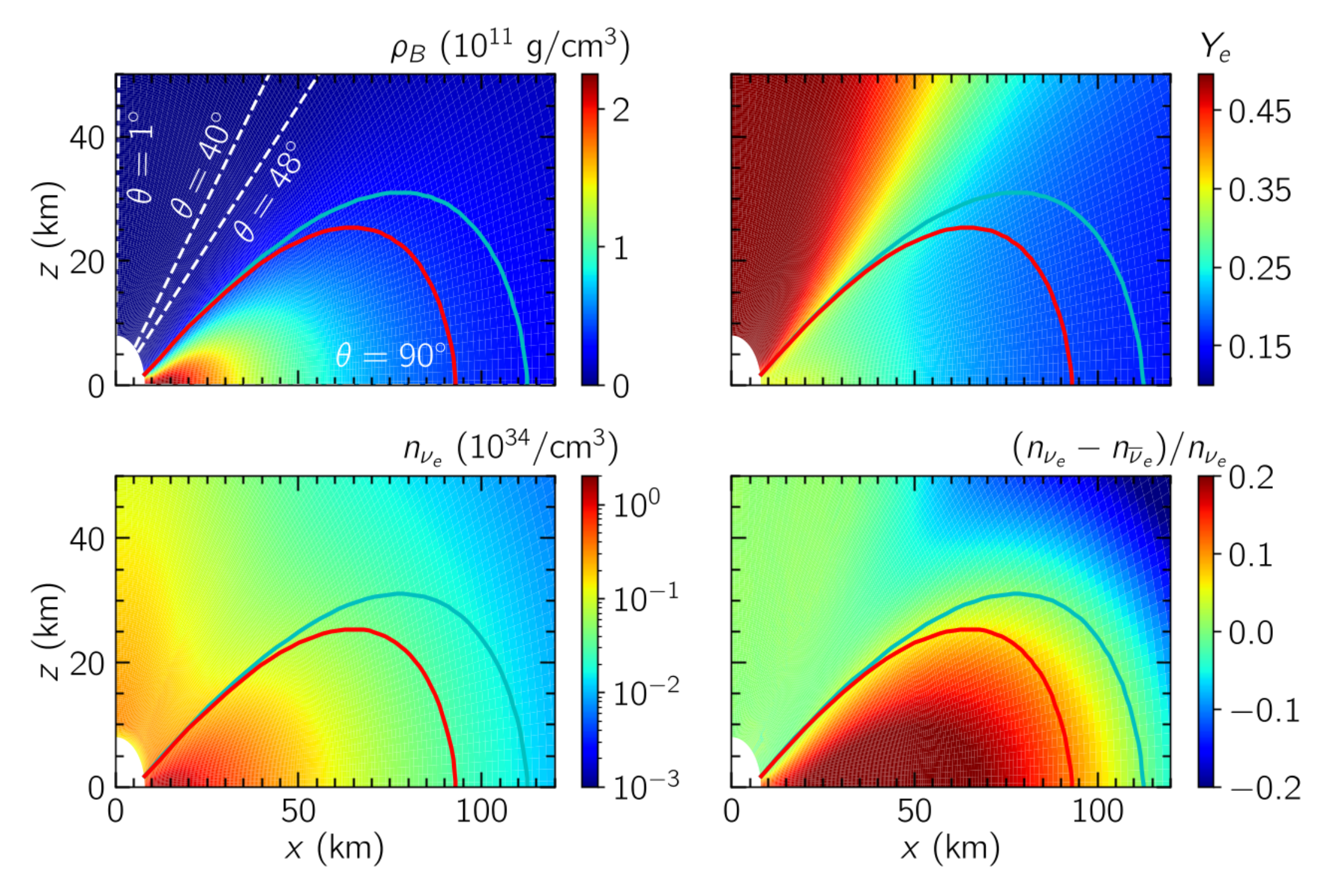}
    \caption{Properties of the BH torus remnant at $25$~ms in the plane spanned by $x$ and $z$, under the assumption of cylindrical symmetry around the $z$ axis:  Baryon mass density (top left panel), electron fraction (top right panel), number density of electron neutrinos (bottom left panel), and relative difference between electron neutrinos and antineutrinos (bottom right panel). The decoupling surfaces of $\nu_e$ and $\overline{\nu}_e$ are  shown in blue and red respectively (see Eq.~\ref{eq:nu_emitting_surf_cond}). In the top left panel, our benchmark neutrino emission directions defined by $\theta = 1^\circ$, $40^\circ$, $48^\circ$ and $90^\circ$ are marked with dashed white lines.
     }
    \label{fig:contour}
\end{figure*}
We rely on  outputs of  a  2D (hereafter, assumed equivalent to a 3D simulation carried out under the assumption of cylindrical symmetry) hydrodynamical simulation of the BH  accretion torus formed in the post-merger phase of a compact binary merger; specifically, we adopt the model M3A8m3a5 presented in Ref.~\cite{Just:2014fka}, based on  subgrid viscosity  and neutrino moment transport, while  neglecting the magnetic field modeling. This simulation  is set up by relying on an idealized
equilibrium torus around a central BH of $3\ M_\odot$; it has dimensionless BH spin parameter $0.8$, and torus of $0.3\ M_\odot$. The total mass of the neutrino-driven  (mostly ejected at early times away from the equator) ejecta is $1.47\times 10^{-3}\ M_\odot$, while the total outflow mass (driven outwards also close to the equator and mostly determined by the viscously driven ejecta)  is $66.2 \times 10^{-3}\ M_\odot$. 
We refer the interested reader to Ref.~\cite{Just:2014fka} for details on the simulation setup.

In this model, as the accretion torus forms, it starts to lose mass while accreting onto the central BH. During the  first $\mathcal{O}(10)$~ms, the environment is optically thick and neutrino cooling is less efficient.  As the density drops, it follows a phase of neutrino-dominated accretion flow, during which neutrino cooling balances viscous heating. 
As the mass and density of the torus decrease, the neutrino production rate is also reduced, until neutrino cooling is not anymore efficient and the torus enters a phase dominated by advection, during which the viscous heating drives the expansion of the torus and launches outflows.

Figure~\ref{fig:contour} illustrates the characteristic properties of our BH torus remnant model and displays the baryon mass density, the electron fraction, as well as the $\nu_e$ number density, and the relative difference between the  $\nu_e$ and $\bar\nu_e$ number densities in the region above the disk. All quantities have been extracted at $25$~ms for representative purposes and are shown in the $x$--$z$ plane, under the assumption of cylindrical symmetry around the $z$ axis. Note that, in the following, we track the flavor conversion physics along a radial direction $r$, defined such that $x = r \cos\theta$ and $z = r \sin\theta$, with $\theta$ being the polar angle measured with respect to the $z$ axis.

One can see that $n_{\bar\nu_e} \simeq n_{\nu_e}$  in the polar region. However, as a function of time, the BH torus evolves from a configuration where $n_{\bar\nu_e} > n_{\nu_e}$ to one with  $n_{\bar\nu_e} < n_{\nu_e}$ in the proximity of the polar axis~\cite{Just:2014fka,Wu:2017drk}. 
On the other hand, non-electron flavors of neutrinos and antineutrinos are 
 thermally produced in small amounts, but can be  
generated through flavor conversion~\cite{Just:2014fka}.

At high densities, neutrinos are  coupled to the matter background. As the matter density decreases, neutrinos decouple from matter and start to free stream. The neutrino energy distributions for the electron flavors  follow  Fermi-Dirac distributions with  non-zero chemical potential in the trapping regime and then tend to become pinched in the free-streaming regime. In the numerical computations, we rely on the numerical energy densities  provided as output of our benchmark NS merger model~\cite{Just:2014fka}.  In order to assess whether the production of sterile particles occurs while the active neutrinos free stream, we estimate the location of the decoupling surfaces by requiring  that the following condition is satisfied for the flux factor~\cite{Wu:2017drk} 
\begin{equation}
    \frac{|\textbf{F}_{\nu_e, \bar\nu_e}|}{n_{\nu_e, \bar\nu_e}} = \frac{1}{3}\ ,
    \label{eq:nu_emitting_surf_cond}
\end{equation}
where $\textbf{F}_{\nu_e, \bar\nu_e}$ is the  number flux and  $n_{\nu_e, \bar\nu_e}$ is the number density of $\nu_e$ or $\bar\nu_e$  extracted from our benchmark hydrodynamical simulation.

Sterile particles could be  produced in the collisional regime (see  Sec.~\ref{sec:flavor_conversion}), hence we compute the mean-free path for the main neutral current (NC) and charged current (CC) interactions (i.e.,~scattering of neutrinos on nucleons, neutrino-(anti)neutrino scattering, Bremsstrahlung processes, and beta reactions) following Refs.~\cite{Leitner:2006sp,Hannestad:1997gc,Strumia:2003zx,Ricciardi:2022pru,Suliga:2020vpz,Suliga:2019bsq}:
\begin{equation}
\label{eq:mfp}
 \lambda_{\nu_e, \bar\nu_e}(E) =  \frac{1}{\sum_{\mathrm{CC, NC}} n_t \sigma(E)}\ ,
\end{equation}
where $\sigma(E)$ is the interaction cross section and $n_t$ the number density of targets. We assume that  Pauli blocking effects are negligible, because  the torus has a mass density much lower than
 the nuclear saturation density ($\rho_B \ll \mathcal{O}(10^{14})$~g$/$cm$^3$, see Fig.~\ref{fig:contour}) and is only moderately degenerate for electrons (see, e.g., Fig.~1 of Ref.~\cite{Wu:2017drk}).

%%%%%%%%%%%%%%%%%%%%%%%%%%%%%%%%%%
%%%%%%%%%%%%%%%%%%%%%%%%%%%%%%%%%%
\section{Active sterile flavor conversion physics}
\label{sec:flavor_conversion}
%%%%%%%%%%%%%%%%%%%%%%%%%%%%%%%%%%
%%%%%%%%%%%%%%%%%%%%%%%%%%%%%%%%%%
In this section, we introduce the equations of motion describing the production of sterile particles. We then investigate the resonant production of sterile particles in NS merger remnants. 

\subsection{Neutrino equations of motion}
\label{sec:EOM}
For simplicity, in this paper, we work in the two-flavor basis $(\nu_e, \nu_s)$ and focus on flavor conversion between electron and sterile flavors. 
In fact, the non-electron flavors are produced through flavor mixing; however,  we neglect flavor conversion among the active flavors. The latter is an approximation, in  light of recent hints supporting evidence for the  development of non-negligible fast neutrino conversion at high densities~\cite{Tamborra:2020cul,Richers:2022zug,Shalgar:2022rjj,Nagakura:2022kic,Shalgar:2022lvv}.
Similar to what shown in Ref.~\cite{Tamborra:2011is}, the production of sterile flavors may further trigger flavor transformation in the active sector, repopulating it. Nevertheless, because of the numerical challenges involved in the modeling of neutrino self-interaction and since we rely on mass and mixing angles between the active and the sterile sectors that are larger than the active sector mixing parameters,  we aim to provide a first explorative glimpse on the production of sterile states in NS merger remnants. An improved modeling of the flavor conversion physics in the presence of sterile neutrinos is left to future work. 

Under the assumption of stationarity, the evolution of the neutrino field in the flavor space is described by the Liouville equation~\cite{Sigl:1993ctk}:
\begin{eqnarray}
\label{eq:eom1}
\partial_r \rho_E &=& - i [H_E, \rho_E] + \mathcal{C}(\rho_E, \bar\rho_E)\ ,\\ \partial_r \bar\rho_E &=& - i [\bar{H}_E, \bar\rho_E] + \bar{\mathcal{C}}(\rho_E, \bar\rho_E)\ ,
\label{eq:eom2}
\end{eqnarray}
where, for each energy mode $E$, $\rho_E$ is a $2\times2$ density matrix, whose diagonal terms are  the neutrino number densities for each flavor: $(n_{\nu_e}, n_{\nu_s})$.  The bar denotes  antineutrino quantities. 
We assume that sterile neutrinos are generated through flavor conversion, i.e.~the initial conditions of our ensemble are such that $\rho_E=\mathrm{diag}(n_{\nu_e}^0,0)$ and $\bar\rho_E=\mathrm{diag}(n_{\bar\nu_e}^0,0)$. The Hamiltonian is 
\begin{equation}
H_E = H_{v, E} + H_{m}\ .
\end{equation}
The vacuum term is a function of the active-sterile mixing angle $\theta_v$ and the vacuum frequency $\omega = \Delta m^2/2E$ (with $\Delta m^2>0$ being  the mass-squared difference):
\begin{eqnarray}
H_{v, E}  = \omega  \begin{pmatrix} 
-\cos 2 \theta_v & \sin 2 \theta_v\\
\sin 2 \theta_v & \cos 2 \theta_v
\end{pmatrix}\ . 
\end{eqnarray}
The vacuum term has opposite sign for neutrinos and antineutrinos. 
The matter term of the Hamiltonian takes into account  the coherent forward scattering on matter
\begin{eqnarray}
H_{m} =  \begin{pmatrix} 
\lambda & 0\\
0 & -\lambda
\end{pmatrix}\ . 
\end{eqnarray}
The effective matter potential is given by~\cite{Nunokawa:1997ct}: 
\begin{equation}
    \lambda= \frac{\sqrt{2} G_F \rho_B}{2m_N}(3 Y_e - 1)\ ,
    \label{eq:eff_pot}
\end{equation}
where $G_F$ is the Fermi constant, $\rho_B$ is the baryon mass density, $m_N$ is the nucleon mass, and $Y_e = (n_{e^-} - n_{e^+})/n_B$ is the electron fraction. 
The terms $\mathcal{C}$ and $\bar{\mathcal{C}}$ in Eqs.~\ref{eq:eom1} and \ref{eq:eom2} represent the collision terms due to the incoherent part of the scattering on the matter background.

Equations~\ref{eq:eom1} and \ref{eq:eom2} assume  
that neutrinos propagate along radial directions ($r$) for simplicity, hence neglecting the neutrino angular distributions. In fact, while the contribution to the flavor conversion history  from neutrinos traveling along non-radial directions should not be negligible, for this explorative work we expect  that the behavior along the radial direction is representative of the flavor transformation phenomenology.

In  dense regions, where the electron flavors are thermally produced, neutrino flavor conversion is suppressed because $\lambda\gg \omega$. As the matter density decreases, sterile flavors can be resonantly produced if the Mikheyev–Smirnov–Wolfenstein (MSW) resonance condition is satisfied~\cite{Mikheev:1986if,PhysRevD.17.2369,osti_5714592}:
\begin{equation}
    \lambda_{\text{res}} = \pm\omega \cos2\theta_v\ ,
    \label{eq:resonance_condition}
\end{equation}
where $+$ applies to neutrinos and $-$ to antineutrinos.

To quantify the amount of flavor conversion at each resonance, we calculate the adiabaticity parameter $\gamma$ at the resonance~\cite{Kim:1987bv}:
\begin{equation}
    \gamma = \frac{\omega}{\pi} \frac{\sin^2 2\theta_v}{\cos 2\theta_v} \left|\frac{{d\lambda/dr}}{\lambda}\right|^{-1}.
    \label{eq:ad_param}
\end{equation}
The corresponding transition probability at the resonance energy $E_{\mathrm{res}}$ is approximated by the Landau-Zener formula~\cite{Kim:1987bv,Parke:1986jy,Blennow:2013rca}~\footnote{Note that, for our cases of interest, $|\lambda|\gg |\omega|$; however, if $\lambda \rightarrow 0$, Eq.~\ref{eq:LZ} holds for $\theta_v\ll 1$~\cite{Parke:1986jy}.}: 
\begin{equation}
    P_{\nu_e \rightarrow \nu_s}(E_{\mathrm{res}}) \approx 1 - \exp\left(-\frac{\pi^2}{2}\gamma\right)\ .
    \label{eq:LZ}
\end{equation}

Resonant conversion between $\overline{\nu}_e$ and $\overline{\nu}_s$  occurs when $\lambda<0$, i.e.~$Y_e \lesssim 1/3$ (see Eq.~\ref{eq:resonance_condition}).  From Fig.~\ref{fig:contour}, it thus becomes evident that $\bar{\nu}_s$'s cannot be produced around the polar region, since $Y_e \gtrsim 1/3$ there.

Sterile particles can also be produced collisionally~\cite{Raffelt:1992bs}.
However, 
for our benchmark NS merger remnant model and sterile mass-mixing parameters, we have verified that the collisional production of sterile (anti)neutrinos is always negligible. An analogous situation occurs in the SN context, where however  keV-mass sterile states can be produced collisionally~\cite{Suliga:2019bsq,Suliga:2020vpz}.

\subsection{Active-sterile flavor conversion}
\begin{figure*}[t]
    \centering
    \includegraphics[width=0.9\textwidth]{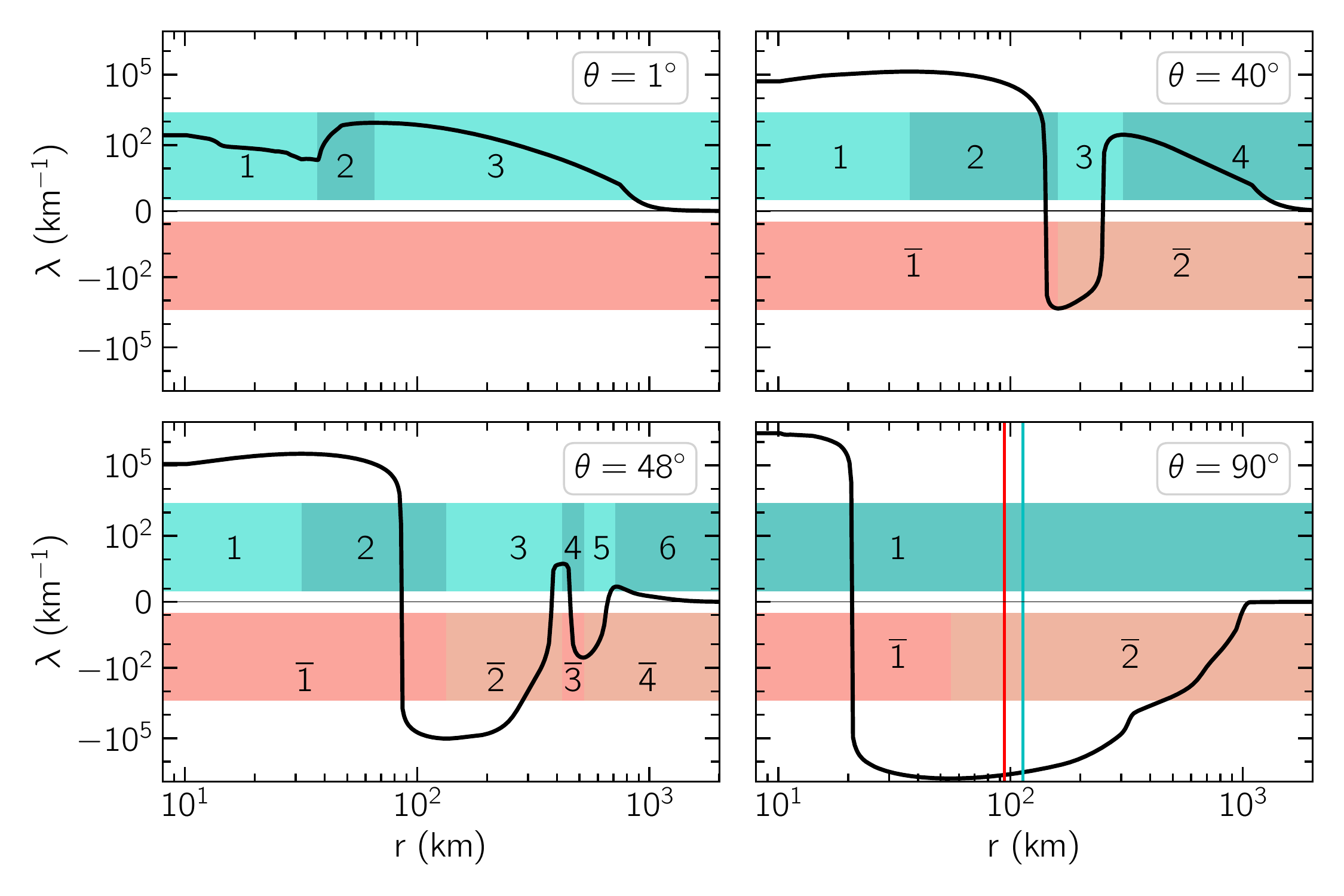}
    \caption{Effective matter potential $\lambda$, extracted at $t=25$~ms, as a function of the radius for our selected radial directions with emission angles: $\theta =1^\circ,  40^\circ, 48^\circ$ and $90^\circ$, from top left to right bottom, respectively (see Fig.~\ref{fig:contour}). The green and red bands show  $\lambda_{\text{res}}$ for  $\left(\sin^2 \theta_v, \Delta m^2\right)= (10^{-2}, 10^{-1}$ eV$^2)$ and $E \in [0.1, 300]$~MeV for neutrinos and antineutrinos, respectively.  Due to the geometry of the torus, the shape and magnitude of $\lambda$ greatly vary  as  functions of $\theta$, and so do the resonance regions. 
    For each $\theta$, the number of MSW resonances occurring for neutrinos and antineutrinos is marked by  different hues. Each resonance  is identified through the change of sign of $d\lambda/dr$.
    For $\theta = 90^{\circ}$, the neutrino and antineutrino decoupling surfaces are plotted as vertical lines in blue and red respectively. The radial directions $\theta = 1^\circ$, $40^\circ$, and $48^\circ$ fall outside the decoupling surfaces.  
    }
    \label{fig:potential}
\end{figure*}
We now intend to investigate the active-sterile flavor conversion phenomenology for our benchmark NS merger remnant model. Figure~\ref{fig:potential} shows the radial neutrino-matter forward scattering potential at $t=25$~ms for representative radial directions with emission angles: $\theta = 1^\circ$, $40^\circ$, $48^\circ$, and $90^\circ$.  The $\theta = 1^\circ$ direction allows to investigate  the flavor conversion physics in the proximity of the polar region and the corresponding $\lambda$ is always positive.  $\theta = 40^{\circ}$  is representative of intermediate directions between the pole and the equator where $\lambda$ is both positive and negative. The $\theta = 48^{\circ}$ potential represents intermediate directions along which $\lambda$ changes sign multiple times, while the $\theta = 90^{\circ}$ potential shows the typical radial evolution of $\lambda$ in the proximity of the equator. Our representative radial directions are chosen to highlight the strong dependence of the flavor conversion phenomenology on the emission angle. All other directions in between these ones have similar features. This is true even though some of our selected radial trajectories originate from regions outside the decoupling sphere. 
For instance, most trajectories originating from the neutrinosphere will eventually evolve outwards nearly radially at large radii and undergo  flavor transformation similar to  one of our represenative  radial directions.  
The magnitude of $\lambda$ is the highest around the equatorial plane (see bottom right panel) and   drops towards the polar axis (see top left panel), see also Eq.~\ref{eq:eff_pot} and Fig.~\ref{fig:contour}.

For $(\sin^2\theta_v, \Delta m^2)= (10^{-2},10^{-1}$ eV$^{2})$ and $E \in [0.1, 300]$~MeV, we can see from Fig.~\ref{fig:potential}  that (anti)neutrinos undergo multiple resonances because of the spatial variations of $Y_e$ and $\rho_B$ as shown in Fig.~\ref{fig:contour}. 
In order for the MSW resonance to occur, the neutrino mean free path (Eq.~\ref{eq:mfp}) has to be  larger than the resonance width, \begin{equation}
\Delta_{\mathrm{res}} = \tan 2\theta_v \left|\frac{d\lambda/dr}{\lambda}\right|^{-1}\ .
\end{equation}
We verified that this is always the case for all scenarios considered in this work.

For the emission directions with $\theta =1^{\circ}$, $40^{\circ}$ and $\theta = 48^{\circ}$,  the first resonance occurs outside the decoupling surfaces (see Fig.~\ref{fig:potential}), where the  production of $\nu_e$'s and $\bar{\nu}_e$'s has essentially stopped and the active flavors  have entered the free-streaming regime. 
This implies that a large production of sterile particles may deplete the active sector (which is not repopulated through thermal processes), with implications on the electron abundance.
On the other hand, for the   $\theta = 90^{\circ}$ direction, the first resonance occurs deep inside in the torus, where $\nu_e$'s and $\bar{\nu}_e$'s are still being thermally produced. 
As a consequence, although flavor conversion at the first resonance may not significantly impact the local number density of $\nu_e$ and $\bar\nu_e$, quickly replenished through thermal processes, it can potentially affect the evolution of the disk. Moreover, for $\theta = 90^{\circ}$, neutrinos undergo MSW resonances before decoupling, which means that not all neutrinos cross the MSW resonances in the forward direction; yet,   neglecting the backward moving neutrinos should have negligible implications on
our qualitative assessment of the impact of the  production of sterile states (see in the following and discussion in Sec.~\ref{sec:mixing_scan}).

\begin{figure*}[t]
    \centering
    \includegraphics[width=0.8\textwidth]{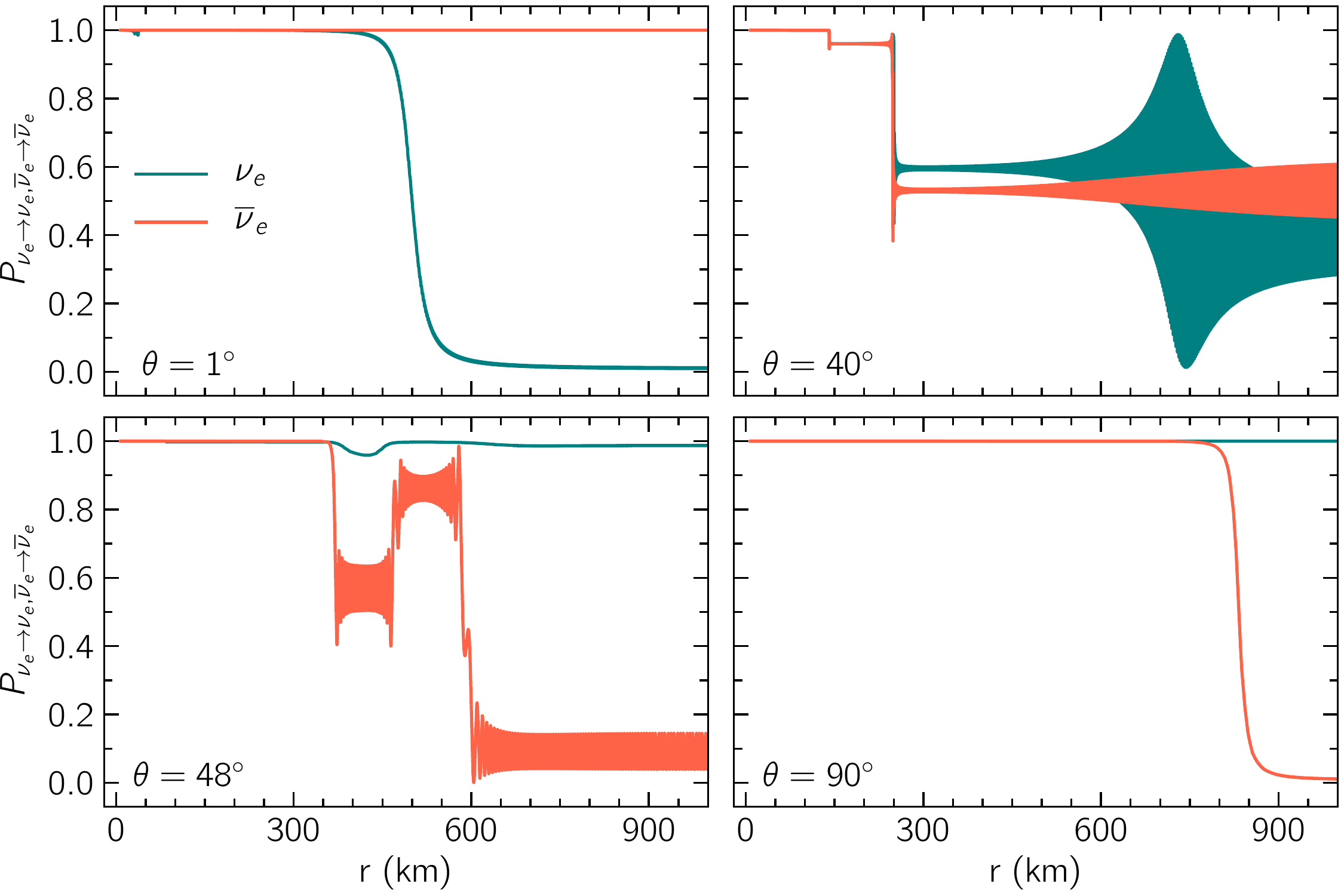}
    \caption{Survival probabilities for $\nu_e$ (in green) and $\overline{\nu}_e$ (in red)  with $E=20$~MeV and $\left(\sin^2 \theta_v, \Delta m^2\right)= (10^{-2}, 10^{-1}$ eV$^2)$ for $\theta = 1^\circ$, $40^\circ$, $48^\circ$, and $90^\circ$ from top left to bottom right, respectively. A varying number of resonances occurs for neutrinos and antineutrinos depending on $\theta$, as visible from Fig.~\ref{fig:potential}. Moreover, because of the spatial variation of the effective matter potential, the adiabaticity of flavor conversion changes as a function of the radius. 
    }
    \label{fig:surv_prob}
\end{figure*}
In order to evaluate the amount of flavor transformation numerically, we introduce the $\nu_e$ survival probability at the resonance radius $r_i$:
\begin{equation}
    P_{\nu_e\rightarrow \nu_e}(E, r_i) = \frac{n_{\nu_e}(E, r_i)}{n_{\nu_e}^0(E, r_i)}\ ,
    \label{eq:Pee}
\end{equation}
where the index $0$ denotes quantities before flavor transformation and $P_{\nu_e\rightarrow \nu_s}(E) = 1- P_{\nu_e\rightarrow \nu_e}(E)$  and $n_{\nu_e}^0(E, r_i)$ is extracted from our benchmark hydrodynamical simulation. A similar expression holds for $P_{\bar\nu_e\rightarrow \bar\nu_e}$. 

Figure~\ref{fig:surv_prob} shows the survival probability of $\nu_e$'s and $\bar\nu_e$'s for $(\sin^2\theta_v,\Delta m^2)=(10^{-2},10^{-1}$~eV$^{2})$
and a selected neutrino energy $E = 20$ MeV, obtained by solving Eqs.~\ref{eq:eom1} and \ref{eq:eom2}  numerically (with initial conditions $n_{\nu_e}^0(E,r_0)$ and $n_{\bar\nu_e}^0(E,r_0)$ extracted from our benchmark merger simulation at  the innermost radius $r_0$) and applying Eq.~\ref{eq:Pee}~\footnote{For $\theta=90^\circ$, neutrinos undergo flavor conversion before decoupling, hence the numerical solution of Eqs.~\ref{eq:eom1} and \ref{eq:eom2}  is approximated since it does not take into account the repopulation term. However, for the mass-mixing parameters adopted in Fig.~\ref{fig:surv_prob} there is no production of sterile states in the first resonance (due to the  highly non-adiabatic MSW resonance) and therefore the result is  independent on the  repopulation effects. The second resonance occurs in the free-streaming regime where Eqs.~\ref{eq:eom1} and \ref{eq:eom2} hold.}. We warn the reader that the resonance radii in Fig.~\ref{fig:surv_prob} depend on the  neutrino energy and the active sterile mixing parameters; MSW resonances at smaller radii should be expected for larger $\Delta m^2$ (see Sec.~\ref{sec:mixing_scan}).

The same computation can be carried out analytically; in this case, the neutrino number density at each resonance radius $r_i$, occurring after neutrino decoupling, is: 
\begin{eqnarray}
    n_{\nu_e}(E, r_i)&=& P_{\nu_e\rightarrow \nu_e}(E, r_i) n_{\nu_e}(E, r_{i-1}) \left(\frac{r_{i-1}}{r_i}\right)^2+\nonumber \\ & & P_{\nu_s\rightarrow \nu_e}(E, r_{i}) n_{\nu_s}(E,r_{i-1}) \left(\frac{r_{i-1}}{r_i}\right)^2\ , 
    \end{eqnarray}
with the   survival probability being computed trough Eq.~\ref{eq:LZ}, and $i-1$ being the former resonance in the event that multiple resonances take place (see Fig.~\ref{fig:potential}).  By comparing with the outputs of our benchmark hydrodynamical simulation, we  verified the neutrino number density at $r_{i-1}$  can be  safely rescaled  by $\sim 1/r^2$ in order to compute the local number density at $r_i$. However,   a special case occurs for the first resonance, where $n_{\nu_e}(E, r_1)= P_{\nu_e\rightarrow \nu_e}(E, r_1) n_{\nu_e}(E, r_1)$ with $n_{\nu_e}(E, r_1)$ being extracted from our benchmark  remnant simulation model. Moreover, for $\theta = 90^\circ$, the first  MSW resonance occurs before neutrino decoupling, hence  $n_{\nu_e}(E, r_2)= P_{\nu_e\rightarrow \nu_e}(E, r_2) n_{\nu_e}(E, r_2) + P_{\nu_s\rightarrow \nu_e}(E, r_{1}) n_{\nu_s}(E,r_{1}) (r_1/r_2)^2$, where $n_{\nu_e}(E, r_2)$ is extracted from our remnant simulation model. Analogous expressions hold for $n_{\nu_s}(E, r_i)$. We find that our analytical computations are in  agreement with the numerical ones (results not shown here).

We can see that the flavor conversion physics is highly dependent on the emission direction and more than two resonances could occur for some directions, as already noticeable from Fig.~\ref{fig:potential} (see, e.g., $\theta = 48^\circ$). As expected, according to the emission direction and for our fixed neutrino energy, the adiabaticity of flavor conversion changes. This has the effect that $\nu_s$'s are minimally produced in the equatorial plane (see $\theta = 90^{\circ}$), whereas in the polar region, no $\bar{\nu}_s$'s are produced since no $\bar{\nu}_e$--$\bar{\nu}_s$ resonances can occur (see $\theta = 1^{\circ}$).  

We warn the reader that, while our choice of the representative radial directions aims to highlight the diversity of the active-sterile flavor conversion phenomenology, not all our benchmark radial directions have comparable relevance for what concerns the astrophysical implications. In fact, for neutrino paths starting within the neutrino decoupling surfaces, $Y_e \lesssim 1/3$ in the innermost regions;
on the other hand, neutrinos  start propagating from a region  with a high value of $\lambda$
(i.e., $Y_e \gtrsim 1/3$) towards regions with  lower $\lambda$ ($Ye \lesssim 1/3$) for $\theta = 40^{\circ}$ and $48^\circ$. As a consequence, for $\theta = 40^{\circ}$ and $48^\circ$, the innermost MSW  resonances may appear peculiar
because  the trajectories start on the black-hole horizon (large $Y_e$). 
However, we note that  neutrinos emitted from the opposite side of the disk and crossing the funnel  would have gone through similar resonances as those represented by $\theta=40^\circ$ and $48^\circ$. 
Thus, our selected radial directions provide a complete picture of the  flavor conversion phenomenology.

Interestingly, due to the torus geometry,  
the flavor conversion phenomenology in NS merger remnants differs from the SN one~\cite{Nunokawa:1997ct,Tamborra:2011is,Wu:2013gxa,Xiong:2019nvw,Pllumbi:2014saa}, where at most two MSW resonances were observed. However, similar to the SN case, the innermost resonances are less adiabatic than the outer ones because of the steep radial profile of $\lambda$.  
Another difference with respect to the SN case is that $\bar\nu_e$'s are naturally more abundant in the compact merger scenario.

\begin{figure*}[t]
    \centering
    \includegraphics[width=0.8\textwidth]{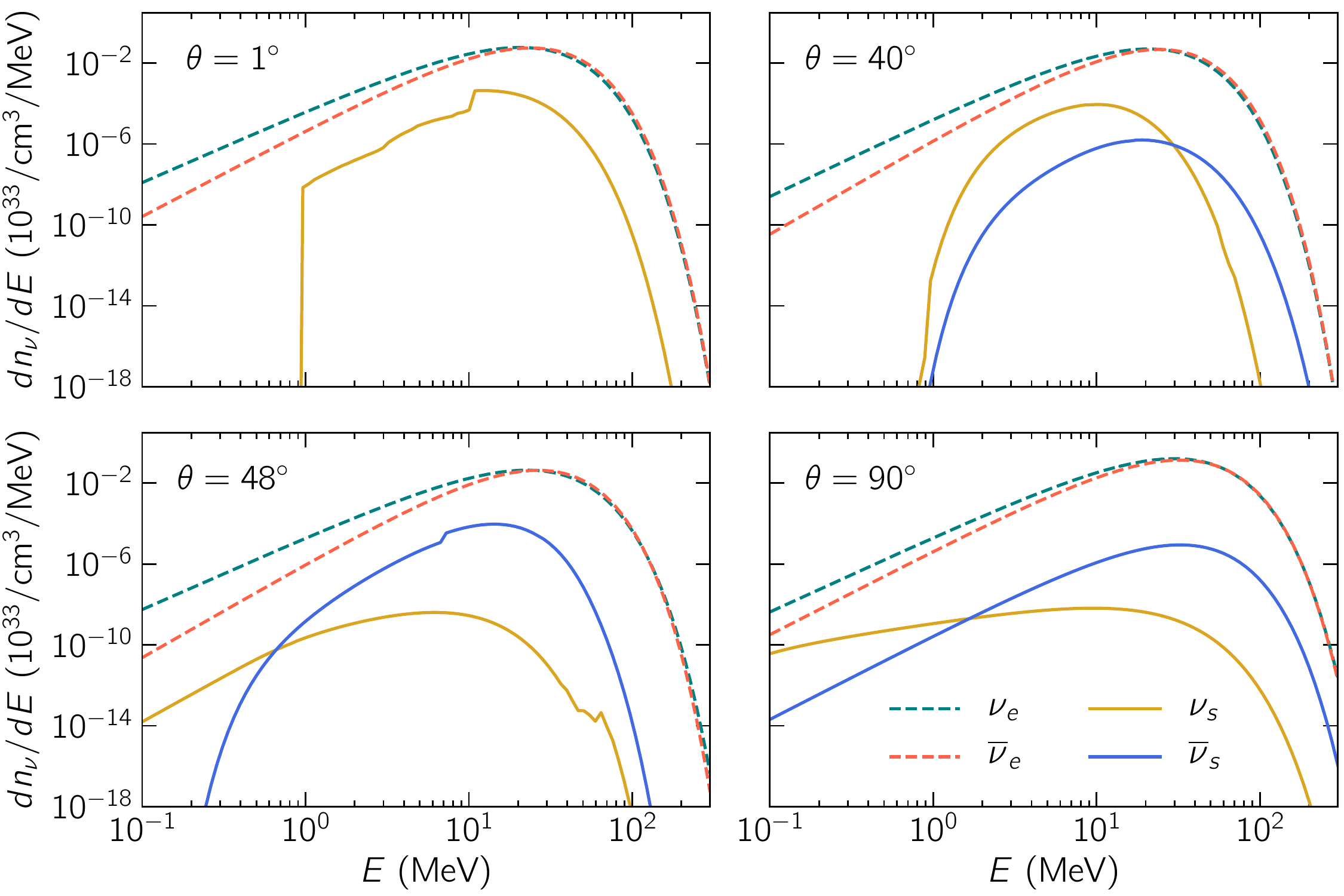}
    \caption{Differential number densities of $\nu_e$ (dashed green line), $\overline{\nu}_e$ (dashed red line), $\nu_s$ (solid ochre line) and $\bar\nu_s$ (solid blue line)  as  functions of energy for the same emission directions and mixing parameters displayed in  Fig.~\ref{fig:surv_prob}.  The dashed lines have been extracted at the innermost radius $8.02$~km and the solid ones at  $1000$~km. Only sterile neutrinos are produced along $\theta=1^\circ$, while more sterile antineutrinos than sterile neutrinos are produced along $\theta=90^\circ$.}
    \label{fig:dens}
\end{figure*}
Figure~\ref{fig:dens} shows the  active and sterile differential number densities as  functions of the energy,  for the same representative mixing parameters and emission directions shown in Fig.~\ref{fig:surv_prob}. We can see how the effects of the varying adiabaticity and the  occurrence of MSW resonances distort the shape of the sterile distributions  at $1000$~km, creating energy-dependent features. In addition, $\bar\nu_s$'s are not produced in the sourroundings of the polar region ($\theta = 1^{\circ}$, top left panel) while more $\bar\nu_s$'s than $\nu_s$'s are produced in the proximity of the equatorial region ($\theta = 90^{\circ}$, bottom right  panel).  

%%%%%%%%%%%%%%%%%%%%%%%%%%%%%%%%%%
%%%%%%%%%%%%%%%%%%%%%%%%%%%%%%%%%%
\section{Dependence of the flavor conversion phenomenology on the sterile mass and mixing parameters}
\label{sec:mixing_scan}
%%%%%%%%%%%%%%%%%%%%%%%%%%%%%%%%%%
%%%%%%%%%%%%%%%%%%%%%%%%%%%%%%%%%%
In this section, we investigate  the physics of flavor conversion and the production of sterile particles as  functions of the sterile mixing parameters. In what follows, we scan the mass-mixing parameter space  considered for light sterile neutrinos~\cite{Boser:2019rta,Acero:2022wqg,Dasgupta:2021ies}; because of the lack of consensus on the mass-mixing sterile parameters necessary to  interpret  the various experimental anomalies (see Sec.~\ref{sec:intro}), we refrain from choosing  benchmark mass-mixing sterile parameters.
\subsection{Occurrence of multiple MSW resonances}
In order to compute the average amount of flavor mixing across  multiple resonances, we introduce the energy averaged survival probability for neutrinos at each resonance:
\begin{equation}
\label{eq:Pave}
\langle P_{\nu_e \rightarrow \nu_e}(r_i) \rangle = \frac{\int dE P_{\nu_e\rightarrow \nu_e}(E, r_i) n_{\nu_e}(E,r_{i-1})}{\int dE n_{\nu_e}(E, r_{i-1})}\ ,
\end{equation}
with $P_{\nu_e \rightarrow \nu_e}(E, r_i)$ defined as in Eq.~\ref{eq:LZ}. 
An analogous expression holds for the survival probability of antineutrinos, $\langle P_{\bar\nu_e \rightarrow \bar\nu_e}(r_i) \rangle$.

\begin{figure*}[t]
    \centering
    \includegraphics[width=\textwidth]{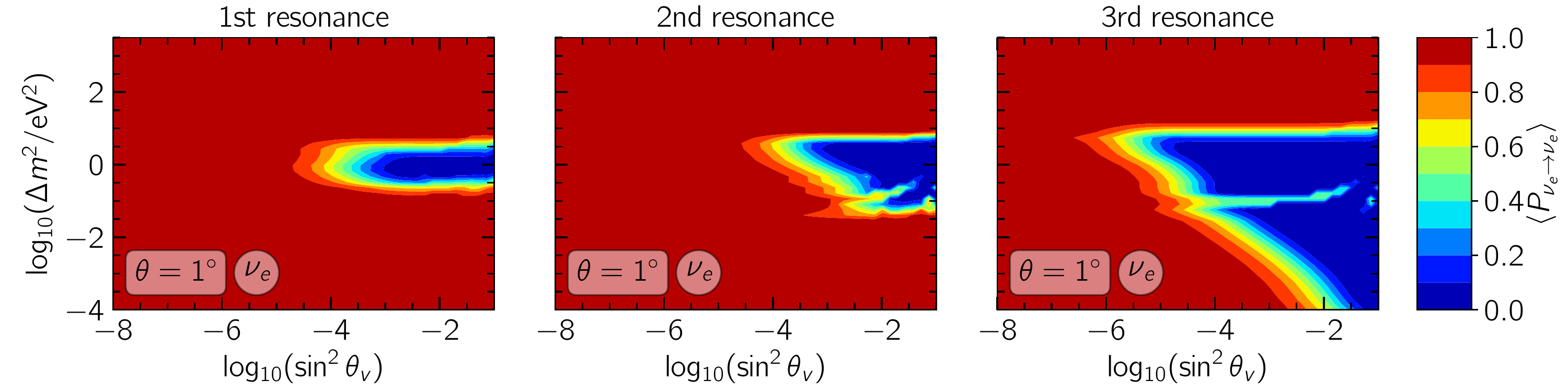}
    \caption{Contour plot of the $\nu_e$ energy averaged survival probability at the three resonances (see Fig.~\ref{fig:contour} and Eq.~\ref{eq:Pave})  for the emission direction $\theta = 1^{\circ}$ in the plane spanned by $\sin^2\theta_v$ and $\Delta m^2$. Only neutrinos undergo resonaces, no MSW resonance occurs for antineutrinos. At the third resonance, the range and slope of the potential (see Fig.~\ref{fig:potential}) allows for a larger region of the mass-mixing parameter space to be affected by full flavor conversion.}
    \label{fig:eng_avg_surv_1_nu}
\end{figure*}
\begin{figure*}[t]
    \centering
    \includegraphics[width=0.7\textwidth]{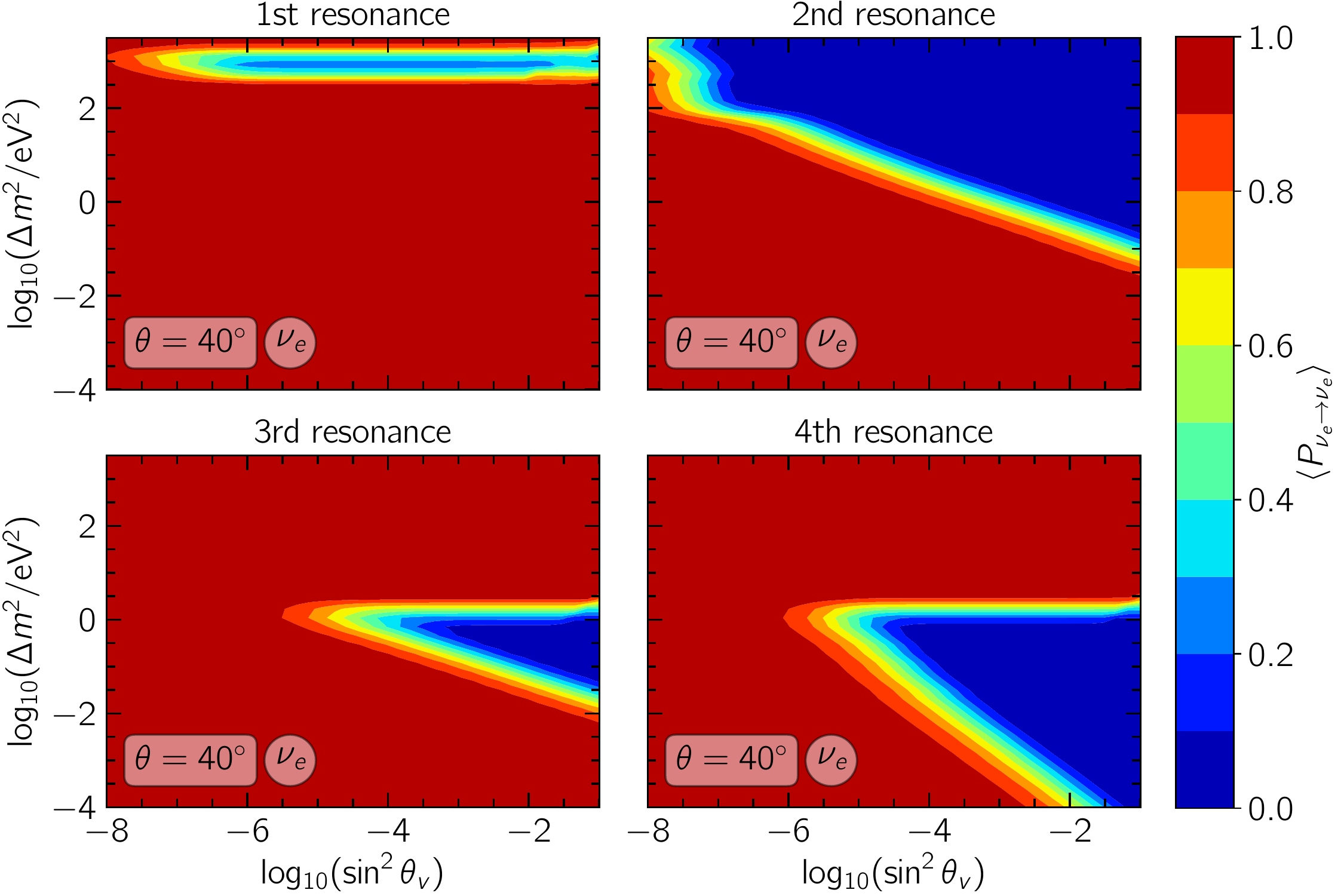}
    \includegraphics[width=0.75\textwidth]{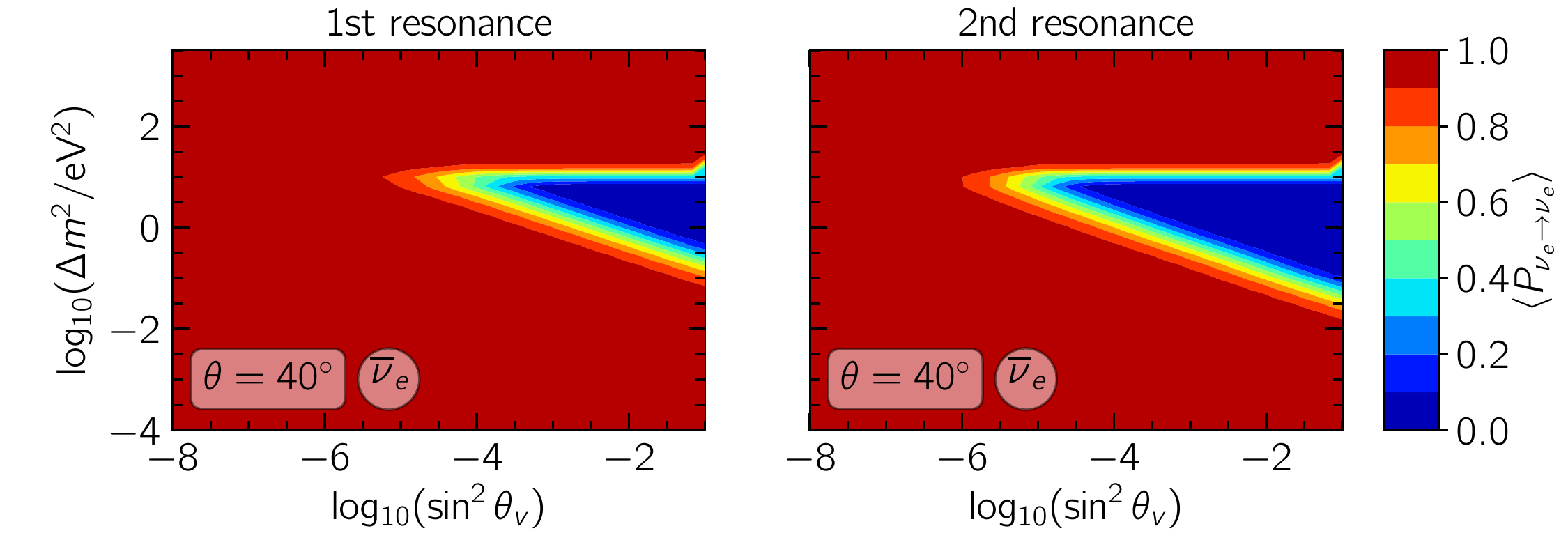}
    \caption{Same as Fig.~\ref{fig:eng_avg_surv_1_nu}, but for $\theta = 40^{\circ}$. Four MSW resonances occur for neutrinos (top panels) and two for antineutrinos (bottom panels). Differently from Fig.~\ref{fig:eng_avg_surv_1_nu}, the region of the parameter space affected by flavor conversion is not always larger as outer resonances are met (and the matter potential becomes shallower).  
    }
    \label{fig:eng_avg_surv_40_nu}
\end{figure*}
\begin{figure*}[t]
    \centering
    \includegraphics[width=\textwidth]{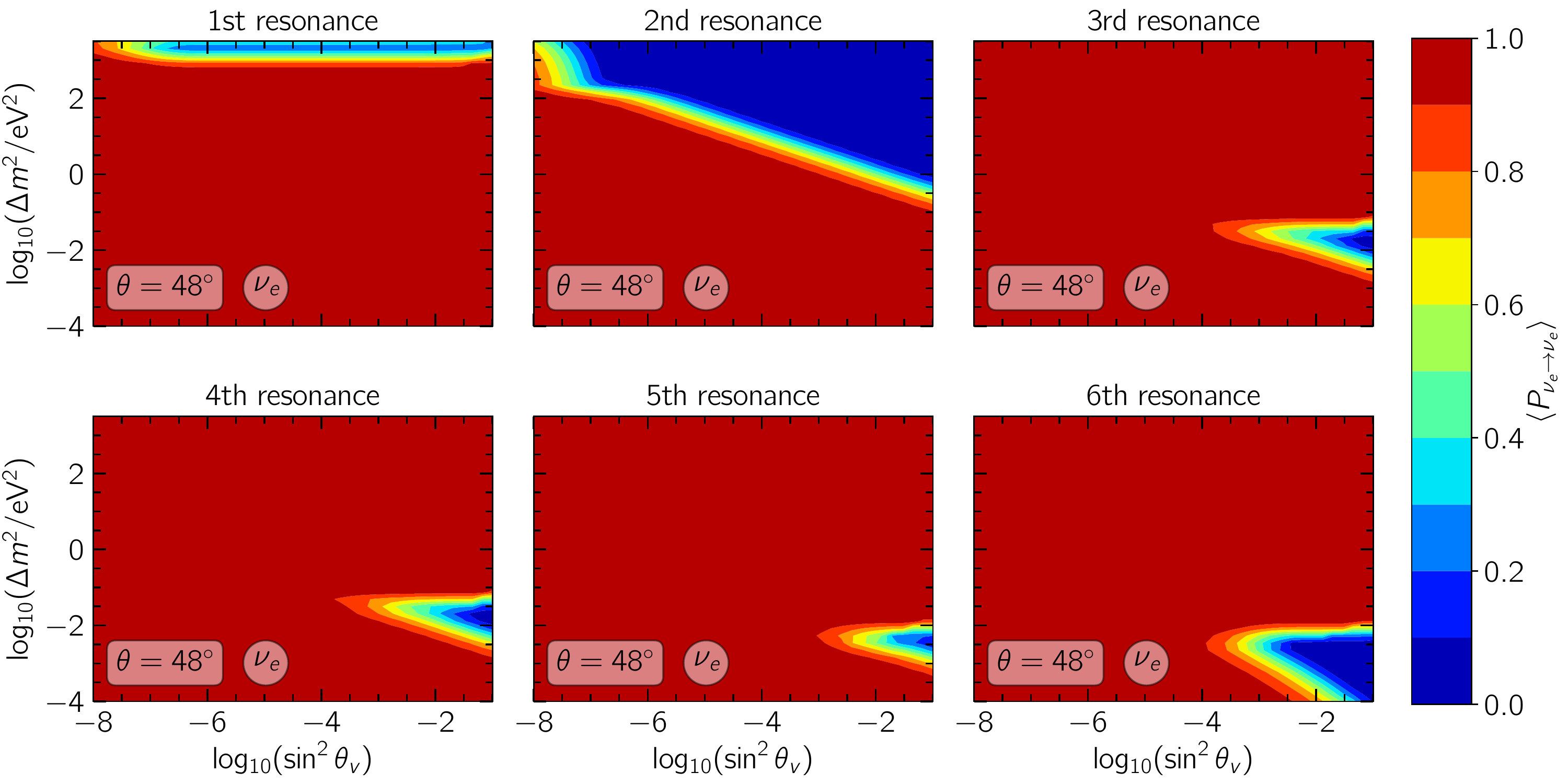}
    \includegraphics[width=0.7\textwidth]{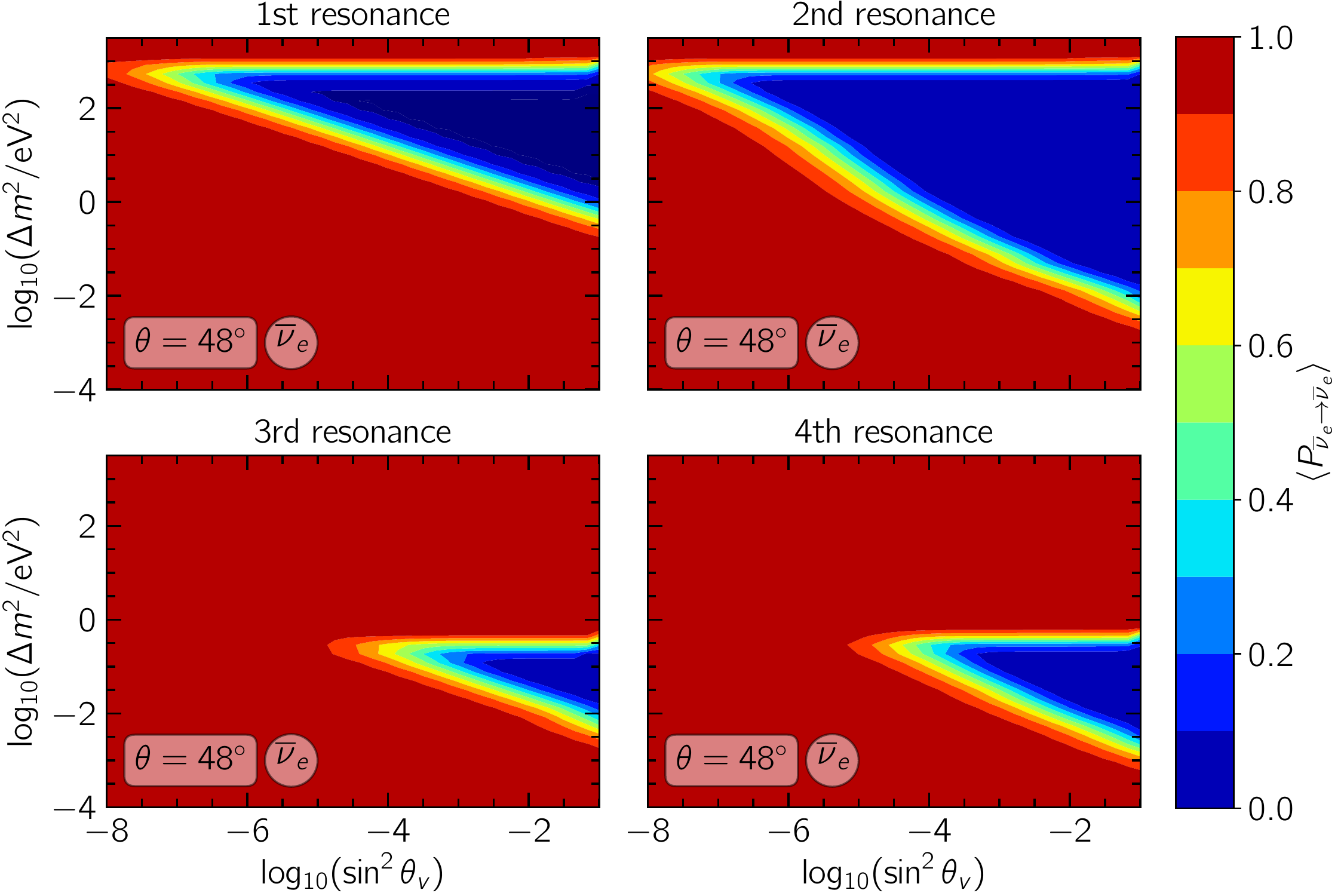}
    \caption{Same as Fig.~\ref{fig:eng_avg_surv_1_nu}, but for $\theta = 48^{\circ}$. Six MSW resonances occur for neutrinos and four for antineutrinos. 
    }
    \label{fig:eng_avg_surv_48_nu}
\end{figure*}
\begin{figure*}[t]
    \centering
    \includegraphics[width=0.4\textwidth]{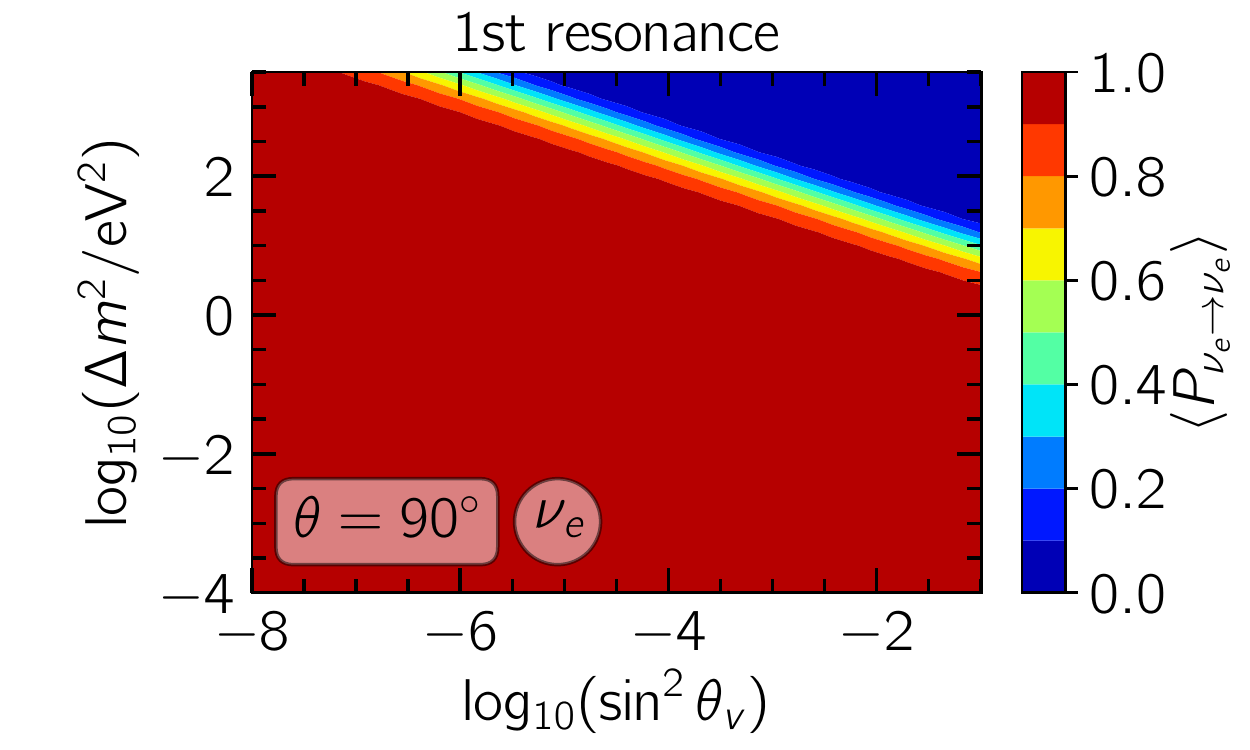}
    \includegraphics[width=0.7\textwidth]{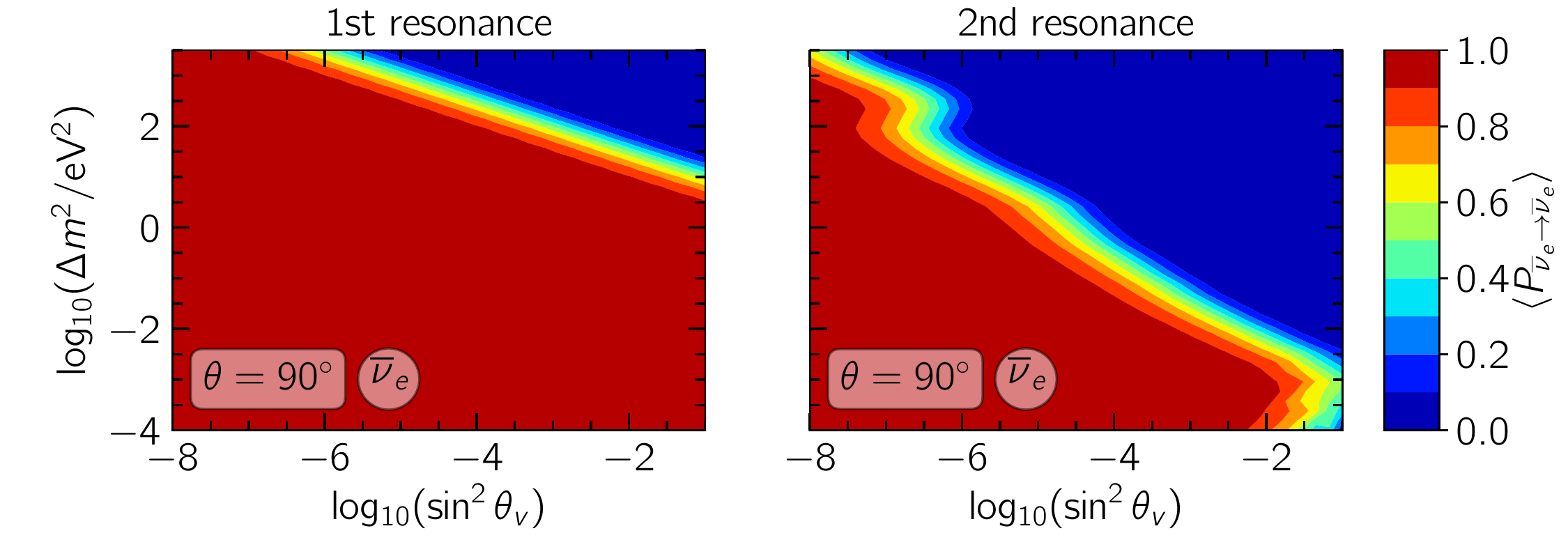}
    \caption{Same as Fig.~\ref{fig:eng_avg_surv_1_nu}, but for $\theta = 90^{\circ}$. One MSW resonance takes place for neutrinos and two for antineutrinos. 
    }
    \label{fig:eng_avg_surv_90_nu}
\end{figure*}
Figures~\ref{fig:eng_avg_surv_1_nu}--\ref{fig:eng_avg_surv_90_nu} show contour plots of $\langle P_{\nu_e \rightarrow \nu_e}(r_i) \rangle$ ($\langle P_{\bar\nu_e \rightarrow \bar\nu_e}(r_i) \rangle$) in the plane spanned by  $(\sin^2 \theta_v, \Delta m^2$) on top (bottom). Each resonance  is identified through the number of times $d{\lambda}/dr$ changes sign, as shown in Fig.~\ref{fig:potential}. 
The amount of flavor transformation that occurs within each resonance region depends on   $\lambda$, which controls which $(\sin^2\theta_v, \Delta m^2)$ undergo MSW resonances. On the other hand, $d{\lambda}/dr$  controls the adiabaticity of each resonance.

There are  general features which most resonance regions display, independently of $\theta$. The  region of partial or full flavor conversion into sterile states has a triangular shape. This is because the adidabaticity of the MSW resonance increases as $\sin^2\theta_v$ and/or $\Delta m^2$ increase (see Eq.~\ref{eq:ad_param}). These triangular regions are bounded from above in most cases since a high $\Delta m^2$  causes $\lambda_{\mathrm{res}}$ to exceed the maximum value of the potential $\lambda$ within that region (or minimum value for antineutrinos), resulting in the absence of resonances for the upper part of the parameter space.

The bottom panels of Fig.~\ref{fig:eng_avg_surv_90_nu} represent the survival probability of antineutrinos for  $\theta = 90^{\circ}$. We see  how the first resonance is much less adiabatic than the second one. This is due to how rapidly $Y_e$ crosses $1/3$, while $\rho_B$ is still very large, causing ${d\lambda/dr}\gg 1$; i.e.,~$\lambda$ is very steep as it changes sign for the first time (see Fig.~\ref{fig:potential}). A similar trend can  be seen in the bottom panels of Figs.~\ref{fig:eng_avg_surv_40_nu} and \ref{fig:eng_avg_surv_48_nu} for $\theta = 40$ and  $48^{\circ}$, though not as prominently. 
Otherwise, the same general features in the neutrino survival probability, also present themsleves in the antineutrino plots.

\subsection{Overall production of sterile neutrinos and antineutrinos}
\begin{figure*}[t]
    \includegraphics[width=0.76\textwidth]{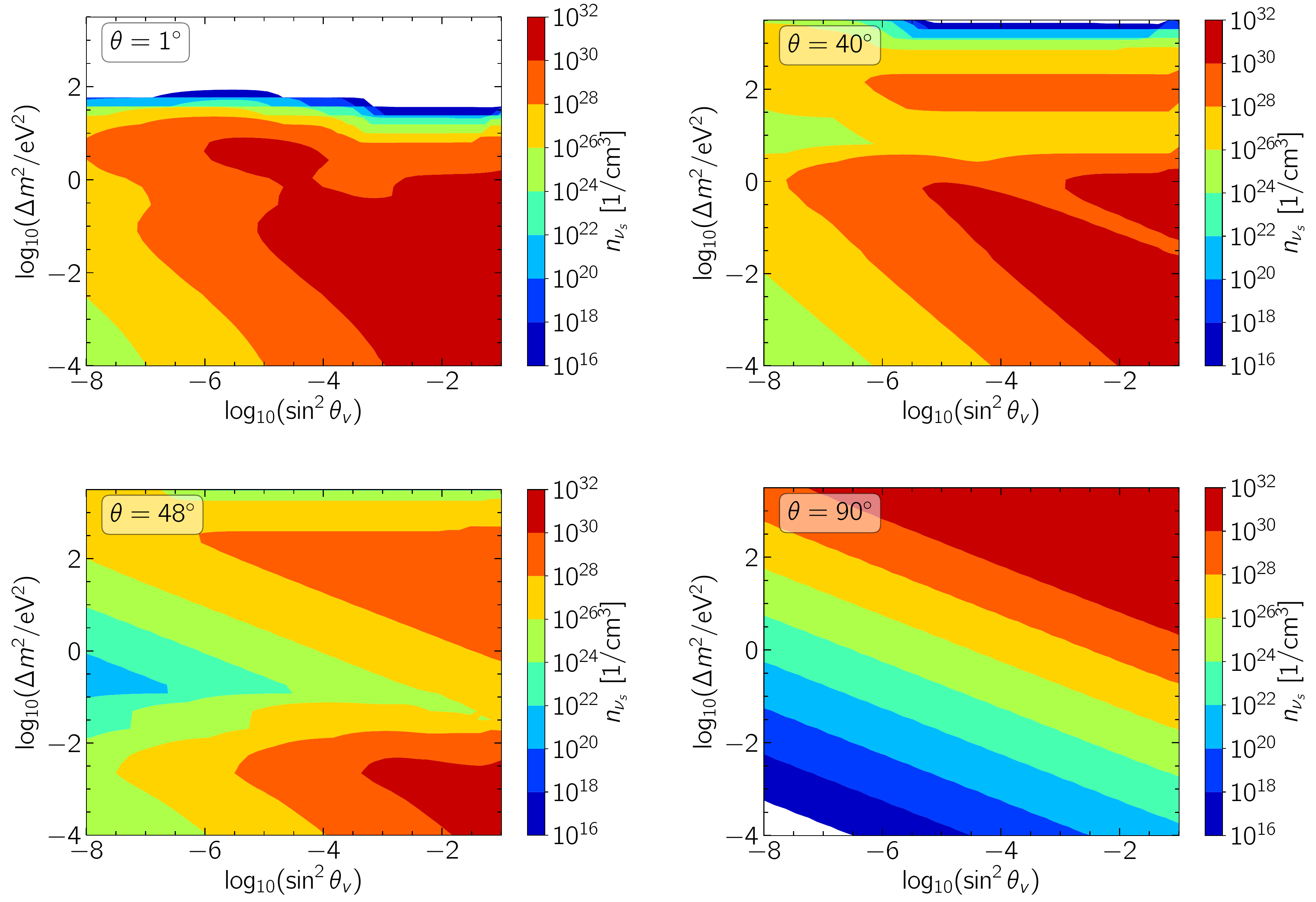}
       \caption{Contour plot of the number density of sterile neutrinos at $1000$~km in the plane spanned by $(\sin^2\theta_V, \Delta m^2)$ for $\theta =1$, $40$, $48$, and $90^\circ$ from top left to bottom right, respectively. The number density of sterile neutrinos is maximal for a large region of the mass-mixing parameter space for $\theta=1^\circ$. 
    }
    \label{fig:n_nu_1_90}
\end{figure*}
\begin{figure*}[t]
      \includegraphics[width=0.76\textwidth]{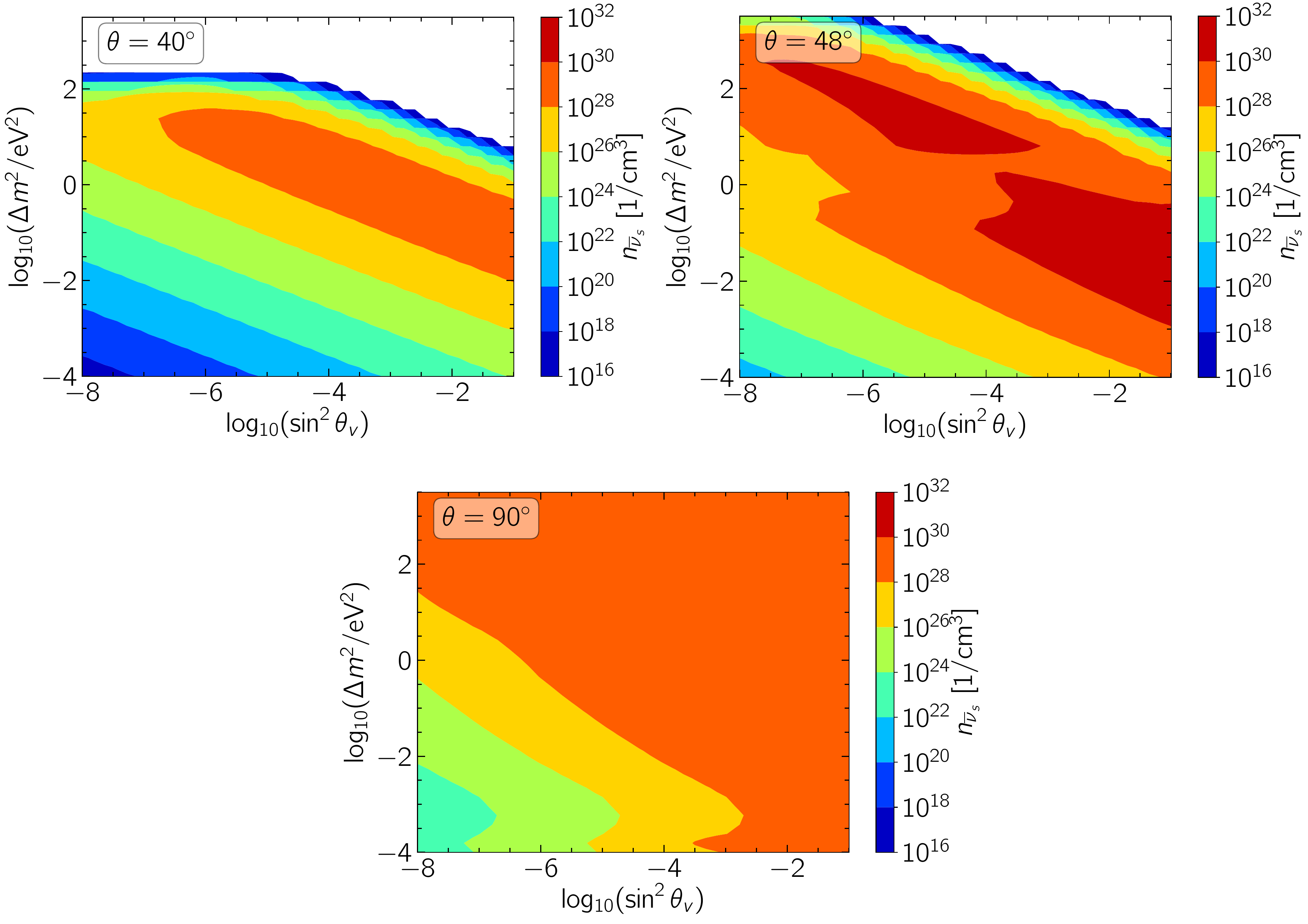}
     \caption{Same as Fig.~\ref{fig:n_nu_1_90}, but for sterile antineutrinos. The contour plot for $\theta=1^\circ$ is omitted as no production of $\bar\nu_s$'s occurs. The number density of sterile antineutrinos is maximal for a large region of the mass-mixing parameter space for $\theta=90^\circ$.    
    }
    \label{fig:n_nu_bar_40_90}
\end{figure*}
Figures~\ref{fig:n_nu_1_90} and \ref{fig:n_nu_bar_40_90}  show the overall number densities of sterile neutrinos and antineutrinos at $r=1000$~km, respectively, produced through  multiple MSW resonances.
The lowest number density of resonantly produced sterile particles is visible in the bottom left corner of the  $(\sin^2\theta_v, \Delta m^2)$ plane for all panels of Figs.~\ref{fig:n_nu_1_90} and \ref{fig:n_nu_bar_40_90}. In general, as 
$\sin^2\theta_v$ increases, the number density of sterile neutrinos in Fig.~\ref{fig:n_nu_1_90} increases.
The  shape of the countours closely follows the patterns shown in Figs.~\ref{fig:eng_avg_surv_1_nu}--\ref{fig:eng_avg_surv_90_nu} and is a  direct consequence of multiple resonances.
As expected, sterile neutrinos are abundantly produced for a larger region of the mass-mixing parameter space for $\theta=1^\circ$.

In the top panels of  Fig.~\ref{fig:n_nu_bar_40_90} 
(for $\theta = 40^{\circ}, 48^{\circ}$), the combined effect of all MSW resonances can be observed in the asymptotic emission of $\overline{\nu}_s$'s. 
On the other hand, in the bottom panel of Fig.~\ref{fig:n_nu_bar_40_90} ($\theta = 90^{\circ}$), the effect of the first resonance is not visible in the top right part of the parameter space. This is because $\bar\nu_s$'s generated adiabatically at the first resonance are reconverted back to $\bar\nu_e$'s at the second resonance. Thus $n_{\bar\nu_s}$ in our plot is determined by the conversions into sterile states occurring at the second resonance; on the other hand, flavor conversion at the first and second resonances leads to an enhanced $n_{\bar\nu_e}$ in the top right corner of the parameter space.

%%%%%%%%%%%%%%%%%%%%%%%%%%%%%%%%%%
%%%%%%%%%%%%%%%%%%%%%%%%%%%%%%%%%%
\section{Active-sterile flavor conversion as a function of the torus evolution}
\label{sec:time}
%%%%%%%%%%%%%%%%%%%%%%%%%%%%%%%%%%
%%%%%%%%%%%%%%%%%%%%%%%%%%%%%%%%%%
\begin{figure*}[t]
    \centering
    \includegraphics[width=0.6\textwidth]{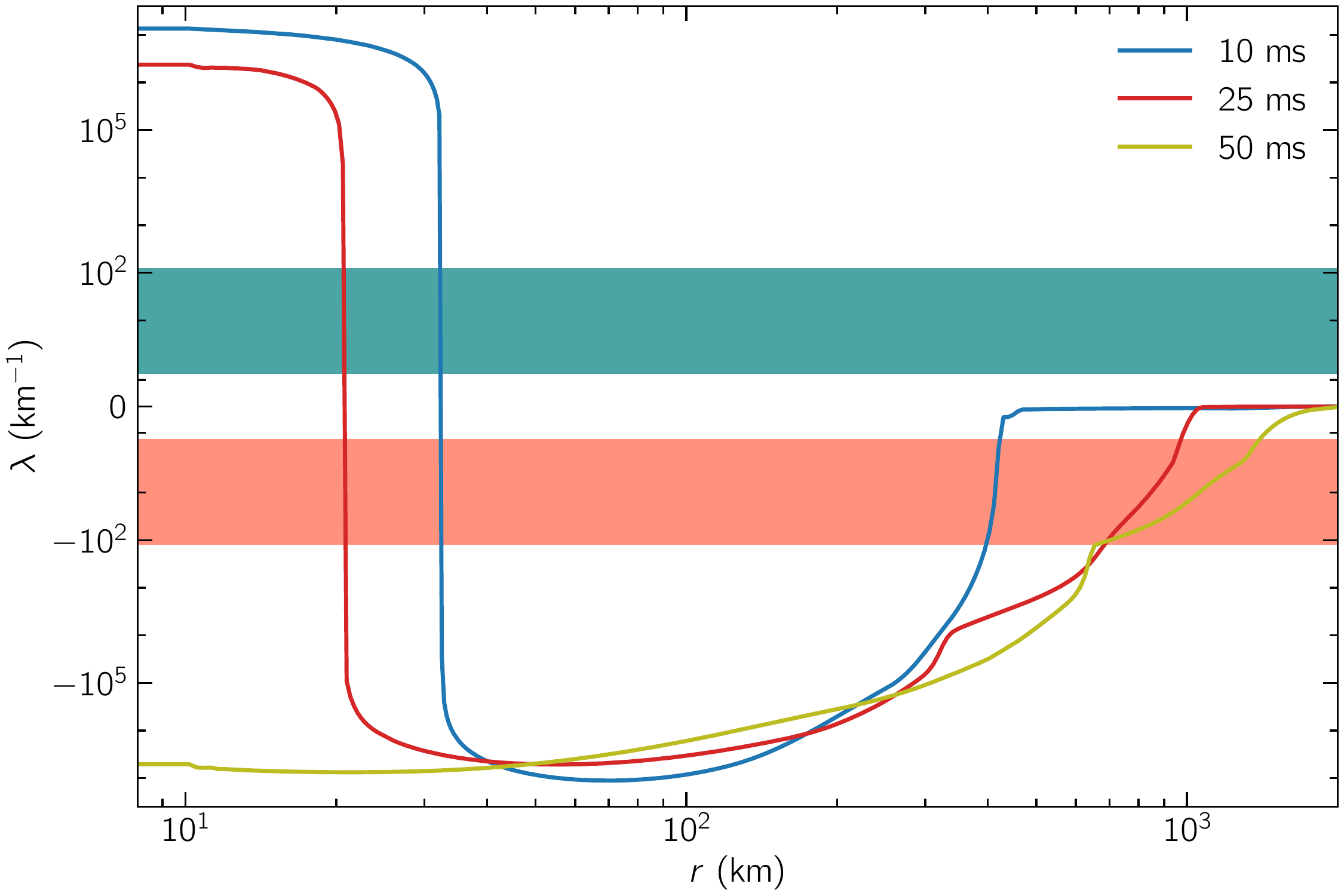}
    \caption{Effective matter potential $\lambda$ as a function of the radius for three different time snapshots, $t= 10, 25$ and $50$ ms in the equatorial plane, i.e.~for $\theta = 90^{\circ}$. The green and red bands show  $\lambda_{\text{res}}$ for  $\left(\sin^2 \theta_v, \Delta m^2\right)= (10^{-2}, 10^{-1}$ eV$^2)$ and $E \in [0.1, 300]$~MeV for neutrinos and antineutrinos, respectively. As $\lambda$ evolves as a function of time, the number of MSW resonances changes.}
    \label{fig:pot_10_25_50}
\end{figure*}
\begin{figure*}[t]
 \centering
  \includegraphics[width=0.36\textwidth]{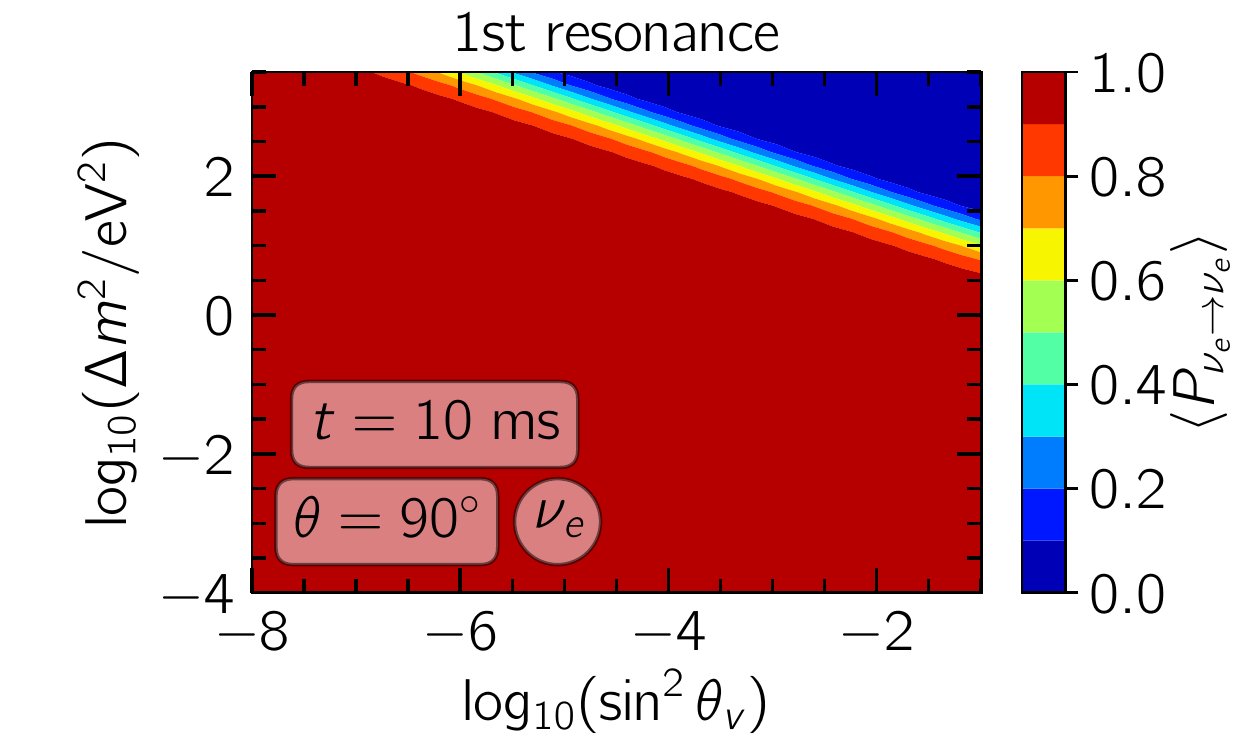}
    \includegraphics[width=0.625\textwidth]{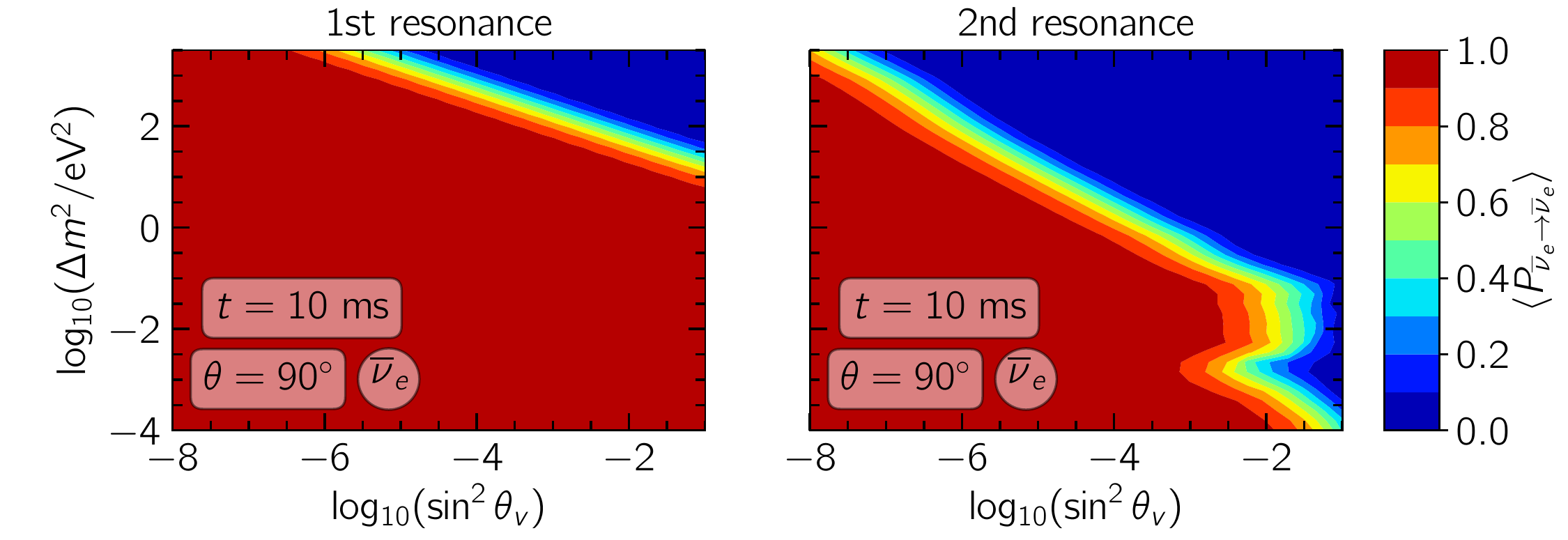}\\[1.5em]
     \includegraphics[width=0.36\textwidth]{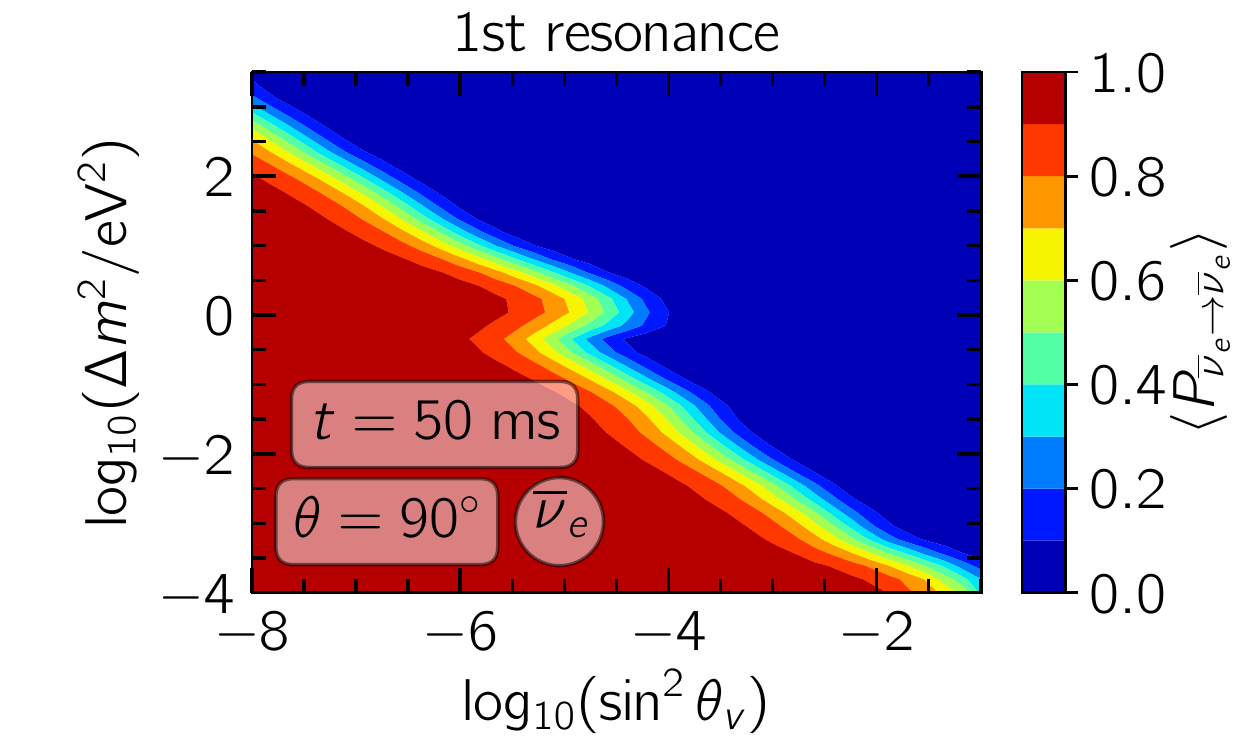}
    \caption{Contour plot of the $\nu_e$ and $\bar\nu_e$ energy averaged survival probabilities for the emission direction $\theta = 90^{\circ}$ at $t=10$ ms (top panels, for $\nu_e$'s and $\bar\nu_e$'s) and $50$ ms (bottom panel, for $\bar\nu_e$'s). As time increases, the region of the mass-mixing parameter space affected by flavor conversion becomes larger.}
    \label{fig:P_nu_10_ms}
\end{figure*}

In this section, we explore the active-sterile flavor conversion physics for three snapshots of the disk evolution ($t=10$, $25$, and $50$~ms) and a fixed emission angle $\theta = 90^\circ$. The  matter potential, $\lambda$, is shown in Fig.~\ref{fig:pot_10_25_50}. One can see that the MSW resonance patterns are similar to the ones investigated in Fig.~\ref{fig:potential} for $t=25$~ms. Similar features to the ones illustrated in Fig.~\ref{fig:potential} can be  found for the radial profiles of $\lambda$ along the polar region and for some intermediate values of $\theta$. However, in the proximity of the equatorial plane ($\theta = 90^{\circ}$), the electron fraction drops causing $\lambda$ to change sign in the innermost regions  for $t>25$~ms. 
As a consequence, no MSW resonances occur for $\nu_e$ at $t=50$~ms and one less MSW resonance takes place for $\bar\nu_e$ with respect to the $t=25$~ms case. Nevertheless, since the innermost resonances were mainly non-adiabatic, we do not find large changes in the overall flavor conversion physics, as shown in Fig.~\ref{fig:P_nu_10_ms} (see Fig.~\ref{fig:eng_avg_surv_90_nu} for comparison). 

As $t$ increases, the matter gradient along the radial directions becomes  gentler, because of the drop in baryon density, resulting in more adiabatic resonances and a larger region of the mass-mixing parameter space affected by flavor conversion at $t=50$~ms than at $t=10$~ms as can be seen from Fig.~\ref{fig:P_nu_10_ms}.

%%%%%%%%%%%%%%%%%%%%%%%%%%%%%%%%%%
%%%%%%%%%%%%%%%%%%%%%%%%%%%%%%%%%%	
\section{Outlook}
\label{sec:outlook}
%%%%%%%%%%%%%%%%%%%%%%%%%%%%%%%%%%
%%%%%%%%%%%%%%%%%%%%%%%%%%%%%%%%%%
By relying on a two-flavor framework ($1$ active $+1$ sterile species), we have explored the active-sterile flavor conversion phenomenology in compact binary merger remnants for the first time. We have investigated the production of sterile states as  a function of the  sterile neutrino mixing parameters, representative radial directions of neutrino emission
from the accretion torus, and temporal evolution of the merger remnant. 

Because of the torus geometry and the neutron richness of the environment, large flavor conversion occurs for antineutrinos. In particular, differently from the SN case, we find that multiple (up to six, see Fig.~\ref{fig:potential}) MSW resonances can take place according to the neutrino emission direction. It is important to stress that, while our representative radial directions highlight differences in the active-sterile flavor conversion phenomenology, the impact of active-sterile neutrino conversion on the physics of the remnant is direction dependent. But,  an assessment of the latter is beyond the scope of this work. The torus 
geometry is responsible for a large production of sterile neutrinos (and no antineutrinos) in the proximity of the polar region and more sterile antineutrinos  than neutrinos in the equatorial region. While we rely on the output of one hydrodynamical simulation of a BH accretion torus~\cite{Just:2014fka}, our main conclusions should be generic as they are fundamentally linked to the  characteristic properties and geometry of binary merger remnants.

As the BH torus evolves, the active-sterile oscillation phenomenology remains unchanged overall. However, the shallower baryon density at later times is responsible for more adiabatic flavor conversion that leads to a larger region of the mass-mixing parameter space being affected by active-sterile flavor conversion.

Our findings rely on a simplified framework for what concerns the modeling of the flavor conversion physics.  We neglect any impact of neutrino-neutrino interaction on the active-sterile neutrino conversion, because of the uncertainties currently involved in our understanding of this phenomenon and the related numerical challenges. Yet, within a simplified framework, it has been proven  that neutrino self-interaction could further affect the active-sterile conversion physics in the SN context~\cite{Tamborra:2011is,Wu:2013gxa,Pllumbi:2014saa,Xiong:2019nvw}.   

Despite the caveats of our modeling, our results robustly suggest that the  non-trivial active-sterile flavor phenomenology occurring in merger remnants can have indirect implications on the disk cooling rate and its outflows. 
For instance,  by relying on the findings of Ref.~\cite{Just:2022flt}, we deduce that the adiabaticity of  $\nu_e$ and $\bar\nu_e$ flavor conversion into sterile states  inside the neutrino sphere (see, e.g., the 1st resonance panel of Fig.~\ref{fig:eng_avg_surv_90_nu} for $\theta=90^\circ$) could potentially accelerate the cooling of the remnant disk and  lower  $Y_e$ in the disk in a similar fashion to what was discussed in Ref.~\cite{Just:2022flt}.
In addition,  flavor conversion occurring in the polar region at radii of $\mathcal{O}(100)$~km (see, e.g., Fig.~\ref{fig:potential} or \ref{fig:surv_prob}) would also reduce the neutrino capture rates by nucleons in the polar outflows where $Y_e$ and the nucleosynthesis outcomes are sensitive to the abundance of the electron flavors, like in the scenario considered  in Ref.~\cite{Wu:2017drk}.

A robust assessment of the impact of the active-sterile flavor conversion physics on  the electromagnetic observables as well as on the disk cooling rate are left to future work, once a reliable modeling of the  active-active  conversion physics in the presence of neutrino self-interactions will be available. In order to place robust constraints on the sterile mixing parameters through future multi-messenger observations, a survey of the flavor conversion phenomenology for various compact binary merger models and  related feedback on the observables will be required. This work proves the  unexplored potential  of upcoming multi-messenger observations of compact binary merger remnants to unveil the existence of sterile neutrinos.

\acknowledgments
We would like to thank Oliver Just  for insightful discussions.
We acknowledge support from the Villum Foundation (Project No.~13164), the Carlsberg Foundation (CF18-0183),  the Deutsche Forschungsgemeinschaft through Sonderforschungbereich
SFB~1258 ``Neutrinos and Dark Matter in Astro- and
Particle Physics'' (NDM), the National Science and Technology Council, Taiwan under Grant Nos.~110-2112-M-001-050, 111-2628-M-001-003-MY4, the Academia Sinica (Project No.~AS-CDA-109-M11), and the Physics Division of the National Center for Theoretical Sciences, Taiwan.

%%%%%%%%%%%%%%%%%%%%%%%%%%%%%%%%%%%%%%%%%%%%%%%%%%%%%%%%%%%%%%%%%%%%%%%%%%%%%%%
%%%%%%%%%%%%%%%%%%%%%%%%%%%%%%%%%%%%%%%%%%%%%%%%%%%%%%%%%%%%%%%%%%%%%%%%%%%%%%%

%%%%%%%%%%%%%%%%%%%%%%%%%%%%%%%%%%%%%%%%%%%%%%%%%%%%%%%%%%%%%%%%%%%%%%%%%%%%%%%
%%%%%%%%%%%%%%%%%%%%%%%%%%%%%%%%%%%%%%%%%%%%%%%%%%%%%%%%%%%%%%%%%%%%%%%%%%%%%%%
\bibliography{bibliography.bib}

\begin{thebibliography}{97}%
\makeatletter
\providecommand \@ifxundefined [1]{%
 \@ifx{#1\undefined}
}%
\providecommand \@ifnum [1]{%
 \ifnum #1\expandafter \@firstoftwo
 \else \expandafter \@secondoftwo
 \fi
}%
\providecommand \@ifx [1]{%
 \ifx #1\expandafter \@firstoftwo
 \else \expandafter \@secondoftwo
 \fi
}%
\providecommand \natexlab [1]{#1}%
\providecommand \enquote  [1]{``#1''}%
\providecommand \bibnamefont  [1]{#1}%
\providecommand \bibfnamefont [1]{#1}%
\providecommand \citenamefont [1]{#1}%
\providecommand \href@noop [0]{\@secondoftwo}%
\providecommand \href [0]{\begingroup \@sanitize@url \@href}%
\providecommand \@href[1]{\@@startlink{#1}\@@href}%
\providecommand \@@href[1]{\endgroup#1\@@endlink}%
\providecommand \@sanitize@url [0]{\catcode `\\12\catcode `\$12\catcode
  `\&12\catcode `\#12\catcode `\^12\catcode `\_12\catcode `\%12\relax}%
\providecommand \@@startlink[1]{}%
\providecommand \@@endlink[0]{}%
\providecommand \url  [0]{\begingroup\@sanitize@url \@url }%
\providecommand \@url [1]{\endgroup\@href {#1}{\urlprefix }}%
\providecommand \urlprefix  [0]{URL }%
\providecommand \Eprint [0]{\href }%
\providecommand \doibase [0]{http://dx.doi.org/}%
\providecommand \selectlanguage [0]{\@gobble}%
\providecommand \bibinfo  [0]{\@secondoftwo}%
\providecommand \bibfield  [0]{\@secondoftwo}%
\providecommand \translation [1]{[#1]}%
\providecommand \BibitemOpen [0]{}%
\providecommand \bibitemStop [0]{}%
\providecommand \bibitemNoStop [0]{.\EOS\space}%
\providecommand \EOS [0]{\spacefactor3000\relax}%
\providecommand \BibitemShut  [1]{\csname bibitem#1\endcsname}%
\let\auto@bib@innerbib\@empty
%</preamble>
\bibitem [{\citenamefont {Margutti}\ and\ \citenamefont
  {Chornock}(2021)}]{Margutti:2020xbo}%
  \BibitemOpen
  \bibfield  {author} {\bibinfo {author} {\bibfnamefont {Raffaella}\
  \bibnamefont {Margutti}}\ and\ \bibinfo {author} {\bibfnamefont {Ryan}\
  \bibnamefont {Chornock}},\ }\bibfield  {title} {\enquote {\bibinfo {title}
  {{First Multimessenger Observations of a Neutron Star Merger}},}\ }\href
  {\doibase 10.1146/annurev-astro-112420-030742} {\bibfield  {journal}
  {\bibinfo  {journal} {Ann. Rev. Astron. Astrophys.}\ }\textbf {\bibinfo
  {volume} {59}},\ \bibinfo {pages} {155--202} (\bibinfo {year} {2021})},\
  \Eprint {http://arxiv.org/abs/2012.04810} {arXiv:2012.04810 [astro-ph.HE]}
  \BibitemShut {NoStop}%
\bibitem [{\citenamefont {Abbott}\ \emph
  {et~al.}(2017{\natexlab{a}})\citenamefont {Abbott} \emph
  {et~al.}}]{LIGOScientific:2017vwq}%
  \BibitemOpen
  \bibfield  {author} {\bibinfo {author} {\bibfnamefont {Benjamin~P.}\
  \bibnamefont {Abbott}} \emph {et~al.} (\bibinfo {collaboration} {LIGO
  Scientific, Virgo}),\ }\bibfield  {title} {\enquote {\bibinfo {title}
  {{GW170817: Observation of Gravitational Waves from a Binary Neutron Star
  Inspiral}},}\ }\href {\doibase 10.1103/PhysRevLett.119.161101} {\bibfield
  {journal} {\bibinfo  {journal} {Phys. Rev. Lett.}\ }\textbf {\bibinfo
  {volume} {119}},\ \bibinfo {pages} {161101} (\bibinfo {year}
  {2017}{\natexlab{a}})},\ \Eprint {http://arxiv.org/abs/1710.05832}
  {arXiv:1710.05832 [gr-qc]} \BibitemShut {NoStop}%
\bibitem [{\citenamefont {Abbott}\ \emph
  {et~al.}(2017{\natexlab{b}})\citenamefont {Abbott} \emph
  {et~al.}}]{LIGOScientific:2017zic}%
  \BibitemOpen
  \bibfield  {author} {\bibinfo {author} {\bibfnamefont {Benjamin~P.}\
  \bibnamefont {Abbott}} \emph {et~al.} (\bibinfo {collaboration} {LIGO
  Scientific, Virgo, Fermi-GBM, INTEGRAL}),\ }\bibfield  {title} {\enquote
  {\bibinfo {title} {{Gravitational Waves and Gamma-rays from a Binary Neutron
  Star Merger: GW170817 and GRB 170817A}},}\ }\href {\doibase
  10.3847/2041-8213/aa920c} {\bibfield  {journal} {\bibinfo  {journal}
  {Astrophys. J. Lett.}\ }\textbf {\bibinfo {volume} {848}},\ \bibinfo {pages}
  {L13} (\bibinfo {year} {2017}{\natexlab{b}})},\ \Eprint
  {http://arxiv.org/abs/1710.05834} {arXiv:1710.05834 [astro-ph.HE]}
  \BibitemShut {NoStop}%
\bibitem [{\citenamefont {Abbott}\ \emph
  {et~al.}(2017{\natexlab{c}})\citenamefont {Abbott} \emph
  {et~al.}}]{LIGOScientific:2017ync}%
  \BibitemOpen
  \bibfield  {author} {\bibinfo {author} {\bibfnamefont {Benjamin~P.}\
  \bibnamefont {Abbott}} \emph {et~al.} (\bibinfo {collaboration} {LIGO
  Scientific, Virgo, Fermi GBM, INTEGRAL, IceCube, AstroSat Cadmium Zinc
  Telluride Imager Team, IPN, Insight-Hxmt, ANTARES, Swift, AGILE Team, 1M2H
  Team, Dark Energy Camera GW-EM, DES, DLT40, GRAWITA, Fermi-LAT, ATCA, ASKAP,
  Las Cumbres Observatory Group, OzGrav, DWF (Deeper Wider Faster Program),
  AST3, CAASTRO, VINROUGE, MASTER, J-GEM, GROWTH, JAGWAR, CaltechNRAO,
  TTU-NRAO, NuSTAR, Pan-STARRS, MAXI Team, TZAC Consortium, KU, Nordic Optical
  Telescope, ePESSTO, GROND, Texas Tech University, SALT Group, TOROS, BOOTES,
  MWA, CALET, IKI-GW Follow-up, H.E.S.S., LOFAR, LWA, HAWC, Pierre Auger, ALMA,
  Euro VLBI Team, Pi of Sky, Chandra Team at McGill University, DFN, ATLAS
  Telescopes, High Time Resolution Universe Survey, RIMAS, RATIR, SKA South
  Africa/MeerKAT}),\ }\bibfield  {title} {\enquote {\bibinfo {title}
  {{Multi-messenger Observations of a Binary Neutron Star Merger}},}\ }\href
  {\doibase 10.3847/2041-8213/aa91c9} {\bibfield  {journal} {\bibinfo
  {journal} {Astrophys. J. Lett.}\ }\textbf {\bibinfo {volume} {848}},\
  \bibinfo {pages} {L12} (\bibinfo {year} {2017}{\natexlab{c}})},\ \Eprint
  {http://arxiv.org/abs/1710.05833} {arXiv:1710.05833 [astro-ph.HE]}
  \BibitemShut {NoStop}%
\bibitem [{\citenamefont {Goldstein}\ \emph {et~al.}(2017)\citenamefont
  {Goldstein} \emph {et~al.}}]{Goldstein:2017mmi}%
  \BibitemOpen
  \bibfield  {author} {\bibinfo {author} {\bibfnamefont {Adam}\ \bibnamefont
  {Goldstein}} \emph {et~al.},\ }\bibfield  {title} {\enquote {\bibinfo {title}
  {{An Ordinary Short Gamma-Ray Burst with Extraordinary Implications:
  Fermi-GBM Detection of GRB 170817A}},}\ }\href {\doibase
  10.3847/2041-8213/aa8f41} {\bibfield  {journal} {\bibinfo  {journal}
  {Astrophys. J. Lett.}\ }\textbf {\bibinfo {volume} {848}},\ \bibinfo {pages}
  {L14} (\bibinfo {year} {2017})},\ \Eprint {http://arxiv.org/abs/1710.05446}
  {arXiv:1710.05446 [astro-ph.HE]} \BibitemShut {NoStop}%
\bibitem [{\citenamefont {Savchenko}\ \emph {et~al.}(2017)\citenamefont
  {Savchenko} \emph {et~al.}}]{Savchenko:2017ffs}%
  \BibitemOpen
  \bibfield  {author} {\bibinfo {author} {\bibfnamefont {Volodymyr}\
  \bibnamefont {Savchenko}} \emph {et~al.},\ }\bibfield  {title} {\enquote
  {\bibinfo {title} {{INTEGRAL Detection of the First Prompt Gamma-Ray Signal
  Coincident with the Gravitational-wave Event GW170817}},}\ }\href {\doibase
  10.3847/2041-8213/aa8f94} {\bibfield  {journal} {\bibinfo  {journal}
  {Astrophys. J. Lett.}\ }\textbf {\bibinfo {volume} {848}},\ \bibinfo {pages}
  {L15} (\bibinfo {year} {2017})},\ \Eprint {http://arxiv.org/abs/1710.05449}
  {arXiv:1710.05449 [astro-ph.HE]} \BibitemShut {NoStop}%
\bibitem [{\citenamefont {Margutti}\ \emph {et~al.}(2017)\citenamefont
  {Margutti} \emph {et~al.}}]{Margutti:2017cjl}%
  \BibitemOpen
  \bibfield  {author} {\bibinfo {author} {\bibfnamefont {Raffaella}\
  \bibnamefont {Margutti}} \emph {et~al.},\ }\bibfield  {title} {\enquote
  {\bibinfo {title} {{The Electromagnetic Counterpart of the Binary Neutron
  Star Merger LIGO/VIRGO GW170817. V. Rising X-ray Emission from an Off-Axis
  Jet}},}\ }\href {\doibase 10.3847/2041-8213/aa9057} {\bibfield  {journal}
  {\bibinfo  {journal} {Astrophys. J. Lett.}\ }\textbf {\bibinfo {volume}
  {848}},\ \bibinfo {pages} {L20} (\bibinfo {year} {2017})},\ \Eprint
  {http://arxiv.org/abs/1710.05431} {arXiv:1710.05431 [astro-ph.HE]}
  \BibitemShut {NoStop}%
\bibitem [{\citenamefont {Troja}\ \emph {et~al.}(2017)\citenamefont {Troja}
  \emph {et~al.}}]{Troja:2017nqp}%
  \BibitemOpen
  \bibfield  {author} {\bibinfo {author} {\bibfnamefont {Eleonora}\
  \bibnamefont {Troja}} \emph {et~al.},\ }\bibfield  {title} {\enquote
  {\bibinfo {title} {{The X-ray counterpart to the gravitational wave event GW
  170817}},}\ }\href {\doibase 10.1038/nature24290} {\bibfield  {journal}
  {\bibinfo  {journal} {Nature}\ }\textbf {\bibinfo {volume} {551}},\ \bibinfo
  {pages} {71--74} (\bibinfo {year} {2017})},\ \Eprint
  {http://arxiv.org/abs/1710.05433} {arXiv:1710.05433 [astro-ph.HE]}
  \BibitemShut {NoStop}%
\bibitem [{\citenamefont {Kasen}\ \emph {et~al.}(2017)\citenamefont {Kasen},
  \citenamefont {Metzger}, \citenamefont {Barnes}, \citenamefont {Quataert},\
  and\ \citenamefont {Ramirez-Ruiz}}]{Kasen:2017sxr}%
  \BibitemOpen
  \bibfield  {author} {\bibinfo {author} {\bibfnamefont {Daniel}\ \bibnamefont
  {Kasen}}, \bibinfo {author} {\bibfnamefont {Brian}\ \bibnamefont {Metzger}},
  \bibinfo {author} {\bibfnamefont {Jennifer}\ \bibnamefont {Barnes}}, \bibinfo
  {author} {\bibfnamefont {Eliot}\ \bibnamefont {Quataert}}, \ and\ \bibinfo
  {author} {\bibfnamefont {Enrico}\ \bibnamefont {Ramirez-Ruiz}},\ }\bibfield
  {title} {\enquote {\bibinfo {title} {{Origin of the heavy elements in binary
  neutron-star mergers from a gravitational wave event}},}\ }\href {\doibase
  10.1038/nature24453} {\bibfield  {journal} {\bibinfo  {journal} {Nature}\
  }\textbf {\bibinfo {volume} {551}},\ \bibinfo {pages} {80} (\bibinfo {year}
  {2017})},\ \Eprint {http://arxiv.org/abs/1710.05463} {arXiv:1710.05463
  [astro-ph.HE]} \BibitemShut {NoStop}%
\bibitem [{\citenamefont {Drout}\ \emph {et~al.}(2017)\citenamefont {Drout}
  \emph {et~al.}}]{Drout:2017ijr}%
  \BibitemOpen
  \bibfield  {author} {\bibinfo {author} {\bibfnamefont {Maria~R.}\
  \bibnamefont {Drout}} \emph {et~al.},\ }\bibfield  {title} {\enquote
  {\bibinfo {title} {{Light Curves of the Neutron Star Merger GW170817/SSS17a:
  Implications for R-Process Nucleosynthesis}},}\ }\href {\doibase
  10.1126/science.aaq0049} {\bibfield  {journal} {\bibinfo  {journal}
  {Science}\ }\textbf {\bibinfo {volume} {358}},\ \bibinfo {pages} {1570--1574}
  (\bibinfo {year} {2017})},\ \Eprint {http://arxiv.org/abs/1710.05443}
  {arXiv:1710.05443 [astro-ph.HE]} \BibitemShut {NoStop}%
\bibitem [{\citenamefont {Cowperthwaite}\ \emph {et~al.}(2017)\citenamefont
  {Cowperthwaite} \emph {et~al.}}]{Cowperthwaite:2017dyu}%
  \BibitemOpen
  \bibfield  {author} {\bibinfo {author} {\bibfnamefont {Philip~S.}\
  \bibnamefont {Cowperthwaite}} \emph {et~al.},\ }\bibfield  {title} {\enquote
  {\bibinfo {title} {{The Electromagnetic Counterpart of the Binary Neutron
  Star Merger LIGO/Virgo GW170817. II. UV, Optical, and Near-infrared Light
  Curves and Comparison to Kilonova Models}},}\ }\href {\doibase
  10.3847/2041-8213/aa8fc7} {\bibfield  {journal} {\bibinfo  {journal}
  {Astrophys. J. Lett.}\ }\textbf {\bibinfo {volume} {848}},\ \bibinfo {pages}
  {L17} (\bibinfo {year} {2017})},\ \Eprint {http://arxiv.org/abs/1710.05840}
  {arXiv:1710.05840 [astro-ph.HE]} \BibitemShut {NoStop}%
\bibitem [{\citenamefont {Villar}\ \emph {et~al.}(2017)\citenamefont {Villar}
  \emph {et~al.}}]{Villar:2017wcc}%
  \BibitemOpen
  \bibfield  {author} {\bibinfo {author} {\bibfnamefont {V.~Ashley}\
  \bibnamefont {Villar}} \emph {et~al.},\ }\bibfield  {title} {\enquote
  {\bibinfo {title} {{The Combined Ultraviolet, Optical, and Near-Infrared
  Light Curves of the Kilonova Associated with the Binary Neutron Star Merger
  GW170817: Unified Data Set, Analytic Models, and Physical Implications}},}\
  }\href {\doibase 10.3847/2041-8213/aa9c84} {\bibfield  {journal} {\bibinfo
  {journal} {Astrophys. J. Lett.}\ }\textbf {\bibinfo {volume} {851}},\
  \bibinfo {pages} {L21} (\bibinfo {year} {2017})},\ \Eprint
  {http://arxiv.org/abs/1710.11576} {arXiv:1710.11576 [astro-ph.HE]}
  \BibitemShut {NoStop}%
\bibitem [{\citenamefont {Shibata}\ \emph {et~al.}(2017)\citenamefont
  {Shibata}, \citenamefont {Fujibayashi}, \citenamefont {Hotokezaka},
  \citenamefont {Kiuchi}, \citenamefont {Kyutoku}, \citenamefont {Sekiguchi},\
  and\ \citenamefont {Tanaka}}]{Shibata:2017xdx}%
  \BibitemOpen
  \bibfield  {author} {\bibinfo {author} {\bibfnamefont {Masaru}\ \bibnamefont
  {Shibata}}, \bibinfo {author} {\bibfnamefont {Sho}\ \bibnamefont
  {Fujibayashi}}, \bibinfo {author} {\bibfnamefont {Kenta}\ \bibnamefont
  {Hotokezaka}}, \bibinfo {author} {\bibfnamefont {Kenta}\ \bibnamefont
  {Kiuchi}}, \bibinfo {author} {\bibfnamefont {Koutarou}\ \bibnamefont
  {Kyutoku}}, \bibinfo {author} {\bibfnamefont {Yuichiro}\ \bibnamefont
  {Sekiguchi}}, \ and\ \bibinfo {author} {\bibfnamefont {Masaomi}\ \bibnamefont
  {Tanaka}},\ }\bibfield  {title} {\enquote {\bibinfo {title} {{Modeling
  GW170817 based on numerical relativity and its implications}},}\ }\href
  {\doibase 10.1103/PhysRevD.96.123012} {\bibfield  {journal} {\bibinfo
  {journal} {Phys. Rev. D}\ }\textbf {\bibinfo {volume} {96}},\ \bibinfo
  {pages} {123012} (\bibinfo {year} {2017})},\ \Eprint
  {http://arxiv.org/abs/1710.07579} {arXiv:1710.07579 [astro-ph.HE]}
  \BibitemShut {NoStop}%
\bibitem [{\citenamefont {Metzger}(2020)}]{Metzger:2019zeh}%
  \BibitemOpen
  \bibfield  {author} {\bibinfo {author} {\bibfnamefont {Brian~D.}\
  \bibnamefont {Metzger}},\ }\bibfield  {title} {\enquote {\bibinfo {title}
  {{Kilonovae}},}\ }\href {\doibase 10.1007/s41114-019-0024-0} {\bibfield
  {journal} {\bibinfo  {journal} {Living Rev. Rel.}\ }\textbf {\bibinfo
  {volume} {23}},\ \bibinfo {pages} {1} (\bibinfo {year} {2020})},\ \Eprint
  {http://arxiv.org/abs/1910.01617} {arXiv:1910.01617 [astro-ph.HE]}
  \BibitemShut {NoStop}%
\bibitem [{\citenamefont {Veske}\ \emph {et~al.}(2020)\citenamefont {Veske},
  \citenamefont {M\'arka}, \citenamefont {Bartos},\ and\ \citenamefont
  {M\'arka}}]{Veske:2020yjt}%
  \BibitemOpen
  \bibfield  {author} {\bibinfo {author} {\bibfnamefont {Do\u{g}a}\
  \bibnamefont {Veske}}, \bibinfo {author} {\bibfnamefont {Zsuzsa}\
  \bibnamefont {M\'arka}}, \bibinfo {author} {\bibfnamefont {Imre}\
  \bibnamefont {Bartos}}, \ and\ \bibinfo {author} {\bibfnamefont {Szabolcs}\
  \bibnamefont {M\'arka}},\ }\bibfield  {title} {\enquote {\bibinfo {title}
  {{Neutrino emission upper limits with maximum likelihood estimators for joint
  astrophysical neutrino searches with large sky localizations}},}\ }\href
  {\doibase 10.1088/1475-7516/2020/05/016} {\bibfield  {journal} {\bibinfo
  {journal} {JCAP}\ }\textbf {\bibinfo {volume} {05}},\ \bibinfo {pages} {016}
  (\bibinfo {year} {2020})},\ \Eprint {http://arxiv.org/abs/2001.00566}
  {arXiv:2001.00566 [astro-ph.HE]} \BibitemShut {NoStop}%
\bibitem [{\citenamefont {Aartsen}\ \emph {et~al.}(2020)\citenamefont {Aartsen}
  \emph {et~al.}}]{IceCube:2020xks}%
  \BibitemOpen
  \bibfield  {author} {\bibinfo {author} {\bibfnamefont {Mark~G.}\ \bibnamefont
  {Aartsen}} \emph {et~al.} (\bibinfo {collaboration} {IceCube}),\ }\bibfield
  {title} {\enquote {\bibinfo {title} {{IceCube Search for Neutrinos Coincident
  with Compact Binary Mergers from LIGO-Virgo\textquoteright{}s First
  Gravitational-wave Transient Catalog}},}\ }\href {\doibase
  10.3847/2041-8213/ab9d24} {\bibfield  {journal} {\bibinfo  {journal}
  {Astrophys. J. Lett.}\ }\textbf {\bibinfo {volume} {898}},\ \bibinfo {pages}
  {L10} (\bibinfo {year} {2020})},\ \Eprint {http://arxiv.org/abs/2004.02910}
  {arXiv:2004.02910 [astro-ph.HE]} \BibitemShut {NoStop}%
\bibitem [{\citenamefont {Hayato}\ \emph {et~al.}(2018)\citenamefont {Hayato}
  \emph {et~al.}}]{Super-Kamiokande:2018dbf}%
  \BibitemOpen
  \bibfield  {author} {\bibinfo {author} {\bibfnamefont {Yoshinari}\
  \bibnamefont {Hayato}} \emph {et~al.} (\bibinfo {collaboration}
  {Super-Kamiokande}),\ }\bibfield  {title} {\enquote {\bibinfo {title}
  {{Search for Neutrinos in Super-Kamiokande Associated with the GW170817
  Neutron-star Merger}},}\ }\href {\doibase 10.3847/2041-8213/aabaca}
  {\bibfield  {journal} {\bibinfo  {journal} {Astrophys. J. Lett.}\ }\textbf
  {\bibinfo {volume} {857}},\ \bibinfo {pages} {L4} (\bibinfo {year} {2018})},\
  \Eprint {http://arxiv.org/abs/1802.04379} {arXiv:1802.04379 [astro-ph.HE]}
  \BibitemShut {NoStop}%
\bibitem [{\citenamefont {Foucart}\ \emph {et~al.}(2015)\citenamefont
  {Foucart}, \citenamefont {O'Connor}, \citenamefont {Roberts}, \citenamefont
  {Duez}, \citenamefont {Haas}, \citenamefont {Kidder}, \citenamefont {Ott},
  \citenamefont {Pfeiffer}, \citenamefont {Scheel},\ and\ \citenamefont
  {Szilagyi}}]{Foucart:2015vpa}%
  \BibitemOpen
  \bibfield  {author} {\bibinfo {author} {\bibfnamefont {Francois}\
  \bibnamefont {Foucart}}, \bibinfo {author} {\bibfnamefont {Evan}\
  \bibnamefont {O'Connor}}, \bibinfo {author} {\bibfnamefont {Luke}\
  \bibnamefont {Roberts}}, \bibinfo {author} {\bibfnamefont {Matthew~D.}\
  \bibnamefont {Duez}}, \bibinfo {author} {\bibfnamefont {Roland}\ \bibnamefont
  {Haas}}, \bibinfo {author} {\bibfnamefont {Lawrence~E.}\ \bibnamefont
  {Kidder}}, \bibinfo {author} {\bibfnamefont {Christian~D.}\ \bibnamefont
  {Ott}}, \bibinfo {author} {\bibfnamefont {Harald~P.}\ \bibnamefont
  {Pfeiffer}}, \bibinfo {author} {\bibfnamefont {Mark~A.}\ \bibnamefont
  {Scheel}}, \ and\ \bibinfo {author} {\bibfnamefont {Bela}\ \bibnamefont
  {Szilagyi}},\ }\bibfield  {title} {\enquote {\bibinfo {title} {{Post-merger
  evolution of a neutron star-black hole binary with neutrino transport}},}\
  }\href {\doibase 10.1103/PhysRevD.91.124021} {\bibfield  {journal} {\bibinfo
  {journal} {Phys. Rev. D}\ }\textbf {\bibinfo {volume} {91}},\ \bibinfo
  {pages} {124021} (\bibinfo {year} {2015})},\ \Eprint
  {http://arxiv.org/abs/1502.04146} {arXiv:1502.04146 [astro-ph.HE]}
  \BibitemShut {NoStop}%
\bibitem [{\citenamefont {Ruffert}\ \emph {et~al.}(1997)\citenamefont
  {Ruffert}, \citenamefont {Janka}, \citenamefont {Takahashi},\ and\
  \citenamefont {Schaefer}}]{Ruffert:1996by}%
  \BibitemOpen
  \bibfield  {author} {\bibinfo {author} {\bibfnamefont {Maximilian}\
  \bibnamefont {Ruffert}}, \bibinfo {author} {\bibfnamefont {H.-Thomas}\
  \bibnamefont {Janka}}, \bibinfo {author} {\bibfnamefont {Kazuya}\
  \bibnamefont {Takahashi}}, \ and\ \bibinfo {author} {\bibfnamefont {Gerhard}\
  \bibnamefont {Schaefer}},\ }\bibfield  {title} {\enquote {\bibinfo {title}
  {{Coalescing neutron stars: A Step towards physical models. 2. Neutrino
  emission, neutron tori, and gamma-ray bursts}},}\ }\href@noop {} {\bibfield
  {journal} {\bibinfo  {journal} {Astron. Astrophys.}\ }\textbf {\bibinfo
  {volume} {319}},\ \bibinfo {pages} {122--153} (\bibinfo {year} {1997})},\
  \Eprint {http://arxiv.org/abs/astro-ph/9606181} {arXiv:astro-ph/9606181}
  \BibitemShut {NoStop}%
\bibitem [{\citenamefont {Metzger}\ and\ \citenamefont
  {Fern\'andez}(2014)}]{Metzger:2014ila}%
  \BibitemOpen
  \bibfield  {author} {\bibinfo {author} {\bibfnamefont {Brian~D.}\
  \bibnamefont {Metzger}}\ and\ \bibinfo {author} {\bibfnamefont {Rodrigo}\
  \bibnamefont {Fern\'andez}},\ }\bibfield  {title} {\enquote {\bibinfo {title}
  {{Red or blue? A potential kilonova imprint of the delay until black hole
  formation following a neutron star merger}},}\ }\href {\doibase
  10.1093/mnras/stu802} {\bibfield  {journal} {\bibinfo  {journal} {Mon. Not.
  Roy. Astron. Soc.}\ }\textbf {\bibinfo {volume} {441}},\ \bibinfo {pages}
  {3444--3453} (\bibinfo {year} {2014})},\ \Eprint
  {http://arxiv.org/abs/1402.4803} {arXiv:1402.4803 [astro-ph.HE]} \BibitemShut
  {NoStop}%
\bibitem [{\citenamefont {Perego}\ \emph {et~al.}(2014)\citenamefont {Perego},
  \citenamefont {Rosswog}, \citenamefont {Cabez\'on}, \citenamefont {Korobkin},
  \citenamefont {K\"appeli}, \citenamefont {Arcones},\ and\ \citenamefont
  {Liebend\"orfer}}]{Perego:2014fma}%
  \BibitemOpen
  \bibfield  {author} {\bibinfo {author} {\bibfnamefont {Albino}\ \bibnamefont
  {Perego}}, \bibinfo {author} {\bibfnamefont {Stephan}\ \bibnamefont
  {Rosswog}}, \bibinfo {author} {\bibfnamefont {Ruben~M.}\ \bibnamefont
  {Cabez\'on}}, \bibinfo {author} {\bibfnamefont {Oleg}\ \bibnamefont
  {Korobkin}}, \bibinfo {author} {\bibfnamefont {Roger}\ \bibnamefont
  {K\"appeli}}, \bibinfo {author} {\bibfnamefont {Almudena}\ \bibnamefont
  {Arcones}}, \ and\ \bibinfo {author} {\bibfnamefont {Matthias}\ \bibnamefont
  {Liebend\"orfer}},\ }\bibfield  {title} {\enquote {\bibinfo {title}
  {{Neutrino-driven winds from neutron star merger remnants}},}\ }\href
  {\doibase 10.1093/mnras/stu1352} {\bibfield  {journal} {\bibinfo  {journal}
  {Mon. Not. Roy. Astron. Soc.}\ }\textbf {\bibinfo {volume} {443}},\ \bibinfo
  {pages} {3134--3156} (\bibinfo {year} {2014})},\ \Eprint
  {http://arxiv.org/abs/1405.6730} {arXiv:1405.6730 [astro-ph.HE]} \BibitemShut
  {NoStop}%
\bibitem [{\citenamefont {Wanajo}\ \emph {et~al.}(2014)\citenamefont {Wanajo},
  \citenamefont {Sekiguchi}, \citenamefont {Nishimura}, \citenamefont {Kiuchi},
  \citenamefont {Kyutoku},\ and\ \citenamefont {Shibata}}]{Wanajo:2014wha}%
  \BibitemOpen
  \bibfield  {author} {\bibinfo {author} {\bibfnamefont {Shinya}\ \bibnamefont
  {Wanajo}}, \bibinfo {author} {\bibfnamefont {Yuichiro}\ \bibnamefont
  {Sekiguchi}}, \bibinfo {author} {\bibfnamefont {Nobuya}\ \bibnamefont
  {Nishimura}}, \bibinfo {author} {\bibfnamefont {Kenta}\ \bibnamefont
  {Kiuchi}}, \bibinfo {author} {\bibfnamefont {Koutarou}\ \bibnamefont
  {Kyutoku}}, \ and\ \bibinfo {author} {\bibfnamefont {Masaru}\ \bibnamefont
  {Shibata}},\ }\bibfield  {title} {\enquote {\bibinfo {title} {{Production of
  all the $r$-process nuclides in the dynamical ejecta of neutron star
  mergers}},}\ }\href {\doibase 10.1088/2041-8205/789/2/L39} {\bibfield
  {journal} {\bibinfo  {journal} {Astrophys. J. Lett.}\ }\textbf {\bibinfo
  {volume} {789}},\ \bibinfo {pages} {L39} (\bibinfo {year} {2014})},\ \Eprint
  {http://arxiv.org/abs/1402.7317} {arXiv:1402.7317 [astro-ph.SR]} \BibitemShut
  {NoStop}%
\bibitem [{\citenamefont {Just}\ \emph {et~al.}(2015)\citenamefont {Just},
  \citenamefont {Bauswein}, \citenamefont {Pulpillo}, \citenamefont {Goriely},\
  and\ \citenamefont {Janka}}]{Just:2014fka}%
  \BibitemOpen
  \bibfield  {author} {\bibinfo {author} {\bibfnamefont {Oliver}\ \bibnamefont
  {Just}}, \bibinfo {author} {\bibfnamefont {Andreas}\ \bibnamefont
  {Bauswein}}, \bibinfo {author} {\bibfnamefont {Ricard~Ardevol}\ \bibnamefont
  {Pulpillo}}, \bibinfo {author} {\bibfnamefont {Stephane}\ \bibnamefont
  {Goriely}}, \ and\ \bibinfo {author} {\bibfnamefont {H.-Thomas}\ \bibnamefont
  {Janka}},\ }\bibfield  {title} {\enquote {\bibinfo {title} {{Comprehensive
  nucleosynthesis analysis for ejecta of compact binary mergers}},}\ }\href
  {\doibase 10.1093/mnras/stv009} {\bibfield  {journal} {\bibinfo  {journal}
  {Mon. Not. Roy. Astron. Soc.}\ }\textbf {\bibinfo {volume} {448}},\ \bibinfo
  {pages} {541--567} (\bibinfo {year} {2015})},\ \Eprint
  {http://arxiv.org/abs/1406.2687} {arXiv:1406.2687 [astro-ph.SR]} \BibitemShut
  {NoStop}%
\bibitem [{\citenamefont {Fujibayashi}\ \emph {et~al.}(2020)\citenamefont
  {Fujibayashi}, \citenamefont {Shibata}, \citenamefont {Wanajo}, \citenamefont
  {Kiuchi}, \citenamefont {Kyutoku},\ and\ \citenamefont
  {Sekiguchi}}]{Fujibayashi:2020qda}%
  \BibitemOpen
  \bibfield  {author} {\bibinfo {author} {\bibfnamefont {Sho}\ \bibnamefont
  {Fujibayashi}}, \bibinfo {author} {\bibfnamefont {Masaru}\ \bibnamefont
  {Shibata}}, \bibinfo {author} {\bibfnamefont {Shinya}\ \bibnamefont
  {Wanajo}}, \bibinfo {author} {\bibfnamefont {Kenta}\ \bibnamefont {Kiuchi}},
  \bibinfo {author} {\bibfnamefont {Koutarou}\ \bibnamefont {Kyutoku}}, \ and\
  \bibinfo {author} {\bibfnamefont {Yuichiro}\ \bibnamefont {Sekiguchi}},\
  }\bibfield  {title} {\enquote {\bibinfo {title} {{Mass ejection from disks
  surrounding a low-mass black hole: Viscous neutrino-radiation hydrodynamics
  simulation in full general relativity}},}\ }\href {\doibase
  10.1103/PhysRevD.101.083029} {\bibfield  {journal} {\bibinfo  {journal}
  {Phys. Rev. D}\ }\textbf {\bibinfo {volume} {101}},\ \bibinfo {pages}
  {083029} (\bibinfo {year} {2020})},\ \Eprint
  {http://arxiv.org/abs/2001.04467} {arXiv:2001.04467 [astro-ph.HE]}
  \BibitemShut {NoStop}%
\bibitem [{\citenamefont {Kullmann}\ \emph {et~al.}(2022)\citenamefont
  {Kullmann}, \citenamefont {Goriely}, \citenamefont {Just}, \citenamefont
  {Ardevol-Pulpillo}, \citenamefont {Bauswein},\ and\ \citenamefont
  {Janka}}]{Kullmann:2021gvo}%
  \BibitemOpen
  \bibfield  {author} {\bibinfo {author} {\bibfnamefont {Ina}\ \bibnamefont
  {Kullmann}}, \bibinfo {author} {\bibfnamefont {Stephan}\ \bibnamefont
  {Goriely}}, \bibinfo {author} {\bibfnamefont {Oliver}\ \bibnamefont {Just}},
  \bibinfo {author} {\bibfnamefont {Ricard}\ \bibnamefont {Ardevol-Pulpillo}},
  \bibinfo {author} {\bibfnamefont {Andreas}\ \bibnamefont {Bauswein}}, \ and\
  \bibinfo {author} {\bibfnamefont {H.-Thomas}\ \bibnamefont {Janka}},\
  }\bibfield  {title} {\enquote {\bibinfo {title} {{Dynamical ejecta of neutron
  star mergers with nucleonic weak processes I: nucleosynthesis}},}\ }\href
  {\doibase 10.1093/mnras/stab3393} {\bibfield  {journal} {\bibinfo  {journal}
  {Mon. Not. Roy. Astron. Soc.}\ }\textbf {\bibinfo {volume} {510}},\ \bibinfo
  {pages} {2804--2819} (\bibinfo {year} {2022})},\ \Eprint
  {http://arxiv.org/abs/2109.02509} {arXiv:2109.02509 [astro-ph.HE]}
  \BibitemShut {NoStop}%
\bibitem [{\citenamefont {Narayan}\ \emph {et~al.}(1992)\citenamefont
  {Narayan}, \citenamefont {Paczynski},\ and\ \citenamefont
  {Piran}}]{Narayan:1992iy}%
  \BibitemOpen
  \bibfield  {author} {\bibinfo {author} {\bibfnamefont {Ramesh}\ \bibnamefont
  {Narayan}}, \bibinfo {author} {\bibfnamefont {Bohdan}\ \bibnamefont
  {Paczynski}}, \ and\ \bibinfo {author} {\bibfnamefont {Tsvi}\ \bibnamefont
  {Piran}},\ }\bibfield  {title} {\enquote {\bibinfo {title} {{Gamma-ray bursts
  as the death throes of massive binary stars}},}\ }\href {\doibase
  10.1086/186493} {\bibfield  {journal} {\bibinfo  {journal} {Astrophys. J.
  Lett.}\ }\textbf {\bibinfo {volume} {395}},\ \bibinfo {pages} {L83--L86}
  (\bibinfo {year} {1992})},\ \Eprint {http://arxiv.org/abs/astro-ph/9204001}
  {arXiv:astro-ph/9204001} \BibitemShut {NoStop}%
\bibitem [{\citenamefont {Berger}(2014)}]{Berger:2013jza}%
  \BibitemOpen
  \bibfield  {author} {\bibinfo {author} {\bibfnamefont {Edo}\ \bibnamefont
  {Berger}},\ }\bibfield  {title} {\enquote {\bibinfo {title} {{Short-Duration
  Gamma-Ray Bursts}},}\ }\href {\doibase 10.1146/annurev-astro-081913-035926}
  {\bibfield  {journal} {\bibinfo  {journal} {Ann. Rev. Astron. Astrophys.}\
  }\textbf {\bibinfo {volume} {52}},\ \bibinfo {pages} {43--105} (\bibinfo
  {year} {2014})},\ \Eprint {http://arxiv.org/abs/1311.2603} {arXiv:1311.2603
  [astro-ph.HE]} \BibitemShut {NoStop}%
\bibitem [{\citenamefont {Just}\ \emph {et~al.}(2016)\citenamefont {Just},
  \citenamefont {Obergaulinger}, \citenamefont {Janka}, \citenamefont
  {Bauswein},\ and\ \citenamefont {Schwarz}}]{Just:2015dba}%
  \BibitemOpen
  \bibfield  {author} {\bibinfo {author} {\bibfnamefont {Oliver}\ \bibnamefont
  {Just}}, \bibinfo {author} {\bibfnamefont {Martin}\ \bibnamefont
  {Obergaulinger}}, \bibinfo {author} {\bibfnamefont {H.-Thomas}\ \bibnamefont
  {Janka}}, \bibinfo {author} {\bibfnamefont {Andreas}\ \bibnamefont
  {Bauswein}}, \ and\ \bibinfo {author} {\bibfnamefont {Nicole}\ \bibnamefont
  {Schwarz}},\ }\bibfield  {title} {\enquote {\bibinfo {title} {{Neutron-star
  merger ejecta as obstacles to neutrino-powered jets of gamma-ray bursts}},}\
  }\href {\doibase 10.3847/2041-8205/816/2/L30} {\bibfield  {journal} {\bibinfo
   {journal} {Astrophys. J. Lett.}\ }\textbf {\bibinfo {volume} {816}},\
  \bibinfo {pages} {L30} (\bibinfo {year} {2016})},\ \Eprint
  {http://arxiv.org/abs/1510.04288} {arXiv:1510.04288 [astro-ph.HE]}
  \BibitemShut {NoStop}%
\bibitem [{\citenamefont {Malkus}\ \emph {et~al.}(2014)\citenamefont {Malkus},
  \citenamefont {Friedland},\ and\ \citenamefont
  {McLaughlin}}]{Malkus:2014iqa}%
  \BibitemOpen
  \bibfield  {author} {\bibinfo {author} {\bibfnamefont {Annelise}\
  \bibnamefont {Malkus}}, \bibinfo {author} {\bibfnamefont {Alexander}\
  \bibnamefont {Friedland}}, \ and\ \bibinfo {author} {\bibfnamefont {Gail~C.}\
  \bibnamefont {McLaughlin}},\ }\bibfield  {title} {\enquote {\bibinfo {title}
  {{Matter-Neutrino Resonance Above Merging Compact Objects}},}\ }\href@noop {}
  {\  (\bibinfo {year} {2014})},\ \Eprint {http://arxiv.org/abs/1403.5797}
  {arXiv:1403.5797 [hep-ph]} \BibitemShut {NoStop}%
\bibitem [{\citenamefont {Malkus}\ \emph {et~al.}(2012)\citenamefont {Malkus},
  \citenamefont {Kneller}, \citenamefont {McLaughlin},\ and\ \citenamefont
  {Surman}}]{Malkus:2012ts}%
  \BibitemOpen
  \bibfield  {author} {\bibinfo {author} {\bibfnamefont {Annelise}\
  \bibnamefont {Malkus}}, \bibinfo {author} {\bibfnamefont {James~P.}\
  \bibnamefont {Kneller}}, \bibinfo {author} {\bibfnamefont {Gail~C.}\
  \bibnamefont {McLaughlin}}, \ and\ \bibinfo {author} {\bibfnamefont
  {Rebecca}\ \bibnamefont {Surman}},\ }\bibfield  {title} {\enquote {\bibinfo
  {title} {{Neutrino oscillations above black hole accretion disks: disks with
  electron-flavor emission}},}\ }\href {\doibase 10.1103/PhysRevD.86.085015}
  {\bibfield  {journal} {\bibinfo  {journal} {Phys. Rev. D}\ }\textbf {\bibinfo
  {volume} {86}},\ \bibinfo {pages} {085015} (\bibinfo {year} {2012})},\
  \Eprint {http://arxiv.org/abs/1207.6648} {arXiv:1207.6648 [hep-ph]}
  \BibitemShut {NoStop}%
\bibitem [{\citenamefont {Wu}\ \emph {et~al.}(2016)\citenamefont {Wu},
  \citenamefont {Duan},\ and\ \citenamefont {Qian}}]{Wu:2015fga}%
  \BibitemOpen
  \bibfield  {author} {\bibinfo {author} {\bibfnamefont {Meng-Ru}\ \bibnamefont
  {Wu}}, \bibinfo {author} {\bibfnamefont {Huaiyu}\ \bibnamefont {Duan}}, \
  and\ \bibinfo {author} {\bibfnamefont {Yong-Zhong}\ \bibnamefont {Qian}},\
  }\bibfield  {title} {\enquote {\bibinfo {title} {{Physics of neutrino flavor
  transformation through matter\textendash{}neutrino resonances}},}\ }\href
  {\doibase 10.1016/j.physletb.2015.11.027} {\bibfield  {journal} {\bibinfo
  {journal} {Phys. Lett. B}\ }\textbf {\bibinfo {volume} {752}},\ \bibinfo
  {pages} {89--94} (\bibinfo {year} {2016})},\ \Eprint
  {http://arxiv.org/abs/1509.08975} {arXiv:1509.08975 [hep-ph]} \BibitemShut
  {NoStop}%
\bibitem [{\citenamefont {Zhu}\ \emph {et~al.}(2016)\citenamefont {Zhu},
  \citenamefont {Perego},\ and\ \citenamefont {McLaughlin}}]{Zhu:2016mwa}%
  \BibitemOpen
  \bibfield  {author} {\bibinfo {author} {\bibfnamefont {Yong-Lin}\
  \bibnamefont {Zhu}}, \bibinfo {author} {\bibfnamefont {Albino}\ \bibnamefont
  {Perego}}, \ and\ \bibinfo {author} {\bibfnamefont {Gail~C.}\ \bibnamefont
  {McLaughlin}},\ }\bibfield  {title} {\enquote {\bibinfo {title} {{Matter
  Neutrino Resonance Transitions above a Neutron Star Merger Remnant}},}\
  }\href {\doibase 10.1103/PhysRevD.94.105006} {\bibfield  {journal} {\bibinfo
  {journal} {Phys. Rev. D}\ }\textbf {\bibinfo {volume} {94}},\ \bibinfo
  {pages} {105006} (\bibinfo {year} {2016})},\ \Eprint
  {http://arxiv.org/abs/1607.04671} {arXiv:1607.04671 [hep-ph]} \BibitemShut
  {NoStop}%
\bibitem [{\citenamefont {Frensel}\ \emph {et~al.}(2017)\citenamefont
  {Frensel}, \citenamefont {Wu}, \citenamefont {Volpe},\ and\ \citenamefont
  {Perego}}]{Frensel:2016fge}%
  \BibitemOpen
  \bibfield  {author} {\bibinfo {author} {\bibfnamefont {Maik}\ \bibnamefont
  {Frensel}}, \bibinfo {author} {\bibfnamefont {Meng-Ru}\ \bibnamefont {Wu}},
  \bibinfo {author} {\bibfnamefont {Cristina}\ \bibnamefont {Volpe}}, \ and\
  \bibinfo {author} {\bibfnamefont {Albino}\ \bibnamefont {Perego}},\
  }\bibfield  {title} {\enquote {\bibinfo {title} {{Neutrino Flavor Evolution
  in Binary Neutron Star Merger Remnants}},}\ }\href {\doibase
  10.1103/PhysRevD.95.023011} {\bibfield  {journal} {\bibinfo  {journal} {Phys.
  Rev. D}\ }\textbf {\bibinfo {volume} {95}},\ \bibinfo {pages} {023011}
  (\bibinfo {year} {2017})},\ \Eprint {http://arxiv.org/abs/1607.05938}
  {arXiv:1607.05938 [astro-ph.HE]} \BibitemShut {NoStop}%
\bibitem [{\citenamefont {Tian}\ \emph {et~al.}(2017)\citenamefont {Tian},
  \citenamefont {Patwardhan},\ and\ \citenamefont {Fuller}}]{Tian:2017xbr}%
  \BibitemOpen
  \bibfield  {author} {\bibinfo {author} {\bibfnamefont {James~Y.}\
  \bibnamefont {Tian}}, \bibinfo {author} {\bibfnamefont {Amol~V.}\
  \bibnamefont {Patwardhan}}, \ and\ \bibinfo {author} {\bibfnamefont
  {George~M.}\ \bibnamefont {Fuller}},\ }\bibfield  {title} {\enquote {\bibinfo
  {title} {{Neutrino Flavor Evolution in Neutron Star Mergers}},}\ }\href
  {\doibase 10.1103/PhysRevD.96.043001} {\bibfield  {journal} {\bibinfo
  {journal} {Phys. Rev. D}\ }\textbf {\bibinfo {volume} {96}},\ \bibinfo
  {pages} {043001} (\bibinfo {year} {2017})},\ \Eprint
  {http://arxiv.org/abs/1703.03039} {arXiv:1703.03039 [astro-ph.HE]}
  \BibitemShut {NoStop}%
\bibitem [{\citenamefont {Shalgar}(2018)}]{Shalgar:2017pzd}%
  \BibitemOpen
  \bibfield  {author} {\bibinfo {author} {\bibfnamefont {Shashank}\
  \bibnamefont {Shalgar}},\ }\bibfield  {title} {\enquote {\bibinfo {title}
  {{Multi-angle calculation of the matter-neutrino resonance near an accretion
  disk}},}\ }\href {\doibase 10.1088/1475-7516/2018/02/010} {\bibfield
  {journal} {\bibinfo  {journal} {JCAP}\ }\textbf {\bibinfo {volume} {02}},\
  \bibinfo {pages} {010} (\bibinfo {year} {2018})},\ \Eprint
  {http://arxiv.org/abs/1707.07692} {arXiv:1707.07692 [hep-ph]} \BibitemShut
  {NoStop}%
\bibitem [{\citenamefont {Tamborra}\ and\ \citenamefont
  {Shalgar}(2021)}]{Tamborra:2020cul}%
  \BibitemOpen
  \bibfield  {author} {\bibinfo {author} {\bibfnamefont {Irene}\ \bibnamefont
  {Tamborra}}\ and\ \bibinfo {author} {\bibfnamefont {Shashank}\ \bibnamefont
  {Shalgar}},\ }\bibfield  {title} {\enquote {\bibinfo {title} {{New
  Developments in Flavor Evolution of a Dense Neutrino Gas}},}\ }\href
  {\doibase 10.1146/annurev-nucl-102920-050505} {\bibfield  {journal} {\bibinfo
   {journal} {Ann. Rev. Nucl. Part. Sci.}\ }\textbf {\bibinfo {volume} {71}},\
  \bibinfo {pages} {165--188} (\bibinfo {year} {2021})},\ \Eprint
  {http://arxiv.org/abs/2011.01948} {arXiv:2011.01948 [astro-ph.HE]}
  \BibitemShut {NoStop}%
\bibitem [{\citenamefont {Wu}\ and\ \citenamefont
  {Tamborra}(2017)}]{Wu:2017qpc}%
  \BibitemOpen
  \bibfield  {author} {\bibinfo {author} {\bibfnamefont {Meng-Ru}\ \bibnamefont
  {Wu}}\ and\ \bibinfo {author} {\bibfnamefont {Irene}\ \bibnamefont
  {Tamborra}},\ }\bibfield  {title} {\enquote {\bibinfo {title} {{Fast neutrino
  conversions: Ubiquitous in compact binary merger remnants}},}\ }\href
  {\doibase 10.1103/PhysRevD.95.103007} {\bibfield  {journal} {\bibinfo
  {journal} {Phys. Rev. D}\ }\textbf {\bibinfo {volume} {95}},\ \bibinfo
  {pages} {103007} (\bibinfo {year} {2017})},\ \Eprint
  {http://arxiv.org/abs/1701.06580} {arXiv:1701.06580 [astro-ph.HE]}
  \BibitemShut {NoStop}%
\bibitem [{\citenamefont {Wu}\ \emph {et~al.}(2017)\citenamefont {Wu},
  \citenamefont {Tamborra}, \citenamefont {Just},\ and\ \citenamefont
  {Janka}}]{Wu:2017drk}%
  \BibitemOpen
  \bibfield  {author} {\bibinfo {author} {\bibfnamefont {Meng-Ru}\ \bibnamefont
  {Wu}}, \bibinfo {author} {\bibfnamefont {Irene}\ \bibnamefont {Tamborra}},
  \bibinfo {author} {\bibfnamefont {Oliver}\ \bibnamefont {Just}}, \ and\
  \bibinfo {author} {\bibfnamefont {H.-Thomas}\ \bibnamefont {Janka}},\
  }\bibfield  {title} {\enquote {\bibinfo {title} {{Imprints of neutrino-pair
  flavor conversions on nucleosynthesis in ejecta from neutron-star merger
  remnants}},}\ }\href {\doibase 10.1103/PhysRevD.96.123015} {\bibfield
  {journal} {\bibinfo  {journal} {Phys. Rev. D}\ }\textbf {\bibinfo {volume}
  {96}},\ \bibinfo {pages} {123015} (\bibinfo {year} {2017})},\ \Eprint
  {http://arxiv.org/abs/1711.00477} {arXiv:1711.00477 [astro-ph.HE]}
  \BibitemShut {NoStop}%
\bibitem [{\citenamefont {George}\ \emph {et~al.}(2020)\citenamefont {George},
  \citenamefont {Wu}, \citenamefont {Tamborra}, \citenamefont
  {Ardevol-Pulpillo},\ and\ \citenamefont {Janka}}]{George:2020veu}%
  \BibitemOpen
  \bibfield  {author} {\bibinfo {author} {\bibfnamefont {Manu}\ \bibnamefont
  {George}}, \bibinfo {author} {\bibfnamefont {Meng-Ru}\ \bibnamefont {Wu}},
  \bibinfo {author} {\bibfnamefont {Irene}\ \bibnamefont {Tamborra}}, \bibinfo
  {author} {\bibfnamefont {Ricard}\ \bibnamefont {Ardevol-Pulpillo}}, \ and\
  \bibinfo {author} {\bibfnamefont {H.-Thomas}\ \bibnamefont {Janka}},\
  }\bibfield  {title} {\enquote {\bibinfo {title} {{Fast neutrino flavor
  conversion, ejecta properties, and nucleosynthesis in newly-formed
  hypermassive remnants of neutron-star mergers}},}\ }\href {\doibase
  10.1103/PhysRevD.102.103015} {\bibfield  {journal} {\bibinfo  {journal}
  {Phys. Rev. D}\ }\textbf {\bibinfo {volume} {102}},\ \bibinfo {pages}
  {103015} (\bibinfo {year} {2020})},\ \Eprint
  {http://arxiv.org/abs/2009.04046} {arXiv:2009.04046 [astro-ph.HE]}
  \BibitemShut {NoStop}%
\bibitem [{\citenamefont {Just}\ \emph {et~al.}(2022)\citenamefont {Just},
  \citenamefont {Abbar}, \citenamefont {Wu}, \citenamefont {Tamborra},
  \citenamefont {Janka},\ and\ \citenamefont {Capozzi}}]{Just:2022flt}%
  \BibitemOpen
  \bibfield  {author} {\bibinfo {author} {\bibfnamefont {Oliver}\ \bibnamefont
  {Just}}, \bibinfo {author} {\bibfnamefont {Sajad}\ \bibnamefont {Abbar}},
  \bibinfo {author} {\bibfnamefont {Meng-Ru}\ \bibnamefont {Wu}}, \bibinfo
  {author} {\bibfnamefont {Irene}\ \bibnamefont {Tamborra}}, \bibinfo {author}
  {\bibfnamefont {H.-Thomas}\ \bibnamefont {Janka}}, \ and\ \bibinfo {author}
  {\bibfnamefont {Francesco}\ \bibnamefont {Capozzi}},\ }\bibfield  {title}
  {\enquote {\bibinfo {title} {{Fast neutrino conversion in hydrodynamic
  simulations of neutrino-cooled accretion disks}},}\ }\href {\doibase
  10.1103/PhysRevD.105.083024} {\bibfield  {journal} {\bibinfo  {journal}
  {Phys. Rev. D}\ }\textbf {\bibinfo {volume} {105}},\ \bibinfo {pages}
  {083024} (\bibinfo {year} {2022})},\ \Eprint
  {http://arxiv.org/abs/2203.16559} {arXiv:2203.16559 [astro-ph.HE]}
  \BibitemShut {NoStop}%
\bibitem [{\citenamefont {Li}\ and\ \citenamefont {Siegel}(2021)}]{Li:2021vqj}%
  \BibitemOpen
  \bibfield  {author} {\bibinfo {author} {\bibfnamefont {Xinyu}\ \bibnamefont
  {Li}}\ and\ \bibinfo {author} {\bibfnamefont {Daniel~M.}\ \bibnamefont
  {Siegel}},\ }\bibfield  {title} {\enquote {\bibinfo {title} {{Neutrino Fast
  Flavor Conversions in Neutron-Star Postmerger Accretion Disks}},}\ }\href
  {\doibase 10.1103/PhysRevLett.126.251101} {\bibfield  {journal} {\bibinfo
  {journal} {Phys. Rev. Lett.}\ }\textbf {\bibinfo {volume} {126}},\ \bibinfo
  {pages} {251101} (\bibinfo {year} {2021})},\ \Eprint
  {http://arxiv.org/abs/2103.02616} {arXiv:2103.02616 [astro-ph.HE]}
  \BibitemShut {NoStop}%
\bibitem [{\citenamefont {Fern\'andez}\ \emph {et~al.}(2022)\citenamefont
  {Fern\'andez}, \citenamefont {Richers}, \citenamefont {Mulyk},\ and\
  \citenamefont {Fahlman}}]{Fernandez:2022yyv}%
  \BibitemOpen
  \bibfield  {author} {\bibinfo {author} {\bibfnamefont {Rodrigo}\ \bibnamefont
  {Fern\'andez}}, \bibinfo {author} {\bibfnamefont {Sherwood}\ \bibnamefont
  {Richers}}, \bibinfo {author} {\bibfnamefont {Nicole}\ \bibnamefont {Mulyk}},
  \ and\ \bibinfo {author} {\bibfnamefont {Steven}\ \bibnamefont {Fahlman}},\
  }\bibfield  {title} {\enquote {\bibinfo {title} {{The Fast Flavor Instability
  in Hypermassive Neutron Star Disk Outflows}},}\ }\href@noop {} {\  (\bibinfo
  {year} {2022})},\ \Eprint {http://arxiv.org/abs/2207.10680} {arXiv:2207.10680
  [astro-ph.HE]} \BibitemShut {NoStop}%
\bibitem [{\citenamefont {Alves~Batista}\ \emph {et~al.}(2021)\citenamefont
  {Alves~Batista} \emph {et~al.}}]{AlvesBatista:2021gzc}%
  \BibitemOpen
  \bibfield  {author} {\bibinfo {author} {\bibfnamefont {Rafael}\ \bibnamefont
  {Alves~Batista}} \emph {et~al.},\ }\bibfield  {title} {\enquote {\bibinfo
  {title} {{EuCAPT White Paper: Opportunities and Challenges for Theoretical
  Astroparticle Physics in the Next Decade}},}\ }\href@noop {} {\  (\bibinfo
  {year} {2021})},\ \Eprint {http://arxiv.org/abs/2110.10074} {arXiv:2110.10074
  [astro-ph.HE]} \BibitemShut {NoStop}%
\bibitem [{\citenamefont {Burns}(2020)}]{Burns:2019byj}%
  \BibitemOpen
  \bibfield  {author} {\bibinfo {author} {\bibfnamefont {Eric}\ \bibnamefont
  {Burns}},\ }\bibfield  {title} {\enquote {\bibinfo {title} {{Neutron Star
  Mergers and How to Study Them}},}\ }\href {\doibase
  10.1007/s41114-020-00028-7} {\bibfield  {journal} {\bibinfo  {journal}
  {Living Rev. Rel.}\ }\textbf {\bibinfo {volume} {23}},\ \bibinfo {pages} {4}
  (\bibinfo {year} {2020})},\ \Eprint {http://arxiv.org/abs/1909.06085}
  {arXiv:1909.06085 [astro-ph.HE]} \BibitemShut {NoStop}%
\bibitem [{\citenamefont {Burns}\ \emph {et~al.}(2019)\citenamefont {Burns}
  \emph {et~al.}}]{Burns:2019tqz}%
  \BibitemOpen
  \bibfield  {author} {\bibinfo {author} {\bibfnamefont {Eric}\ \bibnamefont
  {Burns}} \emph {et~al.},\ }\bibfield  {title} {\enquote {\bibinfo {title} {{A
  Summary of Multimessenger Science with Neutron Star Mergers}},}\ }\href@noop
  {} {\  (\bibinfo {year} {2019})},\ \Eprint {http://arxiv.org/abs/1903.03582}
  {arXiv:1903.03582 [astro-ph.HE]} \BibitemShut {NoStop}%
\bibitem [{\citenamefont {Aggarwal}\ \emph {et~al.}(2021)\citenamefont
  {Aggarwal} \emph {et~al.}}]{Aggarwal:2020olq}%
  \BibitemOpen
  \bibfield  {author} {\bibinfo {author} {\bibfnamefont {Nancy}\ \bibnamefont
  {Aggarwal}} \emph {et~al.},\ }\bibfield  {title} {\enquote {\bibinfo {title}
  {{Challenges and opportunities of gravitational-wave searches at MHz to GHz
  frequencies}},}\ }\href {\doibase 10.1007/s41114-021-00032-5} {\bibfield
  {journal} {\bibinfo  {journal} {Living Rev. Rel.}\ }\textbf {\bibinfo
  {volume} {24}},\ \bibinfo {pages} {4} (\bibinfo {year} {2021})},\ \Eprint
  {http://arxiv.org/abs/2011.12414} {arXiv:2011.12414 [gr-qc]} \BibitemShut
  {NoStop}%
\bibitem [{\citenamefont {Diamond}\ and\ \citenamefont
  {Marques-Tavares}(2022)}]{Diamond:2021ekg}%
  \BibitemOpen
  \bibfield  {author} {\bibinfo {author} {\bibfnamefont {Melissa~D.}\
  \bibnamefont {Diamond}}\ and\ \bibinfo {author} {\bibfnamefont {Gustavo}\
  \bibnamefont {Marques-Tavares}},\ }\bibfield  {title} {\enquote {\bibinfo
  {title} {{\ensuremath{\gamma}-Ray Flashes from Dark Photons in Neutron Star
  Mergers}},}\ }\href {\doibase 10.1103/PhysRevLett.128.211101} {\bibfield
  {journal} {\bibinfo  {journal} {Phys. Rev. Lett.}\ }\textbf {\bibinfo
  {volume} {128}},\ \bibinfo {pages} {211101} (\bibinfo {year} {2022})},\
  \Eprint {http://arxiv.org/abs/2106.03879} {arXiv:2106.03879 [hep-ph]}
  \BibitemShut {NoStop}%
\bibitem [{\citenamefont {Berezhiani}\ and\ \citenamefont
  {Drago}(2000)}]{Berezhiani:1999qh}%
  \BibitemOpen
  \bibfield  {author} {\bibinfo {author} {\bibfnamefont {Zurab}\ \bibnamefont
  {Berezhiani}}\ and\ \bibinfo {author} {\bibfnamefont {Alessandro}\
  \bibnamefont {Drago}},\ }\bibfield  {title} {\enquote {\bibinfo {title}
  {{Gamma-ray bursts via emission of axion - like particles}},}\ }\href
  {\doibase 10.1016/S0370-2693(99)01449-5} {\bibfield  {journal} {\bibinfo
  {journal} {Phys. Lett. B}\ }\textbf {\bibinfo {volume} {473}},\ \bibinfo
  {pages} {281--290} (\bibinfo {year} {2000})},\ \Eprint
  {http://arxiv.org/abs/hep-ph/9911333} {arXiv:hep-ph/9911333} \BibitemShut
  {NoStop}%
\bibitem [{\citenamefont {Berezhiani}\ \emph {et~al.}(2003)\citenamefont
  {Berezhiani}, \citenamefont {Bombaci}, \citenamefont {Drago}, \citenamefont
  {Frontera},\ and\ \citenamefont {Lavagno}}]{Berezhiani:2002ks}%
  \BibitemOpen
  \bibfield  {author} {\bibinfo {author} {\bibfnamefont {Zurab}\ \bibnamefont
  {Berezhiani}}, \bibinfo {author} {\bibfnamefont {Ignazio}\ \bibnamefont
  {Bombaci}}, \bibinfo {author} {\bibfnamefont {Alessandro}\ \bibnamefont
  {Drago}}, \bibinfo {author} {\bibfnamefont {Filippo}\ \bibnamefont
  {Frontera}}, \ and\ \bibinfo {author} {\bibfnamefont {Andrea}\ \bibnamefont
  {Lavagno}},\ }\bibfield  {title} {\enquote {\bibinfo {title} {{Gamma-ray
  bursts from delayed collapse of neutron stars to quark matter stars}},}\
  }\href {\doibase 10.1086/367756} {\bibfield  {journal} {\bibinfo  {journal}
  {Astrophys. J.}\ }\textbf {\bibinfo {volume} {586}},\ \bibinfo {pages}
  {1250--1253} (\bibinfo {year} {2003})},\ \Eprint
  {http://arxiv.org/abs/astro-ph/0209257} {arXiv:astro-ph/0209257} \BibitemShut
  {NoStop}%
\bibitem [{\citenamefont {Dasgupta}\ and\ \citenamefont
  {Kopp}(2021)}]{Dasgupta:2021ies}%
  \BibitemOpen
  \bibfield  {author} {\bibinfo {author} {\bibfnamefont {Basudeb}\ \bibnamefont
  {Dasgupta}}\ and\ \bibinfo {author} {\bibfnamefont {Joachim}\ \bibnamefont
  {Kopp}},\ }\bibfield  {title} {\enquote {\bibinfo {title} {{Sterile
  Neutrinos}},}\ }\href {\doibase 10.1016/j.physrep.2021.06.002} {\bibfield
  {journal} {\bibinfo  {journal} {Phys. Rept.}\ }\textbf {\bibinfo {volume}
  {928}},\ \bibinfo {pages} {1--63} (\bibinfo {year} {2021})},\ \Eprint
  {http://arxiv.org/abs/2106.05913} {arXiv:2106.05913 [hep-ph]} \BibitemShut
  {NoStop}%
\bibitem [{\citenamefont {Abazajian}\ \emph {et~al.}(2012)\citenamefont
  {Abazajian} \emph {et~al.}}]{Abazajian:2012ys}%
  \BibitemOpen
  \bibfield  {author} {\bibinfo {author} {\bibfnamefont {Kevork~N.}\
  \bibnamefont {Abazajian}} \emph {et~al.},\ }\bibfield  {title} {\enquote
  {\bibinfo {title} {{Light Sterile Neutrinos: A White Paper}},}\ }\href@noop
  {} {\  (\bibinfo {year} {2012})},\ \Eprint {http://arxiv.org/abs/1204.5379}
  {arXiv:1204.5379 [hep-ph]} \BibitemShut {NoStop}%
\bibitem [{\citenamefont {B\"oser}\ \emph {et~al.}(2020)\citenamefont
  {B\"oser}, \citenamefont {Buck}, \citenamefont {Giunti}, \citenamefont
  {Lesgourgues}, \citenamefont {Ludhova}, \citenamefont {Mertens},
  \citenamefont {Schukraft},\ and\ \citenamefont {Wurm}}]{Boser:2019rta}%
  \BibitemOpen
  \bibfield  {author} {\bibinfo {author} {\bibfnamefont {Sebastian}\
  \bibnamefont {B\"oser}}, \bibinfo {author} {\bibfnamefont {Christian}\
  \bibnamefont {Buck}}, \bibinfo {author} {\bibfnamefont {Carlo}\ \bibnamefont
  {Giunti}}, \bibinfo {author} {\bibfnamefont {Julien}\ \bibnamefont
  {Lesgourgues}}, \bibinfo {author} {\bibfnamefont {Livia}\ \bibnamefont
  {Ludhova}}, \bibinfo {author} {\bibfnamefont {Susanne}\ \bibnamefont
  {Mertens}}, \bibinfo {author} {\bibfnamefont {Anne}\ \bibnamefont
  {Schukraft}}, \ and\ \bibinfo {author} {\bibfnamefont {Michael}\ \bibnamefont
  {Wurm}},\ }\bibfield  {title} {\enquote {\bibinfo {title} {{Status of Light
  Sterile Neutrino Searches}},}\ }\href {\doibase 10.1016/j.ppnp.2019.103736}
  {\bibfield  {journal} {\bibinfo  {journal} {Prog. Part. Nucl. Phys.}\
  }\textbf {\bibinfo {volume} {111}},\ \bibinfo {pages} {103736} (\bibinfo
  {year} {2020})},\ \Eprint {http://arxiv.org/abs/1906.01739} {arXiv:1906.01739
  [hep-ex]} \BibitemShut {NoStop}%
\bibitem [{\citenamefont {Acero}\ \emph {et~al.}(2022)\citenamefont {Acero}
  \emph {et~al.}}]{Acero:2022wqg}%
  \BibitemOpen
  \bibfield  {author} {\bibinfo {author} {\bibfnamefont {Mario~A.}\
  \bibnamefont {Acero}} \emph {et~al.},\ }\bibfield  {title} {\enquote
  {\bibinfo {title} {{White Paper on Light Sterile Neutrino Searches and
  Related Phenomenology}},}\ }\href@noop {} {\  (\bibinfo {year} {2022})},\
  \Eprint {http://arxiv.org/abs/2203.07323} {arXiv:2203.07323 [hep-ex]}
  \BibitemShut {NoStop}%
\bibitem [{\citenamefont {Aguilar-Arevalo}\ \emph {et~al.}(2001)\citenamefont
  {Aguilar-Arevalo} \emph {et~al.}}]{LSND:2001aii}%
  \BibitemOpen
  \bibfield  {author} {\bibinfo {author} {\bibfnamefont {Alexis~A.}\
  \bibnamefont {Aguilar-Arevalo}} \emph {et~al.} (\bibinfo {collaboration}
  {LSND}),\ }\bibfield  {title} {\enquote {\bibinfo {title} {{Evidence for
  neutrino oscillations from the observation of $\bar{\nu}_e$ appearance in a
  $\bar{\nu}_\mu$ beam}},}\ }\href {\doibase 10.1103/PhysRevD.64.112007}
  {\bibfield  {journal} {\bibinfo  {journal} {Phys. Rev. D}\ }\textbf {\bibinfo
  {volume} {64}},\ \bibinfo {pages} {112007} (\bibinfo {year} {2001})},\
  \Eprint {http://arxiv.org/abs/hep-ex/0104049} {arXiv:hep-ex/0104049}
  \BibitemShut {NoStop}%
\bibitem [{\citenamefont {Aguilar-Arevalo}\ \emph {et~al.}(2021)\citenamefont
  {Aguilar-Arevalo} \emph {et~al.}}]{MiniBooNE:2020pnu}%
  \BibitemOpen
  \bibfield  {author} {\bibinfo {author} {\bibfnamefont {Alexis~A.}\
  \bibnamefont {Aguilar-Arevalo}} \emph {et~al.} (\bibinfo {collaboration}
  {MiniBooNE}),\ }\bibfield  {title} {\enquote {\bibinfo {title} {{Updated
  MiniBooNE neutrino oscillation results with increased data and new background
  studies}},}\ }\href {\doibase 10.1103/PhysRevD.103.052002} {\bibfield
  {journal} {\bibinfo  {journal} {Phys. Rev. D}\ }\textbf {\bibinfo {volume}
  {103}},\ \bibinfo {pages} {052002} (\bibinfo {year} {2021})},\ \Eprint
  {http://arxiv.org/abs/2006.16883} {arXiv:2006.16883 [hep-ex]} \BibitemShut
  {NoStop}%
\bibitem [{\citenamefont {Mueller}\ \emph {et~al.}(2011)\citenamefont {Mueller}
  \emph {et~al.}}]{Mueller:2011nm}%
  \BibitemOpen
  \bibfield  {author} {\bibinfo {author} {\bibfnamefont {Thomas~A.}\
  \bibnamefont {Mueller}} \emph {et~al.},\ }\bibfield  {title} {\enquote
  {\bibinfo {title} {{Improved Predictions of Reactor Antineutrino Spectra}},}\
  }\href {\doibase 10.1103/PhysRevC.83.054615} {\bibfield  {journal} {\bibinfo
  {journal} {Phys. Rev. C}\ }\textbf {\bibinfo {volume} {83}},\ \bibinfo
  {pages} {054615} (\bibinfo {year} {2011})},\ \Eprint
  {http://arxiv.org/abs/1101.2663} {arXiv:1101.2663 [hep-ex]} \BibitemShut
  {NoStop}%
\bibitem [{\citenamefont {Huber}(2011)}]{Huber:2011wv}%
  \BibitemOpen
  \bibfield  {author} {\bibinfo {author} {\bibfnamefont {Patrick}\ \bibnamefont
  {Huber}},\ }\bibfield  {title} {\enquote {\bibinfo {title} {{On the
  determination of anti-neutrino spectra from nuclear reactors}},}\ }\href
  {\doibase 10.1103/PhysRevC.85.029901} {\bibfield  {journal} {\bibinfo
  {journal} {Phys. Rev. C}\ }\textbf {\bibinfo {volume} {84}},\ \bibinfo
  {pages} {024617} (\bibinfo {year} {2011})},\ \bibinfo {note} {[Erratum:
  Phys.Rev.C 85, 029901 (2012)]},\ \Eprint {http://arxiv.org/abs/1106.0687}
  {arXiv:1106.0687 [hep-ph]} \BibitemShut {NoStop}%
\bibitem [{\citenamefont {Giunti}\ \emph
  {et~al.}(2022{\natexlab{a}})\citenamefont {Giunti}, \citenamefont {Li},
  \citenamefont {Ternes},\ and\ \citenamefont {Xin}}]{Giunti:2021kab}%
  \BibitemOpen
  \bibfield  {author} {\bibinfo {author} {\bibfnamefont {Carlo}\ \bibnamefont
  {Giunti}}, \bibinfo {author} {\bibfnamefont {Yufeng}\ \bibnamefont {Li}},
  \bibinfo {author} {\bibfnamefont {Christoph~A.}\ \bibnamefont {Ternes}}, \
  and\ \bibinfo {author} {\bibfnamefont {Zhao}\ \bibnamefont {Xin}},\
  }\bibfield  {title} {\enquote {\bibinfo {title} {{Reactor antineutrino
  anomaly in light of recent flux model refinements}},}\ }\href {\doibase
  10.1016/j.physletb.2022.137054} {\bibfield  {journal} {\bibinfo  {journal}
  {Phys. Lett. B}\ }\textbf {\bibinfo {volume} {829}},\ \bibinfo {pages}
  {137054} (\bibinfo {year} {2022}{\natexlab{a}})},\ \Eprint
  {http://arxiv.org/abs/2110.06820} {arXiv:2110.06820 [hep-ph]} \BibitemShut
  {NoStop}%
\bibitem [{\citenamefont {Kopeikin}\ \emph {et~al.}(2021)\citenamefont
  {Kopeikin}, \citenamefont {Skorokhvatov},\ and\ \citenamefont
  {Titov}}]{Kopeikin:2021ugh}%
  \BibitemOpen
  \bibfield  {author} {\bibinfo {author} {\bibfnamefont {V.}~\bibnamefont
  {Kopeikin}}, \bibinfo {author} {\bibfnamefont {M.}~\bibnamefont
  {Skorokhvatov}}, \ and\ \bibinfo {author} {\bibfnamefont {O.}~\bibnamefont
  {Titov}},\ }\bibfield  {title} {\enquote {\bibinfo {title} {{Reevaluating
  reactor antineutrino spectra with new measurements of the ratio between U235
  and Pu239 \ensuremath{\beta} spectra}},}\ }\href {\doibase
  10.1103/PhysRevD.104.L071301} {\bibfield  {journal} {\bibinfo  {journal}
  {Phys. Rev. D}\ }\textbf {\bibinfo {volume} {104}},\ \bibinfo {pages}
  {L071301} (\bibinfo {year} {2021})},\ \Eprint
  {http://arxiv.org/abs/2103.01684} {arXiv:2103.01684 [nucl-ex]} \BibitemShut
  {NoStop}%
\bibitem [{\citenamefont {Berryman}\ and\ \citenamefont
  {Huber}(2021)}]{Berryman:2020agd}%
  \BibitemOpen
  \bibfield  {author} {\bibinfo {author} {\bibfnamefont {Jeffrey~M.}\
  \bibnamefont {Berryman}}\ and\ \bibinfo {author} {\bibfnamefont {Patrick}\
  \bibnamefont {Huber}},\ }\bibfield  {title} {\enquote {\bibinfo {title}
  {{Sterile Neutrinos and the Global Reactor Antineutrino Dataset}},}\ }\href
  {\doibase 10.1007/JHEP01(2021)167} {\bibfield  {journal} {\bibinfo  {journal}
  {JHEP}\ }\textbf {\bibinfo {volume} {01}},\ \bibinfo {pages} {167} (\bibinfo
  {year} {2021})},\ \Eprint {http://arxiv.org/abs/2005.01756} {arXiv:2005.01756
  [hep-ph]} \BibitemShut {NoStop}%
\bibitem [{\citenamefont {Andriamirado}\ \emph {et~al.}(2020)\citenamefont
  {Andriamirado} \emph {et~al.}}]{PROSPECT:2020raz}%
  \BibitemOpen
  \bibfield  {author} {\bibinfo {author} {\bibfnamefont {Manoa}\ \bibnamefont
  {Andriamirado}} \emph {et~al.} (\bibinfo {collaboration} {PROSPECT,
  STEREO}),\ }\bibfield  {title} {\enquote {\bibinfo {title} {{Note on
  arXiv:2005.05301, 'Preparation of the Neutrino-4 experiment on search for
  sterile neutrino and the obtained results of measurements'}},}\ }\href@noop
  {} {\  (\bibinfo {year} {2020})},\ \Eprint {http://arxiv.org/abs/2006.13147}
  {arXiv:2006.13147 [hep-ex]} \BibitemShut {NoStop}%
\bibitem [{\citenamefont {Berryman}\ \emph {et~al.}(2022)\citenamefont
  {Berryman}, \citenamefont {Coloma}, \citenamefont {Huber}, \citenamefont
  {Schwetz},\ and\ \citenamefont {Zhou}}]{Berryman:2021yan}%
  \BibitemOpen
  \bibfield  {author} {\bibinfo {author} {\bibfnamefont {Jeffrey~M.}\
  \bibnamefont {Berryman}}, \bibinfo {author} {\bibfnamefont {Pilar}\
  \bibnamefont {Coloma}}, \bibinfo {author} {\bibfnamefont {Patrick}\
  \bibnamefont {Huber}}, \bibinfo {author} {\bibfnamefont {Thomas}\
  \bibnamefont {Schwetz}}, \ and\ \bibinfo {author} {\bibfnamefont {Albert}\
  \bibnamefont {Zhou}},\ }\bibfield  {title} {\enquote {\bibinfo {title}
  {{Statistical significance of the sterile-neutrino hypothesis in the context
  of reactor and gallium data}},}\ }\href {\doibase 10.1007/JHEP02(2022)055}
  {\bibfield  {journal} {\bibinfo  {journal} {JHEP}\ }\textbf {\bibinfo
  {volume} {02}},\ \bibinfo {pages} {055} (\bibinfo {year} {2022})},\ \Eprint
  {http://arxiv.org/abs/2111.12530} {arXiv:2111.12530 [hep-ph]} \BibitemShut
  {NoStop}%
\bibitem [{\citenamefont {Giunti}\ and\ \citenamefont
  {Laveder}(2011)}]{Giunti:2010zu}%
  \BibitemOpen
  \bibfield  {author} {\bibinfo {author} {\bibfnamefont {Carlo}\ \bibnamefont
  {Giunti}}\ and\ \bibinfo {author} {\bibfnamefont {Marco}\ \bibnamefont
  {Laveder}},\ }\bibfield  {title} {\enquote {\bibinfo {title} {{Statistical
  Significance of the Gallium Anomaly}},}\ }\href {\doibase
  10.1103/PhysRevC.83.065504} {\bibfield  {journal} {\bibinfo  {journal} {Phys.
  Rev. C}\ }\textbf {\bibinfo {volume} {83}},\ \bibinfo {pages} {065504}
  (\bibinfo {year} {2011})},\ \Eprint {http://arxiv.org/abs/1006.3244}
  {arXiv:1006.3244 [hep-ph]} \BibitemShut {NoStop}%
\bibitem [{\citenamefont {Barinov}\ \emph {et~al.}(2022)\citenamefont {Barinov}
  \emph {et~al.}}]{Barinov:2021asz}%
  \BibitemOpen
  \bibfield  {author} {\bibinfo {author} {\bibfnamefont {Vladislav~V.}\
  \bibnamefont {Barinov}} \emph {et~al.},\ }\bibfield  {title} {\enquote
  {\bibinfo {title} {{Results from the Baksan Experiment on Sterile Transitions
  (BEST)}},}\ }\href {\doibase 10.1103/PhysRevLett.128.232501} {\bibfield
  {journal} {\bibinfo  {journal} {Phys. Rev. Lett.}\ }\textbf {\bibinfo
  {volume} {128}},\ \bibinfo {pages} {232501} (\bibinfo {year} {2022})},\
  \Eprint {http://arxiv.org/abs/2109.11482} {arXiv:2109.11482 [nucl-ex]}
  \BibitemShut {NoStop}%
\bibitem [{\citenamefont {Barinov}\ and\ \citenamefont
  {Gorbunov}(2022)}]{Barinov:2021mjj}%
  \BibitemOpen
  \bibfield  {author} {\bibinfo {author} {\bibfnamefont {Vladislav~V.}\
  \bibnamefont {Barinov}}\ and\ \bibinfo {author} {\bibfnamefont {Dmitry}\
  \bibnamefont {Gorbunov}},\ }\bibfield  {title} {\enquote {\bibinfo {title}
  {{BEST impact on sterile neutrino hypothesis}},}\ }\href {\doibase
  10.1103/PhysRevD.105.L051703} {\bibfield  {journal} {\bibinfo  {journal}
  {Phys. Rev. D}\ }\textbf {\bibinfo {volume} {105}},\ \bibinfo {pages}
  {L051703} (\bibinfo {year} {2022})},\ \Eprint
  {http://arxiv.org/abs/2109.14654} {arXiv:2109.14654 [hep-ph]} \BibitemShut
  {NoStop}%
\bibitem [{\citenamefont {Dentler}\ \emph {et~al.}(2018)\citenamefont
  {Dentler}, \citenamefont {Hern\'andez-Cabezudo}, \citenamefont {Kopp},
  \citenamefont {Machado}, \citenamefont {Maltoni}, \citenamefont
  {Martinez-Soler},\ and\ \citenamefont {Schwetz}}]{Dentler:2018sju}%
  \BibitemOpen
  \bibfield  {author} {\bibinfo {author} {\bibfnamefont {Mona}\ \bibnamefont
  {Dentler}}, \bibinfo {author} {\bibfnamefont {\'Alvaro}\ \bibnamefont
  {Hern\'andez-Cabezudo}}, \bibinfo {author} {\bibfnamefont {Joachim}\
  \bibnamefont {Kopp}}, \bibinfo {author} {\bibfnamefont {Pedro A.~N.}\
  \bibnamefont {Machado}}, \bibinfo {author} {\bibfnamefont {Michele}\
  \bibnamefont {Maltoni}}, \bibinfo {author} {\bibfnamefont {Ivan}\
  \bibnamefont {Martinez-Soler}}, \ and\ \bibinfo {author} {\bibfnamefont
  {Thomas}\ \bibnamefont {Schwetz}},\ }\bibfield  {title} {\enquote {\bibinfo
  {title} {{Updated Global Analysis of Neutrino Oscillations in the Presence of
  eV-Scale Sterile Neutrinos}},}\ }\href {\doibase 10.1007/JHEP08(2018)010}
  {\bibfield  {journal} {\bibinfo  {journal} {JHEP}\ }\textbf {\bibinfo
  {volume} {08}},\ \bibinfo {pages} {010} (\bibinfo {year} {2018})},\ \Eprint
  {http://arxiv.org/abs/1803.10661} {arXiv:1803.10661 [hep-ph]} \BibitemShut
  {NoStop}%
\bibitem [{\citenamefont {Arg{\"u}elles}\ \emph {et~al.}(2022)\citenamefont
  {Arg{\"u}elles}, \citenamefont {Esteban}, \citenamefont {Hostert},
  \citenamefont {Kelly}, \citenamefont {Kopp}, \citenamefont {Machado},
  \citenamefont {Martinez-Soler},\ and\ \citenamefont
  {Perez-Gonzalez}}]{Arguelles:2021meu}%
  \BibitemOpen
  \bibfield  {author} {\bibinfo {author} {\bibfnamefont {Carlos~A.}\
  \bibnamefont {Arg{\"u}elles}}, \bibinfo {author} {\bibfnamefont {Ivan}\
  \bibnamefont {Esteban}}, \bibinfo {author} {\bibfnamefont {Matheus}\
  \bibnamefont {Hostert}}, \bibinfo {author} {\bibfnamefont {Kevin~J.}\
  \bibnamefont {Kelly}}, \bibinfo {author} {\bibfnamefont {Joachim}\
  \bibnamefont {Kopp}}, \bibinfo {author} {\bibfnamefont {Pedro A.~N.}\
  \bibnamefont {Machado}}, \bibinfo {author} {\bibfnamefont {Ivan}\
  \bibnamefont {Martinez-Soler}}, \ and\ \bibinfo {author} {\bibfnamefont
  {Yuber~F.}\ \bibnamefont {Perez-Gonzalez}},\ }\bibfield  {title} {\enquote
  {\bibinfo {title} {{MicroBooNE and the \ensuremath{\nu}e Interpretation of
  the MiniBooNE Low-Energy Excess}},}\ }\href {\doibase
  10.1103/PhysRevLett.128.241802} {\bibfield  {journal} {\bibinfo  {journal}
  {Phys. Rev. Lett.}\ }\textbf {\bibinfo {volume} {128}},\ \bibinfo {pages}
  {241802} (\bibinfo {year} {2022})},\ \Eprint
  {http://arxiv.org/abs/2111.10359} {arXiv:2111.10359 [hep-ph]} \BibitemShut
  {NoStop}%
\bibitem [{\citenamefont {Giunti}\ \emph
  {et~al.}(2022{\natexlab{b}})\citenamefont {Giunti}, \citenamefont {Li},
  \citenamefont {Ternes}, \citenamefont {Tyagi},\ and\ \citenamefont
  {Xin}}]{Giunti:2022btk}%
  \BibitemOpen
  \bibfield  {author} {\bibinfo {author} {\bibfnamefont {C.}~\bibnamefont
  {Giunti}}, \bibinfo {author} {\bibfnamefont {Y.~F.}\ \bibnamefont {Li}},
  \bibinfo {author} {\bibfnamefont {C.~A.}\ \bibnamefont {Ternes}}, \bibinfo
  {author} {\bibfnamefont {O.}~\bibnamefont {Tyagi}}, \ and\ \bibinfo {author}
  {\bibfnamefont {Z.}~\bibnamefont {Xin}},\ }\bibfield  {title} {\enquote
  {\bibinfo {title} {{Gallium Anomaly: Critical View from the Global Picture of
  $\nu_{e}$ and $\bar\nu_{e}$ Disappearance}},}\ }\href@noop {} {\  (\bibinfo
  {year} {2022}{\natexlab{b}})},\ \Eprint {http://arxiv.org/abs/2209.00916}
  {arXiv:2209.00916 [hep-ph]} \BibitemShut {NoStop}%
\bibitem [{\citenamefont {Hagstotz}\ \emph {et~al.}(2021)\citenamefont
  {Hagstotz}, \citenamefont {de~Salas}, \citenamefont {Gariazzo}, \citenamefont
  {Gerbino}, \citenamefont {Lattanzi}, \citenamefont {Vagnozzi}, \citenamefont
  {Freese},\ and\ \citenamefont {Pastor}}]{Hagstotz:2020ukm}%
  \BibitemOpen
  \bibfield  {author} {\bibinfo {author} {\bibfnamefont {Steffen}\ \bibnamefont
  {Hagstotz}}, \bibinfo {author} {\bibfnamefont {Pablo~F.}\ \bibnamefont
  {de~Salas}}, \bibinfo {author} {\bibfnamefont {Stefano}\ \bibnamefont
  {Gariazzo}}, \bibinfo {author} {\bibfnamefont {Martina}\ \bibnamefont
  {Gerbino}}, \bibinfo {author} {\bibfnamefont {Massimiliano}\ \bibnamefont
  {Lattanzi}}, \bibinfo {author} {\bibfnamefont {Sunny}\ \bibnamefont
  {Vagnozzi}}, \bibinfo {author} {\bibfnamefont {Katherine}\ \bibnamefont
  {Freese}}, \ and\ \bibinfo {author} {\bibfnamefont {Sergio}\ \bibnamefont
  {Pastor}},\ }\bibfield  {title} {\enquote {\bibinfo {title} {{Bounds on light
  sterile neutrino mass and mixing from cosmology and laboratory searches}},}\
  }\href {\doibase 10.1103/PhysRevD.104.123524} {\bibfield  {journal} {\bibinfo
   {journal} {Phys. Rev. D}\ }\textbf {\bibinfo {volume} {104}},\ \bibinfo
  {pages} {123524} (\bibinfo {year} {2021})},\ \Eprint
  {http://arxiv.org/abs/2003.02289} {arXiv:2003.02289 [astro-ph.CO]}
  \BibitemShut {NoStop}%
\bibitem [{\citenamefont {Hannestad}\ \emph {et~al.}(2012)\citenamefont
  {Hannestad}, \citenamefont {Tamborra},\ and\ \citenamefont
  {Tram}}]{Hannestad:2012ky}%
  \BibitemOpen
  \bibfield  {author} {\bibinfo {author} {\bibfnamefont {Steen}\ \bibnamefont
  {Hannestad}}, \bibinfo {author} {\bibfnamefont {Irene}\ \bibnamefont
  {Tamborra}}, \ and\ \bibinfo {author} {\bibfnamefont {Thomas}\ \bibnamefont
  {Tram}},\ }\bibfield  {title} {\enquote {\bibinfo {title} {{Thermalisation of
  light sterile neutrinos in the early universe}},}\ }\href {\doibase
  10.1088/1475-7516/2012/07/025} {\bibfield  {journal} {\bibinfo  {journal}
  {JCAP}\ }\textbf {\bibinfo {volume} {07}},\ \bibinfo {pages} {025} (\bibinfo
  {year} {2012})},\ \Eprint {http://arxiv.org/abs/1204.5861} {arXiv:1204.5861
  [astro-ph.CO]} \BibitemShut {NoStop}%
\bibitem [{\citenamefont {Chu}\ \emph {et~al.}(2018)\citenamefont {Chu},
  \citenamefont {Dasgupta}, \citenamefont {Dentler}, \citenamefont {Kopp},\
  and\ \citenamefont {Saviano}}]{Chu:2018gxk}%
  \BibitemOpen
  \bibfield  {author} {\bibinfo {author} {\bibfnamefont {Xiaoyong}\
  \bibnamefont {Chu}}, \bibinfo {author} {\bibfnamefont {Basudeb}\ \bibnamefont
  {Dasgupta}}, \bibinfo {author} {\bibfnamefont {Mona}\ \bibnamefont
  {Dentler}}, \bibinfo {author} {\bibfnamefont {Joachim}\ \bibnamefont {Kopp}},
  \ and\ \bibinfo {author} {\bibfnamefont {Ninetta}\ \bibnamefont {Saviano}},\
  }\bibfield  {title} {\enquote {\bibinfo {title} {{Sterile neutrinos with
  secret interactions\textemdash{}cosmological discord?}}}\ }\href {\doibase
  10.1088/1475-7516/2018/11/049} {\bibfield  {journal} {\bibinfo  {journal}
  {JCAP}\ }\textbf {\bibinfo {volume} {11}},\ \bibinfo {pages} {049} (\bibinfo
  {year} {2018})},\ \Eprint {http://arxiv.org/abs/1806.10629} {arXiv:1806.10629
  [hep-ph]} \BibitemShut {NoStop}%
\bibitem [{\citenamefont {Archidiacono}\ \emph {et~al.}(2020)\citenamefont
  {Archidiacono}, \citenamefont {Gariazzo}, \citenamefont {Giunti},
  \citenamefont {Hannestad},\ and\ \citenamefont
  {Tram}}]{Archidiacono:2020yey}%
  \BibitemOpen
  \bibfield  {author} {\bibinfo {author} {\bibfnamefont {Maria}\ \bibnamefont
  {Archidiacono}}, \bibinfo {author} {\bibfnamefont {Stefano}\ \bibnamefont
  {Gariazzo}}, \bibinfo {author} {\bibfnamefont {Carlo}\ \bibnamefont
  {Giunti}}, \bibinfo {author} {\bibfnamefont {Steen}\ \bibnamefont
  {Hannestad}}, \ and\ \bibinfo {author} {\bibfnamefont {Thomas}\ \bibnamefont
  {Tram}},\ }\bibfield  {title} {\enquote {\bibinfo {title} {{Sterile neutrino
  self-interactions: $H_0$ tension and short-baseline anomalies}},}\ }\href
  {\doibase 10.1088/1475-7516/2020/12/029} {\bibfield  {journal} {\bibinfo
  {journal} {JCAP}\ }\textbf {\bibinfo {volume} {12}},\ \bibinfo {pages} {029}
  (\bibinfo {year} {2020})},\ \Eprint {http://arxiv.org/abs/2006.12885}
  {arXiv:2006.12885 [astro-ph.CO]} \BibitemShut {NoStop}%
\bibitem [{\citenamefont {Nunokawa}\ \emph {et~al.}(1997)\citenamefont
  {Nunokawa}, \citenamefont {Peltoniemi}, \citenamefont {Rossi},\ and\
  \citenamefont {Valle}}]{Nunokawa:1997ct}%
  \BibitemOpen
  \bibfield  {author} {\bibinfo {author} {\bibfnamefont {Hiroshi}\ \bibnamefont
  {Nunokawa}}, \bibinfo {author} {\bibfnamefont {Juha~T.}\ \bibnamefont
  {Peltoniemi}}, \bibinfo {author} {\bibfnamefont {Anna}\ \bibnamefont
  {Rossi}}, \ and\ \bibinfo {author} {\bibfnamefont {Jose W.~F.}\ \bibnamefont
  {Valle}},\ }\bibfield  {title} {\enquote {\bibinfo {title} {{Supernova bounds
  on resonant active sterile neutrino conversions}},}\ }\href {\doibase
  10.1103/PhysRevD.56.1704} {\bibfield  {journal} {\bibinfo  {journal} {Phys.
  Rev. D}\ }\textbf {\bibinfo {volume} {56}},\ \bibinfo {pages} {1704--1713}
  (\bibinfo {year} {1997})},\ \Eprint {http://arxiv.org/abs/hep-ph/9702372}
  {arXiv:hep-ph/9702372} \BibitemShut {NoStop}%
\bibitem [{\citenamefont {Tamborra}\ \emph {et~al.}(2012)\citenamefont
  {Tamborra}, \citenamefont {Raffelt}, \citenamefont {H{\"u}depohl},\ and\
  \citenamefont {Janka}}]{Tamborra:2011is}%
  \BibitemOpen
  \bibfield  {author} {\bibinfo {author} {\bibfnamefont {Irene}\ \bibnamefont
  {Tamborra}}, \bibinfo {author} {\bibfnamefont {Georg~G.}\ \bibnamefont
  {Raffelt}}, \bibinfo {author} {\bibfnamefont {Lorenz}\ \bibnamefont
  {H{\"u}depohl}}, \ and\ \bibinfo {author} {\bibfnamefont {H.-Thomas}\
  \bibnamefont {Janka}},\ }\bibfield  {title} {\enquote {\bibinfo {title}
  {{Impact of eV-mass sterile neutrinos on neutrino-driven supernova
  outflows}},}\ }\href {\doibase 10.1088/1475-7516/2012/01/013} {\bibfield
  {journal} {\bibinfo  {journal} {JCAP}\ }\textbf {\bibinfo {volume} {01}},\
  \bibinfo {pages} {013} (\bibinfo {year} {2012})},\ \Eprint
  {http://arxiv.org/abs/1110.2104} {arXiv:1110.2104 [astro-ph.SR]} \BibitemShut
  {NoStop}%
\bibitem [{\citenamefont {Wu}\ \emph {et~al.}(2014)\citenamefont {Wu},
  \citenamefont {Fischer}, \citenamefont {Huther}, \citenamefont
  {Mart\'\i{}nez-Pinedo},\ and\ \citenamefont {Qian}}]{Wu:2013gxa}%
  \BibitemOpen
  \bibfield  {author} {\bibinfo {author} {\bibfnamefont {Meng-Ru}\ \bibnamefont
  {Wu}}, \bibinfo {author} {\bibfnamefont {Tobias}\ \bibnamefont {Fischer}},
  \bibinfo {author} {\bibfnamefont {Lutz}\ \bibnamefont {Huther}}, \bibinfo
  {author} {\bibfnamefont {Gabriel}\ \bibnamefont {Mart\'\i{}nez-Pinedo}}, \
  and\ \bibinfo {author} {\bibfnamefont {Yong-Zhong}\ \bibnamefont {Qian}},\
  }\bibfield  {title} {\enquote {\bibinfo {title} {{Impact of active-sterile
  neutrino mixing on supernova explosion and nucleosynthesis}},}\ }\href
  {\doibase 10.1103/PhysRevD.89.061303} {\bibfield  {journal} {\bibinfo
  {journal} {Phys. Rev. D}\ }\textbf {\bibinfo {volume} {89}},\ \bibinfo
  {pages} {061303} (\bibinfo {year} {2014})},\ \Eprint
  {http://arxiv.org/abs/1305.2382} {arXiv:1305.2382 [astro-ph.HE]} \BibitemShut
  {NoStop}%
\bibitem [{\citenamefont {Pllumbi}\ \emph {et~al.}(2015)\citenamefont
  {Pllumbi}, \citenamefont {Tamborra}, \citenamefont {Wanajo}, \citenamefont
  {Janka},\ and\ \citenamefont {H{\"u}depohl}}]{Pllumbi:2014saa}%
  \BibitemOpen
  \bibfield  {author} {\bibinfo {author} {\bibfnamefont {Else}\ \bibnamefont
  {Pllumbi}}, \bibinfo {author} {\bibfnamefont {Irene}\ \bibnamefont
  {Tamborra}}, \bibinfo {author} {\bibfnamefont {Shinya}\ \bibnamefont
  {Wanajo}}, \bibinfo {author} {\bibfnamefont {H.-Thomas}\ \bibnamefont
  {Janka}}, \ and\ \bibinfo {author} {\bibfnamefont {Lorenz}\ \bibnamefont
  {H{\"u}depohl}},\ }\bibfield  {title} {\enquote {\bibinfo {title} {{Impact of
  neutrino flavor oscillations on the neutrino-driven wind nucleosynthesis of
  an electron-capture supernova}},}\ }\href {\doibase
  10.1088/0004-637X/808/2/188} {\bibfield  {journal} {\bibinfo  {journal}
  {Astrophys. J.}\ }\textbf {\bibinfo {volume} {808}},\ \bibinfo {pages} {188}
  (\bibinfo {year} {2015})},\ \Eprint {http://arxiv.org/abs/1406.2596}
  {arXiv:1406.2596 [astro-ph.SR]} \BibitemShut {NoStop}%
\bibitem [{\citenamefont {{Xiong}}\ \emph {et~al.}(2019)\citenamefont
  {{Xiong}}, \citenamefont {{Wu}},\ and\ \citenamefont
  {{Qian}}}]{Xiong:2019nvw}%
  \BibitemOpen
  \bibfield  {author} {\bibinfo {author} {\bibfnamefont {Zewei}\ \bibnamefont
  {{Xiong}}}, \bibinfo {author} {\bibfnamefont {Meng-Ru}\ \bibnamefont {{Wu}}},
  \ and\ \bibinfo {author} {\bibfnamefont {Yong-Zhong}\ \bibnamefont
  {{Qian}}},\ }\bibfield  {title} {\enquote {\bibinfo {title} {{Active-Sterile
  Neutrino Oscillations in Neutrino-driven Winds: Implications for
  Nucleosynthesis}},}\ }\href {\doibase 10.3847/1538-4357/ab2870} {\bibfield
  {journal} {\bibinfo  {journal} {\apj}\ }\textbf {\bibinfo {volume} {880}},\
  \bibinfo {eid} {81} (\bibinfo {year} {2019})},\ \Eprint
  {http://arxiv.org/abs/1904.09371} {arXiv:1904.09371 [astro-ph.HE]}
  \BibitemShut {NoStop}%
\bibitem [{\citenamefont {Esmaili}\ \emph {et~al.}(2014)\citenamefont
  {Esmaili}, \citenamefont {Peres},\ and\ \citenamefont
  {Serpico}}]{Esmaili:2014gya}%
  \BibitemOpen
  \bibfield  {author} {\bibinfo {author} {\bibfnamefont {Arman}\ \bibnamefont
  {Esmaili}}, \bibinfo {author} {\bibfnamefont {Orlando L.~G.}\ \bibnamefont
  {Peres}}, \ and\ \bibinfo {author} {\bibfnamefont {Pasquale~Dario}\
  \bibnamefont {Serpico}},\ }\bibfield  {title} {\enquote {\bibinfo {title}
  {{Impact of sterile neutrinos on the early time flux from a galactic
  supernova}},}\ }\href {\doibase 10.1103/PhysRevD.90.033013} {\bibfield
  {journal} {\bibinfo  {journal} {Phys. Rev. D}\ }\textbf {\bibinfo {volume}
  {90}},\ \bibinfo {pages} {033013} (\bibinfo {year} {2014})},\ \Eprint
  {http://arxiv.org/abs/1402.1453} {arXiv:1402.1453 [hep-ph]} \BibitemShut
  {NoStop}%
\bibitem [{\citenamefont {Tang}\ \emph {et~al.}(2020)\citenamefont {Tang},
  \citenamefont {Wang},\ and\ \citenamefont {Wu}}]{Tang:2020pkp}%
  \BibitemOpen
  \bibfield  {author} {\bibinfo {author} {\bibfnamefont {Jian}\ \bibnamefont
  {Tang}}, \bibinfo {author} {\bibfnamefont {Tsechun}\ \bibnamefont {Wang}}, \
  and\ \bibinfo {author} {\bibfnamefont {Meng-Ru}\ \bibnamefont {Wu}},\
  }\bibfield  {title} {\enquote {\bibinfo {title} {{Constraining sterile
  neutrinos by core-collapse supernovae with multiple detectors}},}\ }\href
  {\doibase 10.1088/1475-7516/2020/10/038} {\bibfield  {journal} {\bibinfo
  {journal} {JCAP}\ }\textbf {\bibinfo {volume} {10}},\ \bibinfo {pages} {038}
  (\bibinfo {year} {2020})},\ \Eprint {http://arxiv.org/abs/2005.09168}
  {arXiv:2005.09168 [hep-ph]} \BibitemShut {NoStop}%
\bibitem [{\citenamefont {Leitner}\ \emph {et~al.}(2006)\citenamefont
  {Leitner}, \citenamefont {Alvarez-Ruso},\ and\ \citenamefont
  {Mosel}}]{Leitner:2006sp}%
  \BibitemOpen
  \bibfield  {author} {\bibinfo {author} {\bibfnamefont {Tina}\ \bibnamefont
  {Leitner}}, \bibinfo {author} {\bibfnamefont {Luis}\ \bibnamefont
  {Alvarez-Ruso}}, \ and\ \bibinfo {author} {\bibfnamefont {Ulrich}\
  \bibnamefont {Mosel}},\ }\bibfield  {title} {\enquote {\bibinfo {title}
  {{Neutral current neutrino-nucleus interactions at intermediate energies}},}\
  }\href {\doibase 10.1103/PhysRevC.74.065502} {\bibfield  {journal} {\bibinfo
  {journal} {Phys. Rev. C}\ }\textbf {\bibinfo {volume} {74}},\ \bibinfo
  {pages} {065502} (\bibinfo {year} {2006})},\ \Eprint
  {http://arxiv.org/abs/nucl-th/0606058} {arXiv:nucl-th/0606058} \BibitemShut
  {NoStop}%
\bibitem [{\citenamefont {Hannestad}\ and\ \citenamefont
  {Raffelt}(1998)}]{Hannestad:1997gc}%
  \BibitemOpen
  \bibfield  {author} {\bibinfo {author} {\bibfnamefont {Steen}\ \bibnamefont
  {Hannestad}}\ and\ \bibinfo {author} {\bibfnamefont {Georg~G.}\ \bibnamefont
  {Raffelt}},\ }\bibfield  {title} {\enquote {\bibinfo {title} {{Supernova
  neutrino opacity from nucleon-nucleon Bremsstrahlung and related
  processes}},}\ }\href {\doibase 10.1086/306303} {\bibfield  {journal}
  {\bibinfo  {journal} {Astrophys. J.}\ }\textbf {\bibinfo {volume} {507}},\
  \bibinfo {pages} {339--352} (\bibinfo {year} {1998})},\ \Eprint
  {http://arxiv.org/abs/astro-ph/9711132} {arXiv:astro-ph/9711132} \BibitemShut
  {NoStop}%
\bibitem [{\citenamefont {Strumia}\ and\ \citenamefont
  {Vissani}(2003)}]{Strumia:2003zx}%
  \BibitemOpen
  \bibfield  {author} {\bibinfo {author} {\bibfnamefont {Alessandro}\
  \bibnamefont {Strumia}}\ and\ \bibinfo {author} {\bibfnamefont {Francesco}\
  \bibnamefont {Vissani}},\ }\bibfield  {title} {\enquote {\bibinfo {title}
  {{Precise quasielastic neutrino/nucleon cross-section}},}\ }\href {\doibase
  10.1016/S0370-2693(03)00616-6} {\bibfield  {journal} {\bibinfo  {journal}
  {Phys. Lett. B}\ }\textbf {\bibinfo {volume} {564}},\ \bibinfo {pages}
  {42--54} (\bibinfo {year} {2003})},\ \Eprint
  {http://arxiv.org/abs/astro-ph/0302055} {arXiv:astro-ph/0302055} \BibitemShut
  {NoStop}%
\bibitem [{\citenamefont {Ricciardi}\ \emph {et~al.}(2022)\citenamefont
  {Ricciardi}, \citenamefont {Vignaroli},\ and\ \citenamefont
  {Vissani}}]{Ricciardi:2022pru}%
  \BibitemOpen
  \bibfield  {author} {\bibinfo {author} {\bibfnamefont {Giulia}\ \bibnamefont
  {Ricciardi}}, \bibinfo {author} {\bibfnamefont {Natascia}\ \bibnamefont
  {Vignaroli}}, \ and\ \bibinfo {author} {\bibfnamefont {Francesco}\
  \bibnamefont {Vissani}},\ }\bibfield  {title} {\enquote {\bibinfo {title}
  {{An accurate evaluation of electron (anti-)neutrino scattering on
  nucleons}},}\ }\href {\doibase 10.1007/JHEP08(2022)212} {\bibfield  {journal}
  {\bibinfo  {journal} {JHEP}\ }\textbf {\bibinfo {volume} {08}},\ \bibinfo
  {pages} {212} (\bibinfo {year} {2022})},\ \Eprint
  {http://arxiv.org/abs/2206.05567} {arXiv:2206.05567 [hep-ph]} \BibitemShut
  {NoStop}%
\bibitem [{\citenamefont {Suliga}\ \emph {et~al.}(2020)\citenamefont {Suliga},
  \citenamefont {Tamborra},\ and\ \citenamefont {Wu}}]{Suliga:2020vpz}%
  \BibitemOpen
  \bibfield  {author} {\bibinfo {author} {\bibfnamefont {Anna~M.}\ \bibnamefont
  {Suliga}}, \bibinfo {author} {\bibfnamefont {Irene}\ \bibnamefont
  {Tamborra}}, \ and\ \bibinfo {author} {\bibfnamefont {Meng-Ru}\ \bibnamefont
  {Wu}},\ }\bibfield  {title} {\enquote {\bibinfo {title} {{Lifting the
  core-collapse supernova bounds on keV-mass sterile neutrinos}},}\ }\href
  {\doibase 10.1088/1475-7516/2020/08/018} {\bibfield  {journal} {\bibinfo
  {journal} {JCAP}\ }\textbf {\bibinfo {volume} {08}},\ \bibinfo {pages} {018}
  (\bibinfo {year} {2020})},\ \Eprint {http://arxiv.org/abs/2004.11389}
  {arXiv:2004.11389 [astro-ph.HE]} \BibitemShut {NoStop}%
\bibitem [{\citenamefont {Suliga}\ \emph {et~al.}(2019)\citenamefont {Suliga},
  \citenamefont {Tamborra},\ and\ \citenamefont {Wu}}]{Suliga:2019bsq}%
  \BibitemOpen
  \bibfield  {author} {\bibinfo {author} {\bibfnamefont {Anna~M.}\ \bibnamefont
  {Suliga}}, \bibinfo {author} {\bibfnamefont {Irene}\ \bibnamefont
  {Tamborra}}, \ and\ \bibinfo {author} {\bibfnamefont {Meng-Ru}\ \bibnamefont
  {Wu}},\ }\bibfield  {title} {\enquote {\bibinfo {title} {{Tau lepton
  asymmetry by sterile neutrino emission -- Moving beyond one-zone supernova
  models}},}\ }\href {\doibase 10.1088/1475-7516/2019/12/019} {\bibfield
  {journal} {\bibinfo  {journal} {JCAP}\ }\textbf {\bibinfo {volume} {12}},\
  \bibinfo {pages} {019} (\bibinfo {year} {2019})},\ \Eprint
  {http://arxiv.org/abs/1908.11382} {arXiv:1908.11382 [astro-ph.HE]}
  \BibitemShut {NoStop}%
\bibitem [{\citenamefont {Richers}\ and\ \citenamefont
  {Sen}(2022)}]{Richers:2022zug}%
  \BibitemOpen
  \bibfield  {author} {\bibinfo {author} {\bibfnamefont {Sherwood}\
  \bibnamefont {Richers}}\ and\ \bibinfo {author} {\bibfnamefont {Manibrata}\
  \bibnamefont {Sen}},\ }\bibfield  {title} {\enquote {\bibinfo {title} {{Fast
  Flavor Transformations}},}\ }\href@noop {} {\  (\bibinfo {year} {2022})},\
  \Eprint {http://arxiv.org/abs/2207.03561} {arXiv:2207.03561 [astro-ph.HE]}
  \BibitemShut {NoStop}%
\bibitem [{\citenamefont {Shalgar}\ and\ \citenamefont
  {Tamborra}(2022{\natexlab{a}})}]{Shalgar:2022rjj}%
  \BibitemOpen
  \bibfield  {author} {\bibinfo {author} {\bibfnamefont {Shashank}\
  \bibnamefont {Shalgar}}\ and\ \bibinfo {author} {\bibfnamefont {Irene}\
  \bibnamefont {Tamborra}},\ }\bibfield  {title} {\enquote {\bibinfo {title}
  {{Supernova Neutrino Decoupling Is Altered by Flavor Conversion}},}\
  }\href@noop {} {\  (\bibinfo {year} {2022}{\natexlab{a}})},\ \Eprint
  {http://arxiv.org/abs/2206.00676} {arXiv:2206.00676 [astro-ph.HE]}
  \BibitemShut {NoStop}%
\bibitem [{\citenamefont {Nagakura}\ and\ \citenamefont
  {Zaizen}(2022)}]{Nagakura:2022kic}%
  \BibitemOpen
  \bibfield  {author} {\bibinfo {author} {\bibfnamefont {Hiroki}\ \bibnamefont
  {Nagakura}}\ and\ \bibinfo {author} {\bibfnamefont {Masamichi}\ \bibnamefont
  {Zaizen}},\ }\bibfield  {title} {\enquote {\bibinfo {title} {{Time-dependent,
  quasi-steady, and global features of fast neutrino-flavor conversion}},}\
  }\href@noop {} {\  (\bibinfo {year} {2022})},\ \Eprint
  {http://arxiv.org/abs/2206.04097} {arXiv:2206.04097 [astro-ph.HE]}
  \BibitemShut {NoStop}%
\bibitem [{\citenamefont {Shalgar}\ and\ \citenamefont
  {Tamborra}(2022{\natexlab{b}})}]{Shalgar:2022lvv}%
  \BibitemOpen
  \bibfield  {author} {\bibinfo {author} {\bibfnamefont {Shashank}\
  \bibnamefont {Shalgar}}\ and\ \bibinfo {author} {\bibfnamefont {Irene}\
  \bibnamefont {Tamborra}},\ }\bibfield  {title} {\enquote {\bibinfo {title}
  {{Neutrino Flavor Conversion, Advection, and Collisions: The Full
  Solution}},}\ }\href@noop {} {\  (\bibinfo {year} {2022}{\natexlab{b}})},\
  \Eprint {http://arxiv.org/abs/2207.04058} {arXiv:2207.04058 [astro-ph.HE]}
  \BibitemShut {NoStop}%
\bibitem [{\citenamefont {Sigl}\ and\ \citenamefont
  {Raffelt}(1993)}]{Sigl:1993ctk}%
  \BibitemOpen
  \bibfield  {author} {\bibinfo {author} {\bibfnamefont {G{\"u}nther}\
  \bibnamefont {Sigl}}\ and\ \bibinfo {author} {\bibfnamefont {Georg~G.}\
  \bibnamefont {Raffelt}},\ }\bibfield  {title} {\enquote {\bibinfo {title}
  {{General kinetic description of relativistic mixed neutrinos}},}\ }\href
  {\doibase 10.1016/0550-3213(93)90175-O} {\bibfield  {journal} {\bibinfo
  {journal} {Nucl. Phys. B}\ }\textbf {\bibinfo {volume} {406}},\ \bibinfo
  {pages} {423--451} (\bibinfo {year} {1993})}\BibitemShut {NoStop}%
\bibitem [{\citenamefont {Mikheev}\ and\ \citenamefont
  {Smirnov}(1986)}]{Mikheev:1986if}%
  \BibitemOpen
  \bibfield  {author} {\bibinfo {author} {\bibfnamefont {S.~P.}\ \bibnamefont
  {Mikheev}}\ and\ \bibinfo {author} {\bibfnamefont {A.~Yu.}\ \bibnamefont
  {Smirnov}},\ }\bibfield  {title} {\enquote {\bibinfo {title} {{Neutrino
  Oscillations in a Variable Density Medium and Neutrino Bursts Due to the
  Gravitational Collapse of Stars}},}\ }\href@noop {} {\bibfield  {journal}
  {\bibinfo  {journal} {Sov. Phys. JETP}\ }\textbf {\bibinfo {volume} {64}},\
  \bibinfo {pages} {4--7} (\bibinfo {year} {1986})},\ \Eprint
  {http://arxiv.org/abs/0706.0454} {arXiv:0706.0454 [hep-ph]} \BibitemShut
  {NoStop}%
\bibitem [{\citenamefont {Wolfenstein}(1978)}]{PhysRevD.17.2369}%
  \BibitemOpen
  \bibfield  {author} {\bibinfo {author} {\bibfnamefont {L.}~\bibnamefont
  {Wolfenstein}},\ }\bibfield  {title} {\enquote {\bibinfo {title} {Neutrino
  oscillations in matter},}\ }\href {\doibase 10.1103/PhysRevD.17.2369}
  {\bibfield  {journal} {\bibinfo  {journal} {Phys. Rev. D}\ }\textbf {\bibinfo
  {volume} {17}},\ \bibinfo {pages} {2369--2374} (\bibinfo {year}
  {1978})}\BibitemShut {NoStop}%
\bibitem [{\citenamefont {Mikheev}\ and\ \citenamefont
  {Smirnov}()}]{osti_5714592}%
  \BibitemOpen
  \bibfield  {author} {\bibinfo {author} {\bibfnamefont {S~P}\ \bibnamefont
  {Mikheev}}\ and\ \bibinfo {author} {\bibfnamefont {A~Y}\ \bibnamefont
  {Smirnov}},\ }\bibfield  {title} {\enquote {\bibinfo {title} {Resonance
  enhancement of oscillations in matter and solar neutrino spectroscopy},}\
  }\href {https://www.osti.gov/biblio/5714592} {\bibinfo  {journal} {Sov. J.
  Nucl. Phys. (Engl. Transl.); (United States)}\ }\BibitemShut {NoStop}%
\bibitem [{\citenamefont {Kim}\ \emph {et~al.}(1988)\citenamefont {Kim},
  \citenamefont {Kim},\ and\ \citenamefont {Sze}}]{Kim:1987bv}%
  \BibitemOpen
\bibfield  {journal} {  }\bibfield  {author} {\bibinfo {author} {\bibfnamefont
  {C.~W.}\ \bibnamefont {Kim}}, \bibinfo {author} {\bibfnamefont {Jewan}\
  \bibnamefont {Kim}}, \ and\ \bibinfo {author} {\bibfnamefont {W.~K.}\
  \bibnamefont {Sze}},\ }\bibfield  {title} {\enquote {\bibinfo {title} {{On
  the Geometrical Representation of Neutrino Oscillations in Vacuum and
  Matter}},}\ }\href {\doibase 10.1103/PhysRevD.37.1072} {\bibfield  {journal}
  {\bibinfo  {journal} {Phys. Rev. D}\ }\textbf {\bibinfo {volume} {37}},\
  \bibinfo {pages} {1072} (\bibinfo {year} {1988})}\BibitemShut {NoStop}%
\bibitem [{\citenamefont {Parke}(1986)}]{Parke:1986jy}%
  \BibitemOpen
  \bibfield  {author} {\bibinfo {author} {\bibfnamefont {Stephen~J.}\
  \bibnamefont {Parke}},\ }\bibfield  {title} {\enquote {\bibinfo {title}
  {{Nonadiabatic Level Crossing in Resonant Neutrino Oscillations}},}\ }\href
  {\doibase 10.1103/PhysRevLett.57.1275} {\bibfield  {journal} {\bibinfo
  {journal} {Phys. Rev. Lett.}\ }\textbf {\bibinfo {volume} {57}},\ \bibinfo
  {pages} {1275--1278} (\bibinfo {year} {1986})}\BibitemShut {NoStop}%
\bibitem [{\citenamefont {Blennow}\ and\ \citenamefont
  {Smirnov}(2013)}]{Blennow:2013rca}%
  \BibitemOpen
  \bibfield  {author} {\bibinfo {author} {\bibfnamefont {Mattias}\ \bibnamefont
  {Blennow}}\ and\ \bibinfo {author} {\bibfnamefont {Alexei~Yu.}\ \bibnamefont
  {Smirnov}},\ }\bibfield  {title} {\enquote {\bibinfo {title} {{Neutrino
  propagation in matter}},}\ }\href {\doibase 10.1155/2013/972485} {\bibfield
  {journal} {\bibinfo  {journal} {Adv. High Energy Phys.}\ }\textbf {\bibinfo
  {volume} {2013}},\ \bibinfo {pages} {972485} (\bibinfo {year} {2013})},\
  \Eprint {http://arxiv.org/abs/1306.2903} {arXiv:1306.2903 [hep-ph]}
  \BibitemShut {NoStop}%
\bibitem [{\citenamefont {Raffelt}\ and\ \citenamefont
  {Sigl}(1993)}]{Raffelt:1992bs}%
  \BibitemOpen
  \bibfield  {author} {\bibinfo {author} {\bibfnamefont {Georg~G.}\
  \bibnamefont {Raffelt}}\ and\ \bibinfo {author} {\bibfnamefont {G{\"u}nther}\
  \bibnamefont {Sigl}},\ }\bibfield  {title} {\enquote {\bibinfo {title}
  {{Neutrino flavor conversion in a supernova core}},}\ }\href {\doibase
  10.1016/0927-6505(93)90020-E} {\bibfield  {journal} {\bibinfo  {journal}
  {Astropart. Phys.}\ }\textbf {\bibinfo {volume} {1}},\ \bibinfo {pages}
  {165--184} (\bibinfo {year} {1993})},\ \Eprint
  {http://arxiv.org/abs/astro-ph/9209005} {arXiv:astro-ph/9209005} \BibitemShut
  {NoStop}%
\end{thebibliography}%
%%%%%%%%%%%%%%%%%%%%%%%%%%%%%%%%%%%%%%%%%%%%%%%%%%%%%%%%%%%%%%%%%%%%%%%%%%%%%%%
%%%%%%%%%%%%%%%%%%%%%%%%%%%%%%%%%%%%%%%%%%%%%%%%%%%%%%%%%%%%%%%%%%%%%%%%%%%%%%%
\end{document}